\documentclass[a4paper,12pt,onecolumn, authoryear]{elsarticle}
\usepackage[english]{babel}
\usepackage{subcaption} 
\usepackage{multirow} 
\usepackage{float} 
\usepackage{amssymb}
\usepackage[intlimits]{amsmath} 
\usepackage{amsthm} 
\usepackage{booktabs} 
\usepackage{geometry}
\usepackage{enumitem}
\usepackage{array}
\usepackage[usenames,dvipsnames]{color}
\usepackage{tikz}

\definecolor{b}{rgb}{0,0,.8}	
\definecolor{g}{rgb}{0,.6,0}	
\definecolor{n}{rgb}{0,0,0}	
\definecolor{h}{rgb}{0.4,0.2,0.2}	
\definecolor{v}{rgb}{0.2,0.6,0}

\newcommand{\C}{{\mathbb C}}

\newcommand{\E}{{\mathbb E}}

\newcommand{\N}{{\mathbb N}}

\renewcommand{\P}{{\mathbb P}}

\newcommand{\R}{{\mathbb R}}

\newcommand{\V}{{\mathbb V}}

\newcommand{\X}{{\mathbb X}}

\newcommand{\Z}{{\mathbb Z}}

\newcommand{\DD}{{\mathcal{D}}}
\newcommand{\EE}{{\mathcal{E}}}

\newcommand{\II}{{\mathcal{I}}}

\newcommand{\NN}{{\mathcal{N}}}

\newcommand{\WW}{{\mathcal{W}}}

\newcommand{\bsb}{\boldsymbol b}

\newcommand{\bsX}{\boldsymbol X}
\newcommand{\bsY}{\boldsymbol Y}

\newcommand{\bsnull}{\boldsymbol 0}

\newcommand{\bsbeta}{\boldsymbol \beta}

\newcommand{\bsmu}{\boldsymbol \mu}
\newcommand{\bseps}{\boldsymbol \varepsilon}

\newcommand{\eps}{{\varepsilon}}

\DeclareMathOperator*{\argmin}{arg\,min}

\newcommand{\ov}\overline
\newcommand{\what}{\widehat}
\newcommand{\wtilde}{\widetilde}

\newcommand{\rig}\right
\newcommand{\lef}\left
\newcommand{\nf}\normalfont

\definecolor{rickgreen}{rgb}{0,0.6,0}

\begin{document}
\title{Electricity Price Forecasting using \\ Sale and Purchase Curves:\\ The X-Model}

\author{Florian~Ziel}
\ead{ziel@europa-uni.de}

\author{Rick~Steinert}
\ead{steinert@europa-uni.de}

\address{Europa-Universit\"at Viadrina, Gro\ss e Scharrnstra\ss e 59, 15230 Frankfurt (Oder), Germany}

\begin{frontmatter}
\begin{abstract}
Our paper aims to model and forecast the electricity price by taking a completely new perspective on the data. It will be the first approach which is able to combine the insights of market structure models with extensive and modern econometric analysis. Instead of directly modeling the electricity price as it is usually done in time series or data mining approaches, we model and utilize its true source: the sale and purchase curves of the electricity exchange. We will refer to this new model as X-Model, as almost every deregulated electricity price is simply the result of the intersection of the electricity supply and demand curve at a certain auction. Therefore we show an approach to deal with a tremendous amount of auction data, using a subtle data processing technique as well as dimension reduction and lasso based estimation methods. We incorporate not only several known features, such as seasonal behavior or the impact of other processes like renewable energy, but also completely new elaborated stylized 
facts of the bidding structure. Our model is able to capture the non-linear behavior of the electricity price, which is especially useful for predicting huge price spikes. Using simulation methods we show how to derive prediction intervals for probabilistic forecasting. We describe and show the proposed methods for the day-ahead EPEX spot price of Germany and Austria.

\end{abstract}
\end{frontmatter}

\section{Introduction} \label{Introduction}
In the recent decades modeling electricity prices have become a complex and broad field of research. Due to the liberalization of markets and increasing disclosure of data, new insights concerning the structure and behavior of the prices were gained. {Researchers pointed out that there are typical characteristics of electricity prices regardless where it has been traded.} These are summarized as the stylized facts of electricity prices, see e.g. \cite{weron2006modeling}. One of these stylized facts concerns tremendous deviations of the price pattern from its mean, called price spikes. This specific feature of electricity prices has huge impacts for research as well as politics and companies. Many electricity companies, e.g. in Germany, are obliged to market some of their electricity at an exchange, which makes their earnings prone to heavy price spikes and creates a complex task for their risk management department. Moreover, many financial contracts such as futures or options are dependent on the 
variance of 
the price process and therefore demand eligible estimation techniques. Also long-term cost calculation for investment projects or political programs like the development of renewable energy are dependent on stable and reliable methods for calculation of electricity prices, which can account for the likelihood of price spikes.
\par
Therefore, a great variety of models for estimating the electricity price occurred during the past decades. Those models are often related to well-known models of the finance literature but can originate from many other fields of research. \cite{weron2014electricity} for instance divides electricity price models into five different groups, multi-agent, fundamental, reduced-form, statistical and computational intelligence models. Besides the multi-agent {and fundamental} approaches all models have in common that they focus on the price itself or related time series like renewable energy or electricity demand. Multi-agent models usually focus on the supply and demand of electricity to obtain prices by equilibrium, optimization or simulation (\cite{ventosa2005electricity}, \cite{liu2012multi}), but hence often do not incorporate the time-series of electricity bids and asks of a real exchange into their approaches. {Fundamental approaches cover a great variety of models but mainly emphasize the basic 
economic and physical relationships of the market \citep{weron2014electricity}.} 

\par
Concerning price spikes, the distinction between different model approaches can be refined when the explicit or implicit incorporation of price spikes is considered. In the area of time series models the usage of specific heteroscedastic models for the variance of the process are typical (e.g. \cite{bowden2008short}, \cite{liu2013applying}). But standard GARCH-type models cannot account for all of the extreme price events within the data (\cite{swider2007extended}). Hence, many researcher developed extended models which can account for severe price movements. These models commonly fall into two main categories. First, there are regime-switching models, which introduce different regimes, usually a base and a spike regime, with different probabilities for a price spike to occur 
(see, for instance \cite{karakatsani2008forecasting}, \cite{janczura2012efficient}, \cite{eichler2013fitting}). Second, there are diffusion models, which add a jump component, e.g. a Poisson process, to allow for price spikes (see, for instance \cite{weron2008market}, \cite{escribano2011modelling}). Rarely there are approaches which focus solely on the price spike itself and try to forecast the event without modeling the whole price time series, e.g. in \cite{christensen2012forecasting}. 
\par 
However, all of these approaches for modeling price spikes have in common that they are focused mainly on the price time series and not of the underlying mechanic which determines the price process. The electricity price can also be seen as the intersection between the part of the electricity supply and demand which was traded at an exchange. The resulting sale and purchase curves, which are also referred to as ask and bid curves or {market} supply and {market} demand curves, contain all the information which is needed to determine the market price but provide even further information on all the other prices for other market volumes. This information can be necessary especially for the estimation of the likelihood of extreme price events, as the elasticity of the price, which can be obtained from the shape of the sale and purchase curves, vastly accounts for price movements. 

\par
{But even though a time-series approach for modeling and especially forecasting auction data is relatively new and has not been applied for electricity price data in a comprehensible manner, modeling the structure of the supply and demand curves in general has been done by some authors, even if very little of them do utilize real auction data. Most of these models belong to the field of fundamental models, but are also often referred to as structural models, as they try to capture the structure of the market. Many of them originate from the field of derivative pricing and do not focus on forecasting the electricity price itself and therefore avoid the uncertainties which come along with it. \cite{barlow2002diffusion} is one of the first authors in electricity price research who formulates a model motivated by real auction data of an electricity market. In his paper he uses a non-linear Ornstein-Uhlenbeck process to obtain a realistic image of the true underlying price process and is also able to capture 
extreme price events. In the book of \cite{eydeland2003energy} in chapter 7 a basic market model approach which maps the energy supply 
to the price of electricity is introduced. They make use of the structure of the market by constructing the so called bid stack, which refers to the marketed aggregated supply of energy for different prices and should, in theory, be equivalent to the sale curve at the investigated auction market.\footnote{{We want to point out that the bid stack is not necessarily the same as the energy supply, as the bid stack includes also e.g. bidding behavior. For more information on this we suggest to read \cite{eydeland2003energy}}} {Given the specific cost functions of energy generators they are able to determine the bid stack function and afterwards the system price of electricity. Another promising approach arose in the working paper of \cite{buzoianu2005dynamic}, who model the marketed supply and demand curves. They assume a linear demand function and a nonlinear supply function to construct a price-quantity model, where the intersection of both curves equals the market clearing price. To approximate the 
market 
curves they use external factors like temperature, gas energy supply and gas price. \cite{boogert2008supply} use a market structure approach which includes the relationship of electricity demand to available capacity to forecast electricity prices and the probability of spikes for the Dutch electricity market. 
Another structural approach can be found in \cite{howison2009stochastic} and \cite{carmona2013electricity} who perform an analysis of the sale and purchase structure and integrate some of its aspects by incorporating the bid stack model. Extensions to basic structural models are often done via the introduction of market specific determinants, as for instance the solar and wind power feed-in as done by \cite{wagner2014residual} or $\text{CO}_2$-emissions as done by \cite{hendricks2013clean}. 

Some of the recent approaches try to capitalize the increasing amount of available data, especially the hourly auction data of the EPEX, which allows for a deep analysis of the real offered volumes for selling and purchasing electricity. As this results usually in a large amount of data and therefore complexity, some researcher tried to simplify the resulting market curves by merging them into a new curve with desirable properties. For instance, \cite{eichler2012new} illustrate in an extended abstract an idea for modeling 
the 
German/Austrian EPEX price using the supply/demand curves. They 
utilize the curves to model a scaled supply and demand spread using an autoregressive time series model with weekday effects. \cite{coulon2014hourly} try to overcome the common issue of the assumption of inelastic demands by constructing a ``price curve'' out of the marketed supply and demand curve for the same hour. The resulting curve exhibits many well-known typical behavioral attributes, e.g. weekday effects. The price curve is then matched with a pseudo-demand curve, which is again a vertical line, where the intersection of both results in the market clearing price. {
A related approach is used by \cite{aneiros2013functional} for the Spanish electricity market.
They consider a functional modelling approach for 
a similar price curve as defined in \cite{coulon2014hourly}, but call it ``residual demand curve''. 
However, in electricity price research the term residual demand curve is usually more common in the framework of market and bidding behavior (as in \cite{hortacsu2008understanding}, \cite{vazquez2014residual} or \cite{portela2016residual}). 
}
\cite{hildmann2015empirical} analyze empirically the impact of renewables to the real auction data of the EPEX, if they were not subsidized by the government. For instance, by manipulating the marketed supply curve accordingly they show that negative prices diminish completely when the wind power feed-in is marketed at its true marginal costs. A more detailed survey on structural models can be found in \cite{carmona2014survey}.}}
\par 

All of these papers have in common, that they exhibit at least one of the following major drawbacks. They do not incorporate real auction data (e.g. \cite{boogert2008supply}), they assume, that the demand is inelastic and therefore focus only on the bid stack (e.g. \cite{eydeland2003energy}, \cite{howison2009stochastic}, \cite{carmona2013electricity})
\footnote{The assumption of an inelastic demand can be justified for some markets. In the case of the electricity market for Germany and Austria on the contrary, where a large proportion of trading is done between different energy companies on a national and international basis, the assumption of inelastic demand is not realistic.},  they use simplifications or modulations which skip the important correlation structure between bids (e.g. \cite{buzoianu2005dynamic}, \cite{coulon2014hourly}) or they are not properly adjusted for forecasting real electricity prices (e.g. \cite{barlow2002diffusion}). Besides electricity price research an econometric time-
series approach which actually covers the  contemporaneous nature of functionally related 
 and time-dependent auction data can be found in \cite{bowsher2004modelling}, who applies  a functional signal plus noise time series model to a security of the FTSE100.

\par

\par
Our idea {aims to fill the gap between research done in time-series analysis, where the structure of the market is usually left out and the research done in structural analysis, where empirical data is utilized very rarely and even less thoroughly. It is especially new in the sense that it gets the best of both ends, it will provide deep inside on the bidding behavior of market participants, while still remaining a high accuracy in probabilistic forecasting of the market price.} We will therefore use the true data generating process, e.g. the sale and purchase curves of the electricity price, to provide better {probabilistic} forecasts for extreme price movements while still modeling the time series of electricity prices by an autoregressive approach. We will use the hourly day-ahead electricity price auction data of Germany and Austria provided by the EPEX Spot, also known as Phelix. It will be shown that incorporating the sale and purchase data yields promising results for forecasting the 
likelihood of extreme price events. Within our approach we will be able to estimate the full prediction density of electricity prices. 
\par 
Our paper is organized as follows. The next section focuses on our idea and will describe the data and our observations for the EPEX Spot day-ahead auctions. We will follow up with a detailed description of our model and its specific setup for the auction data. Afterwards we show the empirical results of our approach. Our last section discusses our findings and will provide insights for possible improvements and future research. During the paper we will use the phrase ``price curves'' for both, the sale and purchase curve.
Every price will be provided in EUR/MWh and every volume in MW, if not specified otherwise.
 {Note that the market clearing volume is reported by the EPEX as energy in MWh. As we will only consider hourly 
 data we denote the volume in MW.} 



\section{Price formation process and price curves structure}
The electricity price of exchanges is the result of competitive bidding and offering. Focusing merely on the time series of prices therefore neglects their true source. If the true sale and purchase curves were known, the price could be solely determined by the intersection of both curves - regardless of any time dependencies between different prices. Many authors point out that the price is driven by external factors, e.g. wind and solar or electricity demand, see for instance \cite{weron2014electricity}. However, taking a closer look on the underlying price process, it can be stated that it is the buyers and sellers on an electricity exchange who are influenced by those factors and therefore adjust their bids. Reasons for that can be e.g. that these market participants are electricity companies who are facing heavy overproduction of electricity due to an unexpected change in wind speed or temperature or an underproduction due to outages of power plants. 
\par
But those market participants are not equal, they can be investment companies, electricity producers or transmission service operators, among others. Also not all electricity producers are equal - they have distinct production portfolios and are therefore more or less likely prone to e.g. heavy weather conditions. An unexpected shift in wind production levels for instance can therefore lead to a little or vast change in prices, dependent on if the equilibrium price of the market was already mainly driven by wind producers. This diversified information is summarized in the sale and purchase curve of electricity prices. Hence, especially for estimating heavy price movements it is essential to know, if the market is capable of adjusting for external shocks easily or if a tremendous price spike will occur. This sensitivity of the intersection price can therefore be obtained by analyzing the original price curves instead of only their outcome as price time series. To motivate 
our 
idea even further, we decided on showcasing the day-ahead price of the 12.04.2015 of the EPEX Spot for Germany and Austria. We will use this day throughout our whole paper, as it provides easily traceable insights for the typical price movement process when an extreme price spike occurs.
\par
\begin{figure}[htb!]
\centering
 \includegraphics[width=.99\textwidth, height=.42\textwidth]{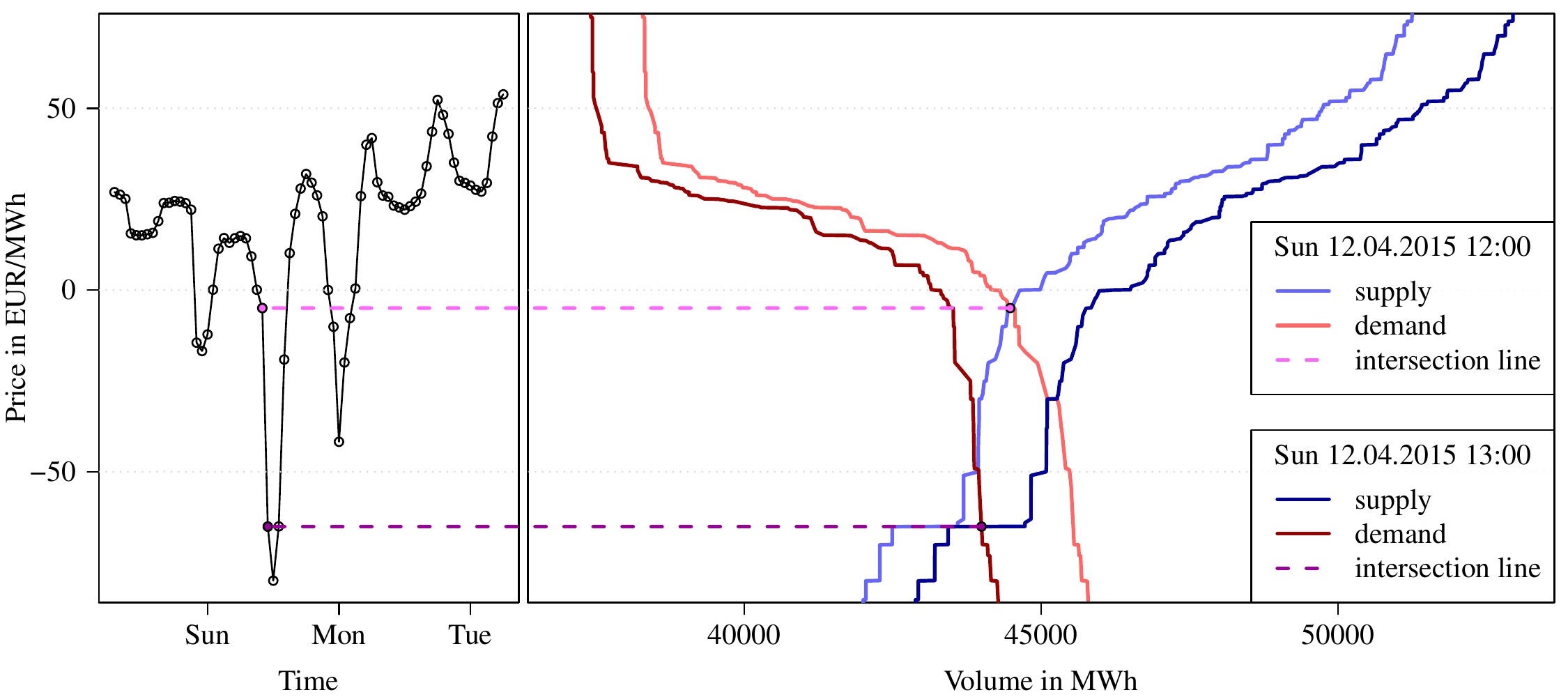} 
 \includegraphics[width=.99\textwidth, height=.42\textwidth]{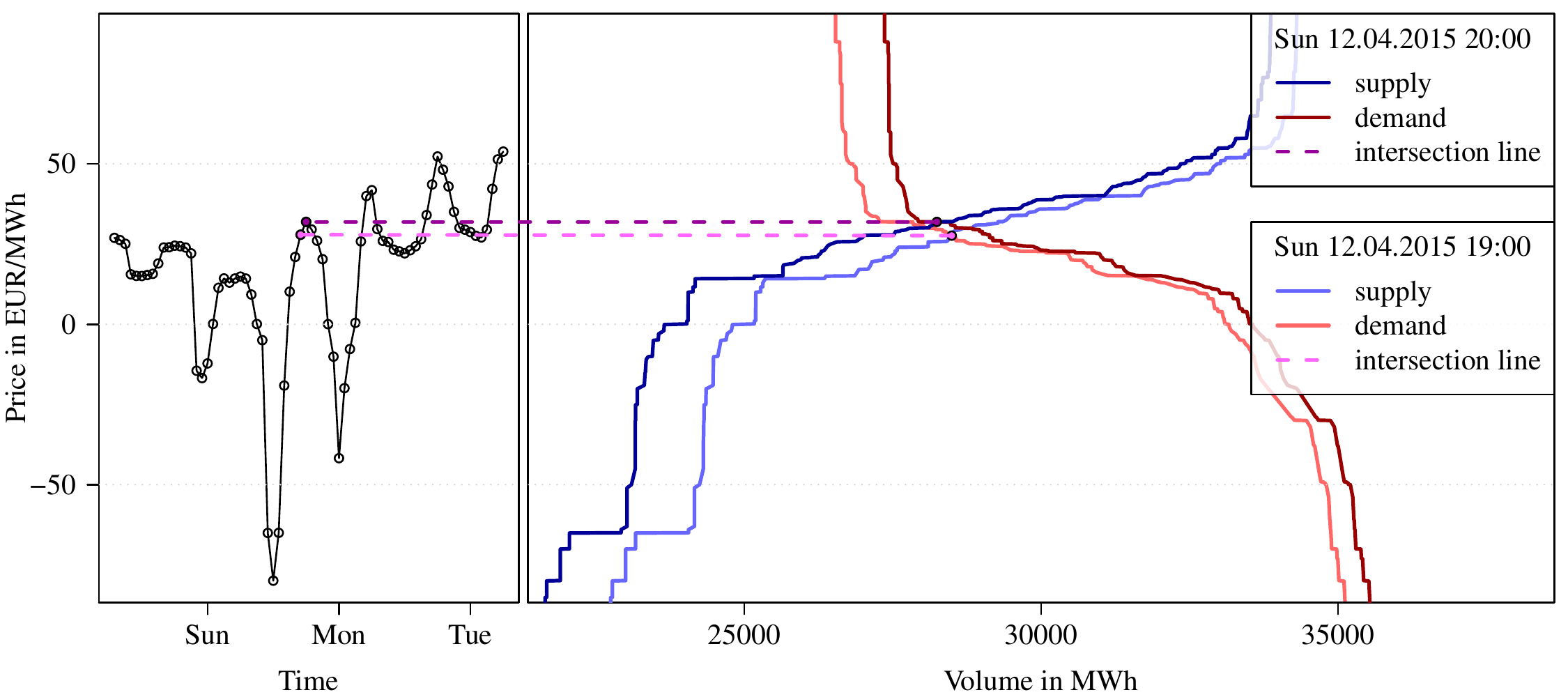} 
\caption{Market clearing price time series with supply and demand curves for 12:00 and 13:00, and 19:00 and 20:00 hours resp. on 12.04.2015 auctions.}
 \label{fig_intro}
\end{figure}
The left-hand side of Figure \ref{fig_intro} shows the day-ahead electricity price of the 12.04.2015. In the upper and lower right-hand side the price curves for 12:00 and 13:00 and 19:00 and 20:00 respectively are provided. The horizontal axis of this area represents the trading volume and the vertical axis the price. It is shown that during the afternoon hours the electricity price heavily declined reaching even negative values. Examining the price curves for 12:00 that day on the upper right-hand side two typical phenomena for such an observation can be seen. First, the traded volume is, in comparison to other hours that day, relatively high. Second, the slope of the demand curve for any price and volume combination with a lower price than the market price is extremely negative. Simultaneously, the slope of the supply curve is extremely positive for price and volume combinations with a lower price than the market price - at least for price combinations close to the actual market price. Monitoring the left-
hand side of the figure shows that the price exhibited a tremendous price decline from 12:00 to 13:00. Taking into account the phenomena mentioned beforehand gives insights on why such a heavy price spike was even possible. The high amount of supplied electricity shifted the price to a level, where usually only a relatively small proportion of bids can be found, e.g. the supply and demand curves exhibited { high ``steps'' (i.e. in horizontal)}. Those ``steps'' result in the second observation of curves having extreme negative or positive slopes close to the market price. This in turn indicates, that the equilibrium price is very sensitive to external shocks. Any sudden decrease in demand which would lead to a left-shift of the demand curve or any sudden increase in production which would lead to a right-shift of the supply curve has a great impact on the price - especially in comparison to other, higher price levels. But we can also see that the supply curve for 13:00 exhibits a slope of almost zero 
around 
the 
intersection price, indicating that any further decrease in demand or increase in supply will not have the same vast effects than before. And indeed, the price movement from 13:00 to 14:00 was much smaller than the one from 12:00 to 13:00. In contrast, the price curves for 19:00 and 20:00 on the lower right-hand side of the figure show the typical behavior of price curves when the market is not 
very prone to extreme events. The slopes of the price curves right from the intersection price seem not to have extremely positive and negative slopes respectively. Only the demand curve left of the intersection price seem to have a very negative slope, but matching it with the supply curve it can be seen that any shifts to the right or left will be captured by the supply curve easily and can therefore not result in a heavy price spike. \par
Under the assumption that not only the price but also the price curves are dependent on time, we will derive a model which is tailor-made for the day-ahead auction market of the EPEX Spot in Germany and Austria for the purpose of estimating the likelihood of heavy price movements. Therefore and in order to introduce our model we need to take a closer look at the EPEX Spot market and the observed bidding structure of their participants. 
\par 
For our summary statistics {and all computation results up to section 4} we use the data from 01.10.2012 to 19.04.2015. 
However, all techniques can be applied to other electricity markets in exactly the same way but under considering of their corresponding market features.

The day-ahead electricity spot price of the EPEX will be traded in daily auctions at 12:00 CET for 
the hours of the next day. So there are in general 24 prices everyday.  Due to the daylight saving time we 
have once a year 23 values in March and 25 values in October. For the 24 auctions on a common day we use the labels 0:00, 1:00, $\ldots$, 23:00 within this paper. 

\begin{figure}[htb!]
\centering
 \includegraphics[width=.99\textwidth]{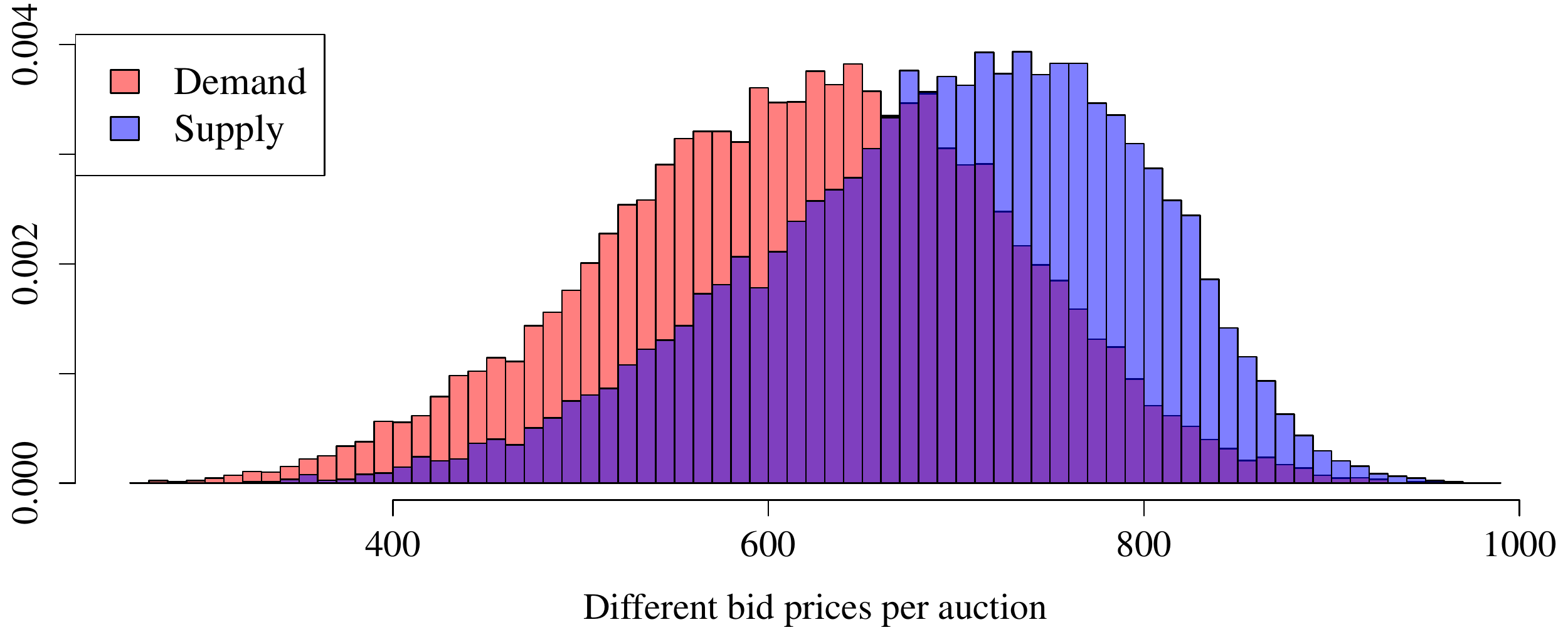} 
\caption{Amount of different bid prices for the auctions.}
 \label{fig_bid_hist}
\end{figure}

Since 2008 the electricity spot price is set to be between $P_{\min}= -500$ and $P_{\max}= 3000$. 
Before that there were no negative prices allowed. 
The traders at the EPEX can make bids for either selling or buying a certain amount of electric power.
By the EPEX regulations, the minimal order size for Germany and Austria is 0.1 MW for a one hour block and the minimal price difference between different orders is 0.1 EUR/MWh. Hence, there are in total 35001 different possible prices on the full price grid $\P = \{-500,-499.9, \ldots, 2999.9, 3000\}$. But in practice not every of those possible prices is utilized. Also, a single trader can only submit up to 256 distinct price and volume combinations as offers. 
Usually there are about 700 different featured prices which construct each curve, which we depicted in Figure \ref{fig_bid_hist}. The illustration shows the histogram for the amount of different prices for both, the demand and the supply curve of electricity at the EPEX Spot. It can be seen that the range of different prices covers approximately 200 to 1000, depending on which side of the market is considered.
In general we have slightly more different prices for the supply side.
In total there were about 31000 bids on distinct prices within the considered period.

{Moreover, {there are other order types allowed, e.g.} {standard block orders, linked block orders and exclusive block orders. 
The underlying market coupling algorithm to derive the market clearing price from all submitted bids is very complex. Since February 2014 the EUPHEMIA (acronym of Pan-European Hybrid Electricity Market Integration Algorithm) is used to compute the market clearing price and volume by maximizing the total welfare. This involves a market coupling of several European markets like the EPEX, APX, Nordpool and OMIE. {However, the EPEX provides only the sale and purchase curves as illustrated in Figure \ref{fig_intro}, not every single bid. Solely this dataset is used for our empirical supply and demand analysis.} {Thus we assume indirectly that{, given a certain hour,} {all underlying bids are standard bids (single 1-hour block). This implies that we neglect the specific impact of potential complex bids like block orders.
}}}}

%

\begin{figure}[htb!]
\centering
\begin{subfigure}[b]{.49\textwidth}
 \includegraphics[width=1\textwidth]{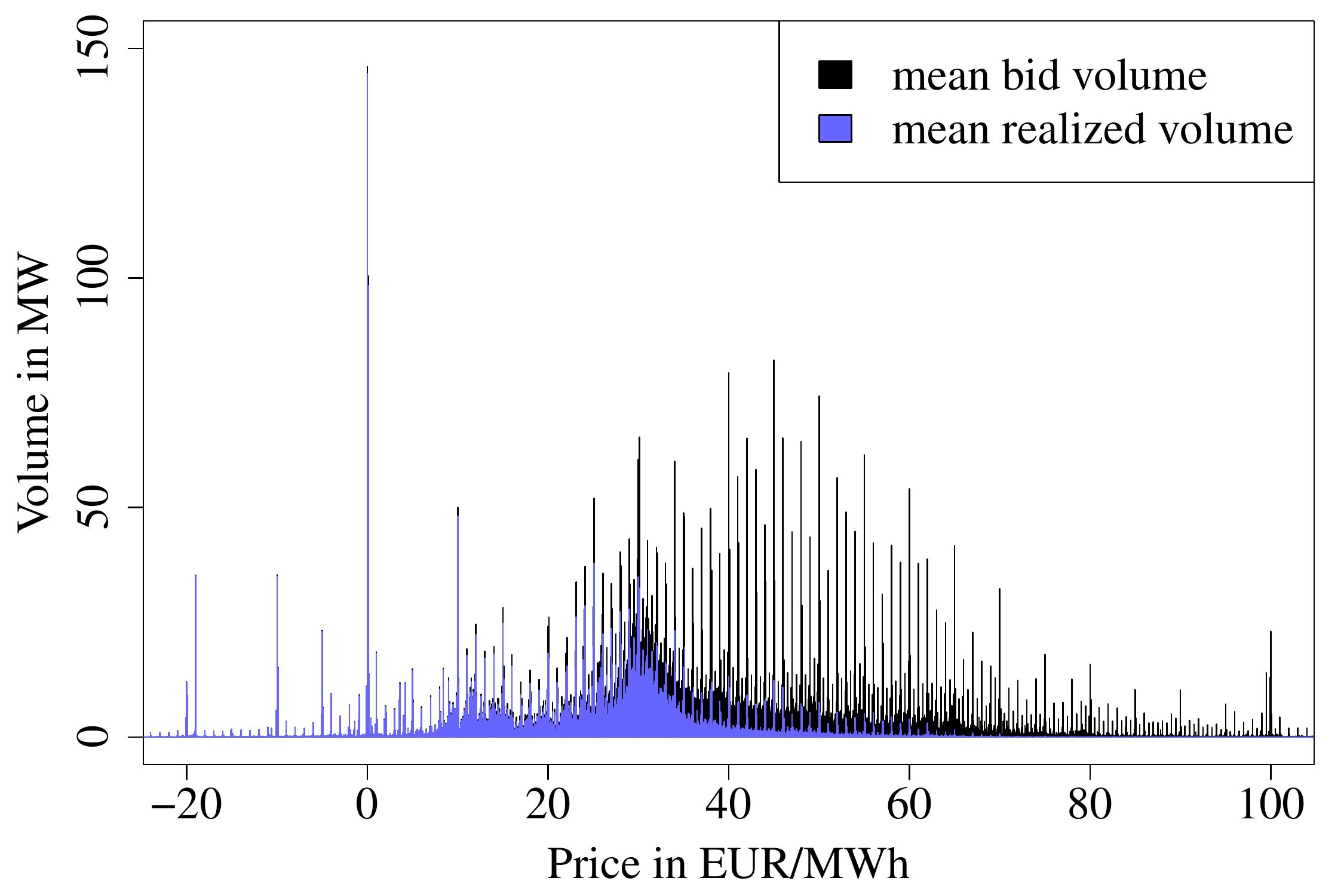} 
  \caption{Bid and realized volumes: supply}
  \label{fig_dm_3}
\end{subfigure}
\begin{subfigure}[b]{.49\textwidth}
 \includegraphics[width=1\textwidth]{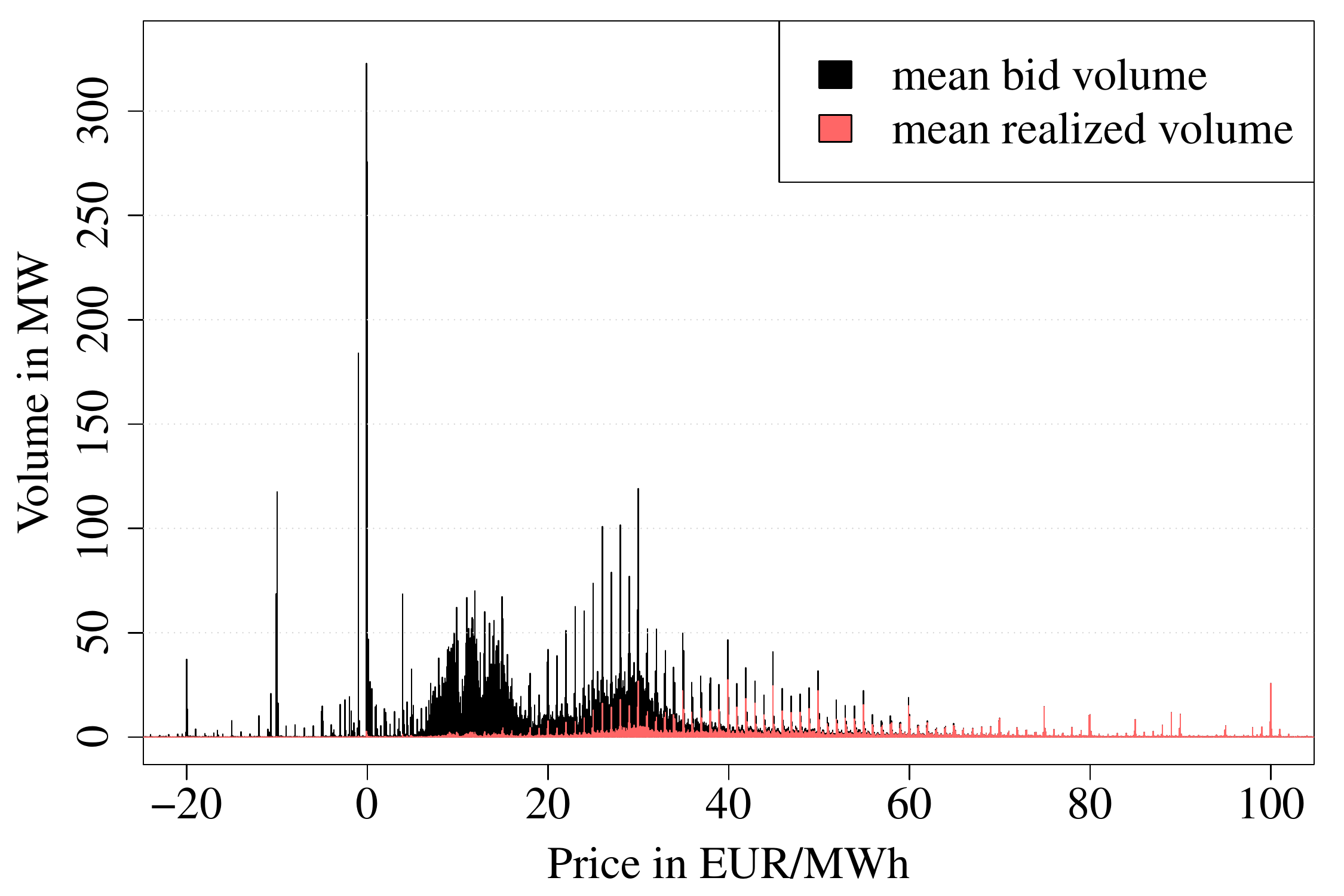} 
  \caption{Bid and realized volumes: demand}
  \label{fig_dm_4}
\end{subfigure}
\caption{Average bid volumes $\ov{V_S}(P)$ and $\ov{V_D}(P)$ for $P\in \P$ in the price range -20 to 100.}
 \label{fig_bid}
\end{figure}

As mentioned before, most of the prices in the price grid $\P$ where not bid at all. 
{ Nevertheless, given the price grid $\P$ we can explore the bid supply and demand volumes $V_{S,t}(P)$ and $V_{D,t}(P)$ at price $P\in \P$ and time $t$.
Later on we will see that especially the prices with an actual bid at time $t$ play an important role in the price coupling algorithm for the shape of the sale and purchase curves. 
Therefore we introduce with $\P_{S,t}$ and $\P_{D,t}$ the bid prices on the supply and demand side at time point $t$. Obviously they are defined to be all prices
with a positive bid volume
\begin{equation}
\P_{S,t} = \{ P \in \P | V_{S,t}(P) > 0 \} \ \ \text{and} \ \
\P_{D,t} = \{ P \in \P | V_{D,t}(P) > 0 \} 
\label{eq_price_grid_t} .
\end{equation}

When the bids $V_{S,t}$ and $V_{D,t}$ are aggregated the well-known price curves for a certain hour can be constructed, which maps a certain amount of supply or demand to a certain price.
The aggregated supply and demand volumes $V_{S,t}(P)$ with $P\in \P_{S,t}$ and $V_{D,t}(P)$ with $\P_{D,t}$ match exactly the 
corresponding points at the sale and purchase curve illustrated in the Figure \ref{fig_intro}.
Mathematically precise the sale and purchase curves are characterized by
\begin{equation}
S_t(P) = \sum_{ \substack{p \in \P_{S,t} \\ p\leq P}} V_{S,t}(p)  \text{ for } P \in \P_{S,t} \ \ \text{and} \ \
D_t(P) = \sum_{ \substack{p \in \P_{D,t} \\ p \geq P}} V_{D,t}(p)   \text{ for } P \in \P_{D,t} 
\label{eq_price_curve} .
\end{equation}
However, equation \eqref{eq_price_curve} defines the supply curve explicitly only on the price grids $\P_{S,t}$ and $\P_{D,t}$. 
As mentioned, according to the operational rules of the EPEX the market clearing price is determined by the EUPHEMIA algorithm which involves complex orders as well.
Nevertheless, it is assumed by the EPEX that the relation of two different bid price and quantity combinations of one market participant is linear. 
Therefore and to simplify the used algorithm we will use linear interpolation between to different price and quantity combinations given by 
$\{ (S_t(P), P) | P \in \P_{S,t} \}$ and $\{ (D_t(P), P) | P \in \P_{D,t} \}$ for the supply and demand. 
The market clearing price will be calculated by the resulting intersection of both price curves{, rounded to two decimal places}. 
Consequently there is sometimes a small but rather negligible difference to the true market price.  
{In particular in 64\% of all cases this price matches the market coupling price, in 89\% of all cases the difference is less than 0.1 EUR/MWh which is the smallest 
bidding unit and in 99.8\% this difference is less than 1 EUR/MWh.
} 
This fact is managed by certain rules for the traders, so that the amount of volume
of the market clearing price must be delivered.

To understand the characteristics of the bid volumes $V_{S,t}(P)$ and $V_{D,t}(P)$ better we ignore the time dependency in a first step.
We evaluate the mean bid volumes 
\begin{equation}
\ov{V_S}(P) = \frac{1}{T}\sum_{t=1}^T V_{S,t}(P) \ \ \text{ and }  \ \ \ov{V_D}(P) = \frac{1}{T}\sum_{t=1}^T V_{D,t}(P) 
\label{eq_mean_volume}
\end{equation}
 for the supply and demand side given a $P\in\P$
 and the number of observations $T$ across all hours in the database. }
Figure \ref{fig_bid} shows the average bid volume of supply and demand volumes $\ov{V_S}(P)$ and $\ov{V_D}(P)$ with $P\in \P$ 
within the price range of -20 to 100.
Additionally we highlighted the realized amount. We can observe that for the supply side almost all 
bid low prices were realized whereas for high prices only a few were. This relation is reversed for the demand side. We also observe some patterns in the bidding behavior of some traders. For example, we have quite large volumes at a price of 0, but very small amounts for e.g. 0.3 or -0.3.
Furthermore we have spikes at multiples of 5, so e.g. 70.0 has a larger value than 69.0 or 71.0. 
This shows that agents seem to prefer to bid on round numbers. This may give indication towards the assumption that at 
least a noticeable amount of trading is done by human decision and not based on algorithmic trading rules.


The large bid volumes at a certain price leads to a price cluster around this price. 
The most intense price cluster can be found at a value of zero, which can be retrieved from Figure \ref{fig_bid}. Even in the small time frame shown in Figure \ref{fig_intro} there are four realized electricity price values very close to zero. 
For the auction at 12:00 we see that the high possibility for a price of zero is mainly driven by the supply side.
Figure \ref{fig_intro} also shows that for the auction at 12:00 there is another price cluster at -65, which is again driven by the supply side.
And again, for the realized price we can observe two values very close to -65, e.g. the realized price for 13:00 is exactly -65.02.

\begin{figure}[htb!]
\centering
 \includegraphics[width=.99\textwidth]{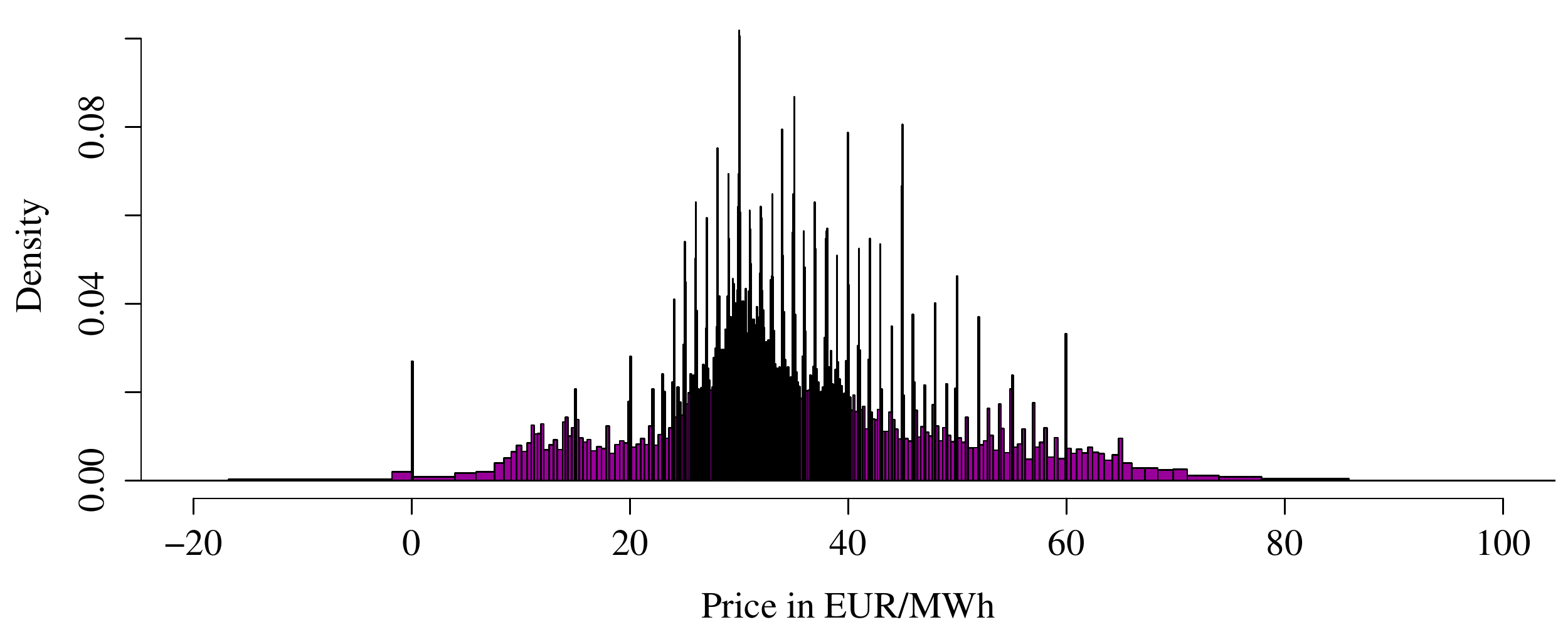} 
\caption{Histogram of realized prices (market clearing prices) in the price range -20 to 100.}
 \label{fig_price_hist}
\end{figure}

In Figure \ref{fig_price_hist} we see the histogram of all market clearing prices. To visualize the 
price cluster effect it is important to choose tiny histogram bins. Here every tiny rectangle represents about $0.33\%$ of the probability mass.
We observe a very spiky histogram 
that exhibits some properties of the bid prices in Figure \ref{fig_bid}. 
For example we see clearly a price cluster 
at zero. Here the relative frequency that the market clearing price is between -0.5 and 0.5 is 
with 0.634\% relatively large. In contrast, the relative frequency to get a price in the neighboring intervals of the same size from -1.5 to -0.5 and 0.5 to 1.5 is only 0.079\% and 0.056\%, so about 10 times smaller. Other price clusters can again be found at all full integers between 
10 and 60, where those clusters that are divisible by 5 are more distinct.  In general the density of the electricity 
price is complicated due to its multi-modal shape. The modi are at the mentioned price clusters. As
far as we know there is no model in electricity price modeling that at least tries to capture this behavior. In contrast to that, our modeling approach will incorporate this effect and thus try to capture the true market behavior more realistically.

\section{Model for the supply and demand curve}
\begin{figure}[htb!]
\centering
\begin{subfigure}[b]{.495\textwidth}
 \includegraphics[width=1\textwidth, height=.65\textwidth]{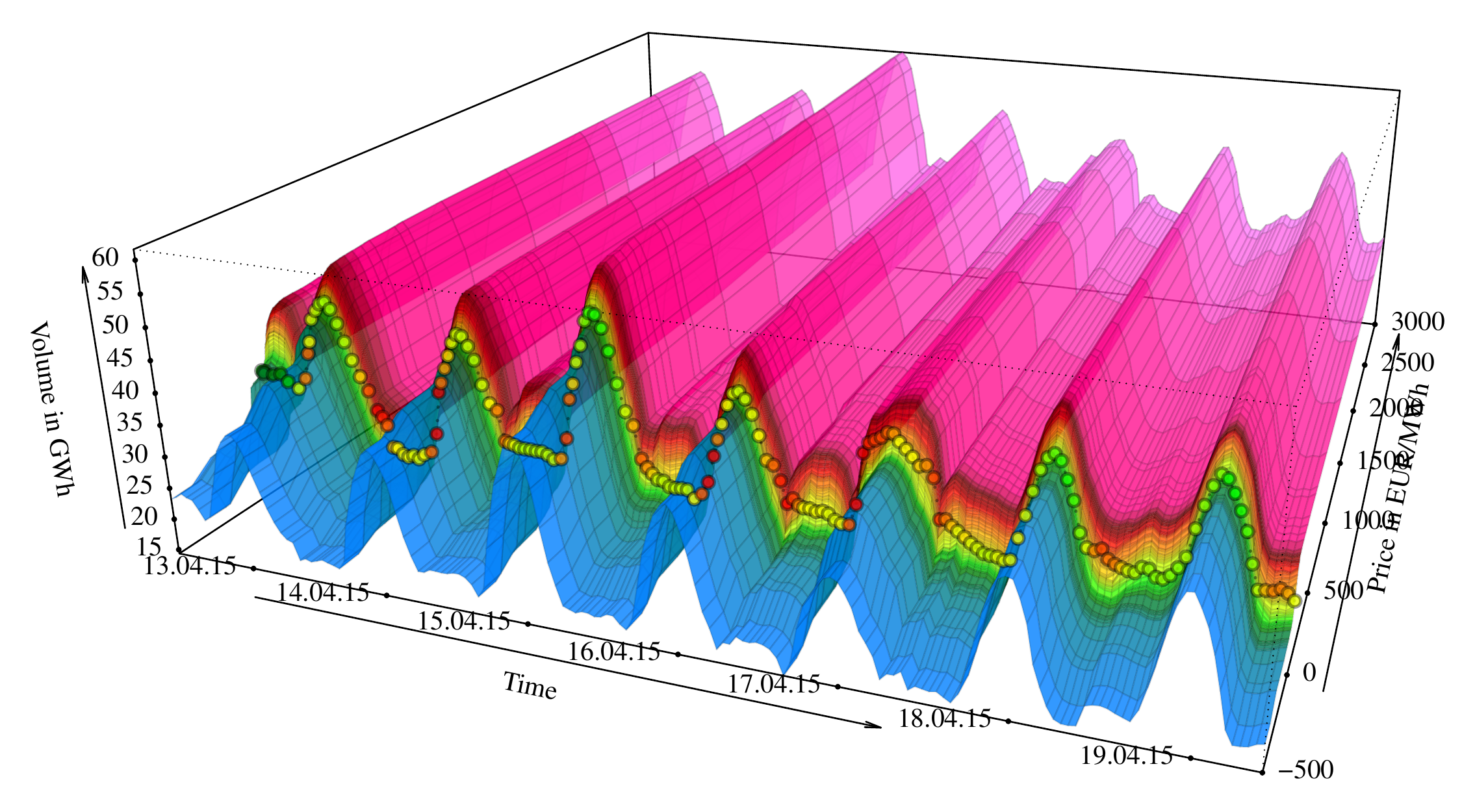} 
  \caption{Supply from -500 to 3000 in EUR/MWh }
  \label{fig_3d_1}
\end{subfigure}
\begin{subfigure}[b]{.495\textwidth}
 \includegraphics[width=1\textwidth, height=.65\textwidth]{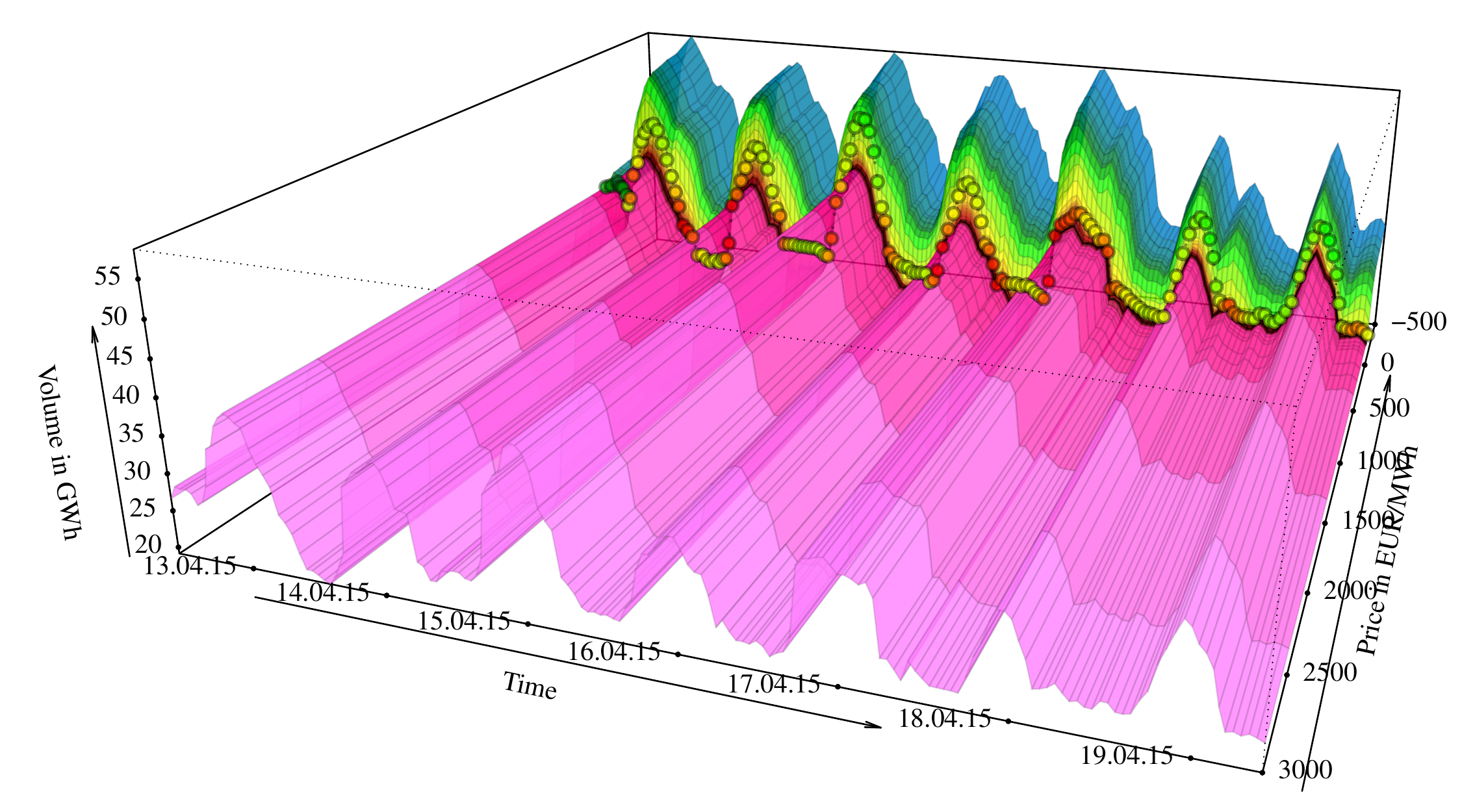} 
  \caption{Demand from -500 to 3000 in EUR/MWh}
  \label{fig_3d_2}
\end{subfigure}
\begin{subfigure}[b]{.495\textwidth}
 \includegraphics[width=1\textwidth, height=.65\textwidth]{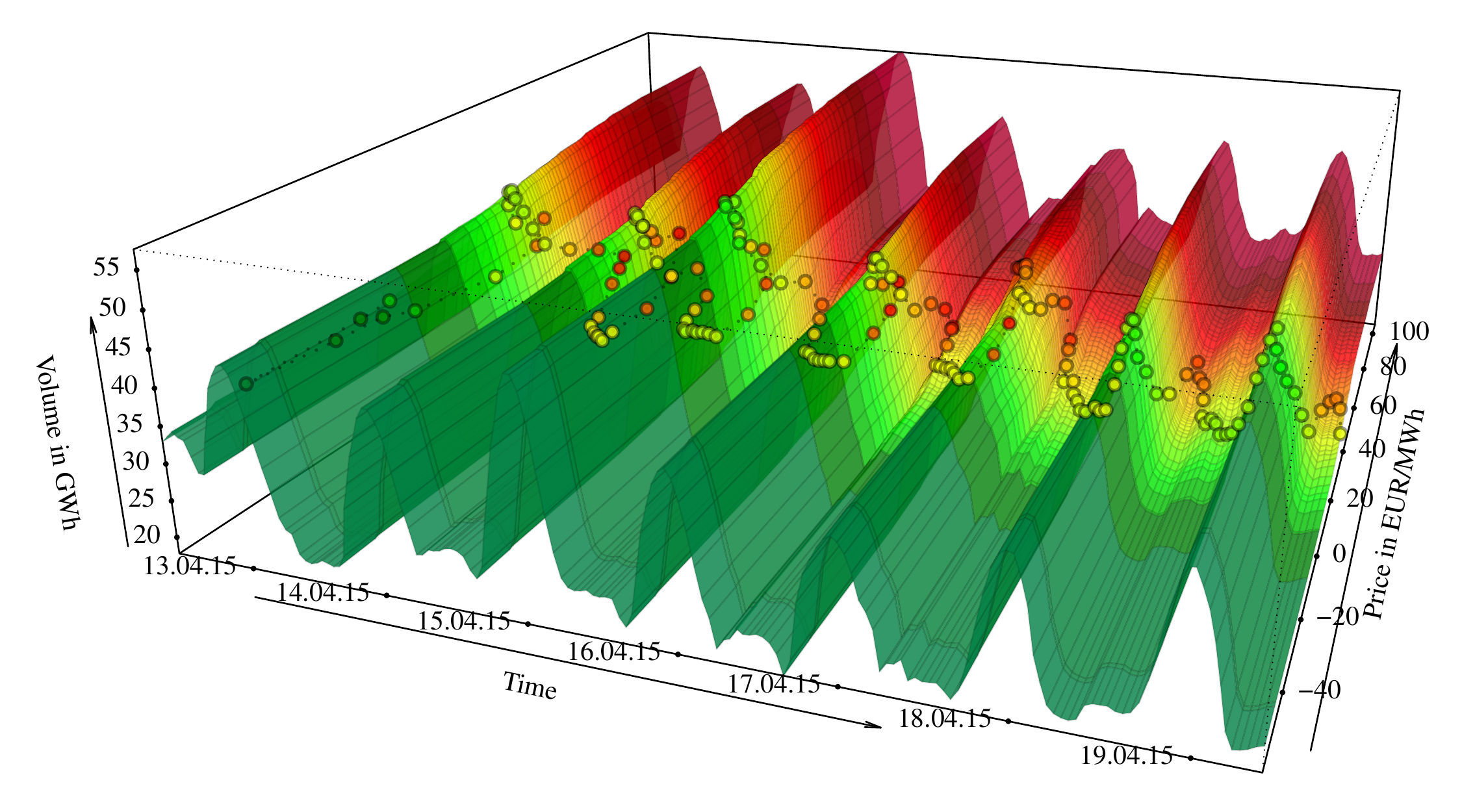} 
  \caption{Supply from -20 to 100 EUR/MWh}
  \label{fig_3d_3}
\end{subfigure}
\begin{subfigure}[b]{.495\textwidth}
 \includegraphics[width=1\textwidth, height=.65\textwidth]{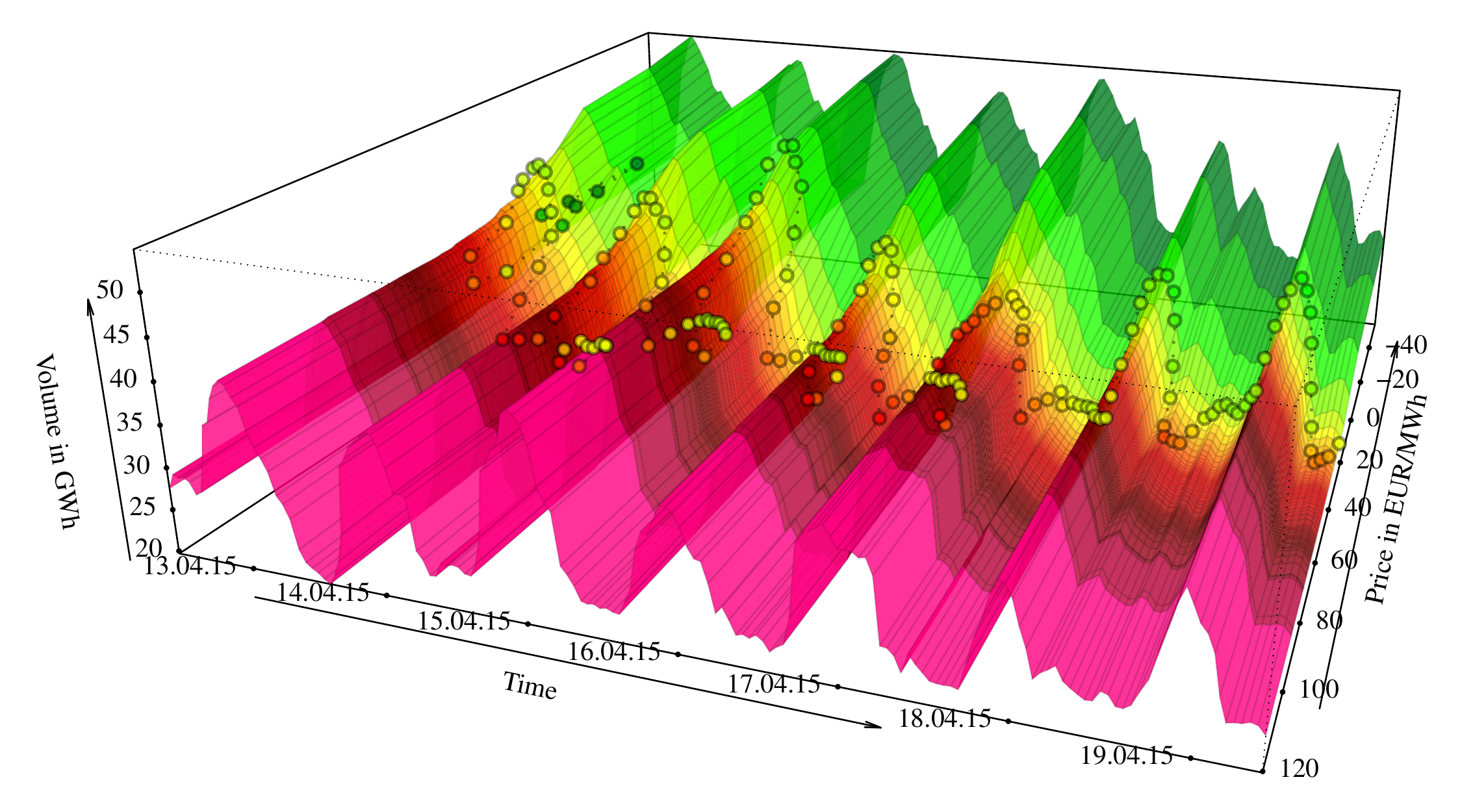} 
  \caption{Demand from -20 to 100 EUR/MWh}
  \label{fig_3d_4}
\end{subfigure}
 \begin{subfigure}[b]{.49\textwidth}
 \includegraphics[width=1\textwidth, height=.25\textwidth]{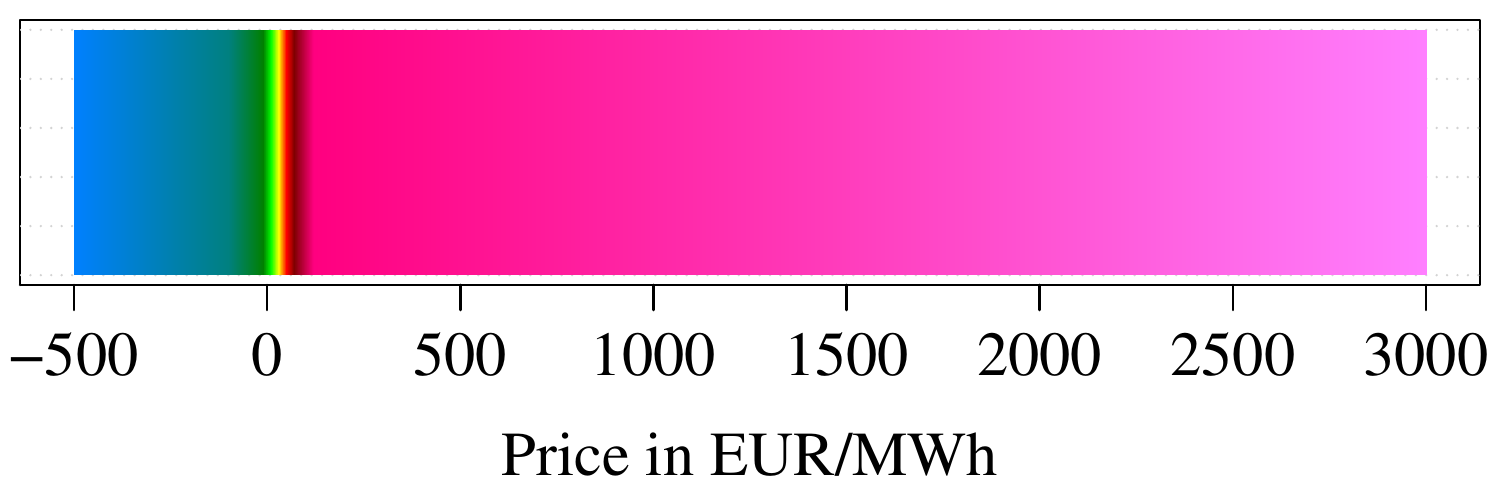} 
  \caption{Legend from -500 to 3000 in EUR/MWh}
  \label{fig_time_leg1}
\end{subfigure}
\begin{subfigure}[b]{.49\textwidth}
 \includegraphics[width=1\textwidth, height=.25\textwidth]{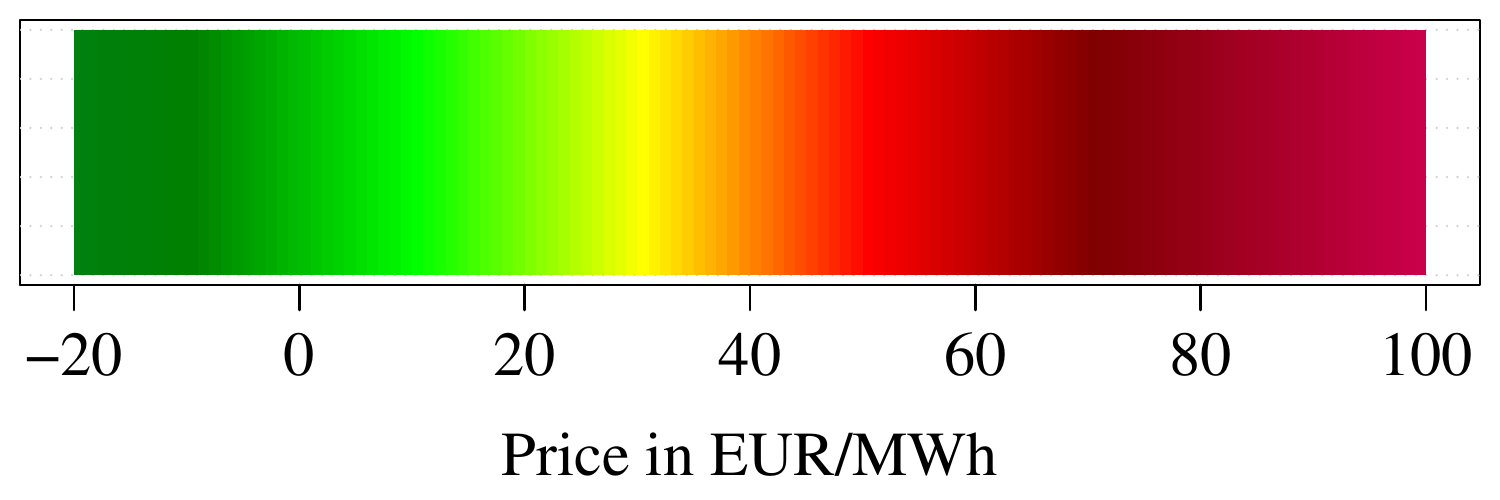} 
  \caption{Legend from -20 to 100 in EUR/MWh}
  \label{fig_time_leg2}
\end{subfigure}
\caption{3d price curves from 13. April 2015 to 19. April 2015 in a price color plot {with color legends}. 
{The colored circles represent the market clearing price and volume, the color matches the price in the color legends.}}
 \label{fig_3d-sd-curve-time-example}

\end{figure}

Modeling the supply and demand curve of electricity prices is a very complex task. Researcher who try to analyze the complex bidding structure of the supply and demand at electricity exchange usually utilize multi-agent models {or fundamental models} (\cite{weron2014electricity}). But those approaches do rarely take into account the real time series of auction data and are therefore unsuitable for giving practical information on short-term forecasts of the electricity price time series. This is especially interesting when taking a closer look at the price curves over time. In Figure \ref{fig_3d-sd-curve-time-example} we show the time series of both price curves from 13.04.2015 to 19.04.2015 in a three-dimensional plot. To put emphasis on the price scale, {which is presented at the y-axis,} we added a colored legend for them which can be found in the lower two pictures of the figure. The upper two pictures show the price curves on the full price grid, whereas the the two pictures in the middle focus on 
the price range close to 
the market clearing price. Judging only by the figure it can be obtained that both, the supply and the demand curve, exhibit a seasonal pattern over time with at least daily dependence.

\par 
However, to our knowledge there is not a single paper for the electricity market which actually models real price curves {and uses them directly} to forecast real electricity price time-series by an econometric approach. Therefore the following sections will describe the necessary setup for such a model highly detailed and use references whenever our model idea makes use of a well-known econometric technique.

For our model we proceed in three steps:

\begin{enumerate}
 \item To overcome the massive amount of data we will organize the {bid volumes} in price classes.  This will be discussed in Section \ref{price_classes}.
 \item We provide a stochastic model to forecast the bid volume of each price class. Section \ref{time_series_model} will cover this step. 
 \item {Given the forecasted bids within each price class we reassemble the precise bidding structure by reconstructing the classes. 
 Then we calculate the supply and demand curves to compute the market clearing price by the intersection of both curves. This will be 
 explained in Section \ref{reconstruction}. }
\end{enumerate}

We will refer to our model for the sale and purchase curves of the electricity price as X-Model throughout the paper.
 We choose the letter X, as it symbolizes visually the intersection of the supply and demand curve.

\subsection{Price classes for bids} \label{price_classes}

As mentioned there are 35001 possible volumes on the full price grid $\P$. Theoretically we could model each of these processes but this 
is almost unfeasible due to computational burdens. Therefore we show how to choose and apply a simple dimension reduction 
procedure to the price formation process that is computational manageable and still balances the related loss of information. 
Therefore, we merge the 35001 prices in $\P$ into a smaller amount of classes. For the bids within a price class we will assume later on that they behave similarly over time.  

\begin{figure}[htb!]
\centering
\begin{subfigure}[b]{.49\textwidth}
 \includegraphics[width=1\textwidth]{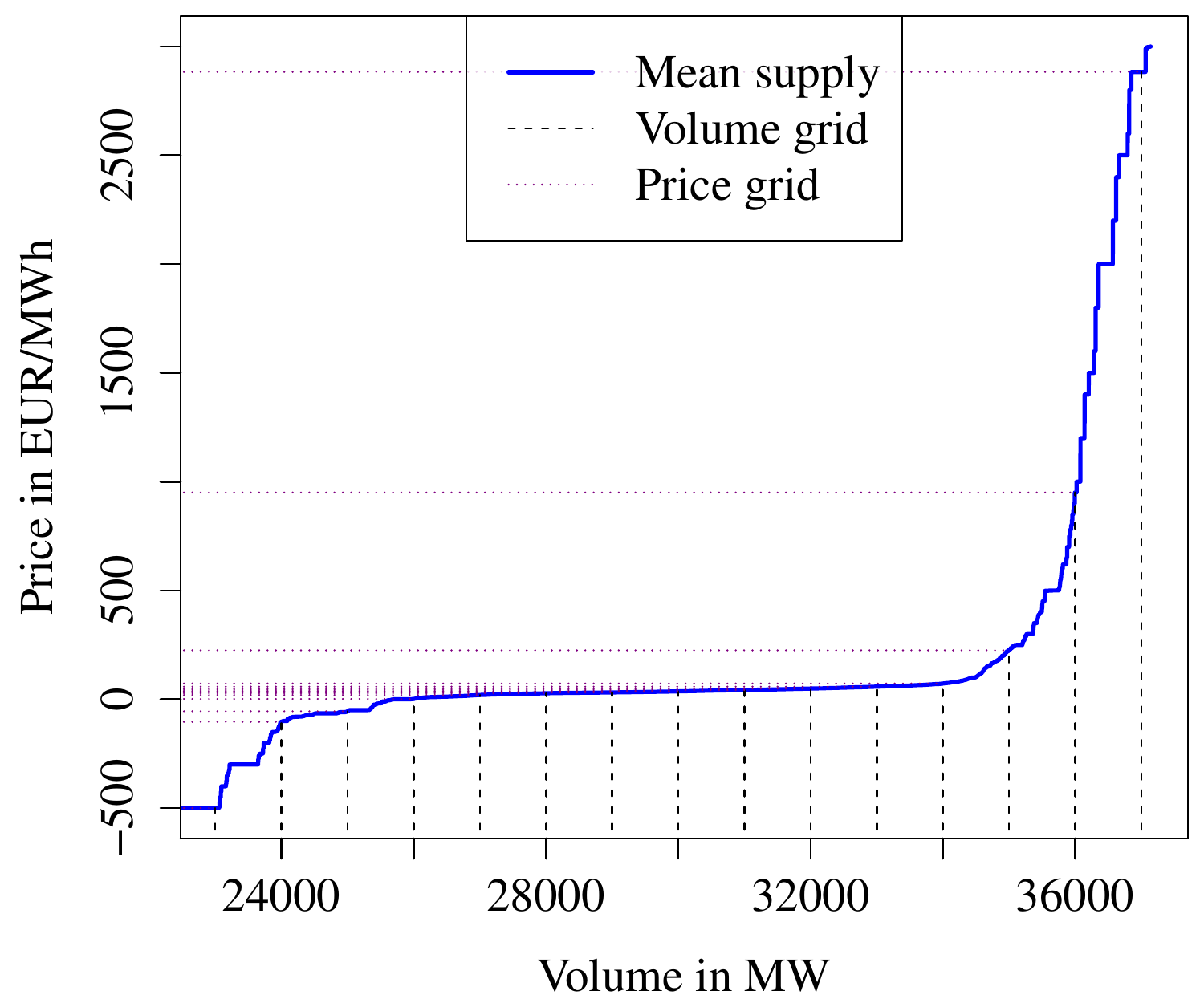} 
  \caption{Mean supply on range -500 EUR/MWh to 3000 EUR/MWh}
  \label{fig_group_sup1}
\end{subfigure}
\begin{subfigure}[b]{.49\textwidth}
 \includegraphics[width=1\textwidth]{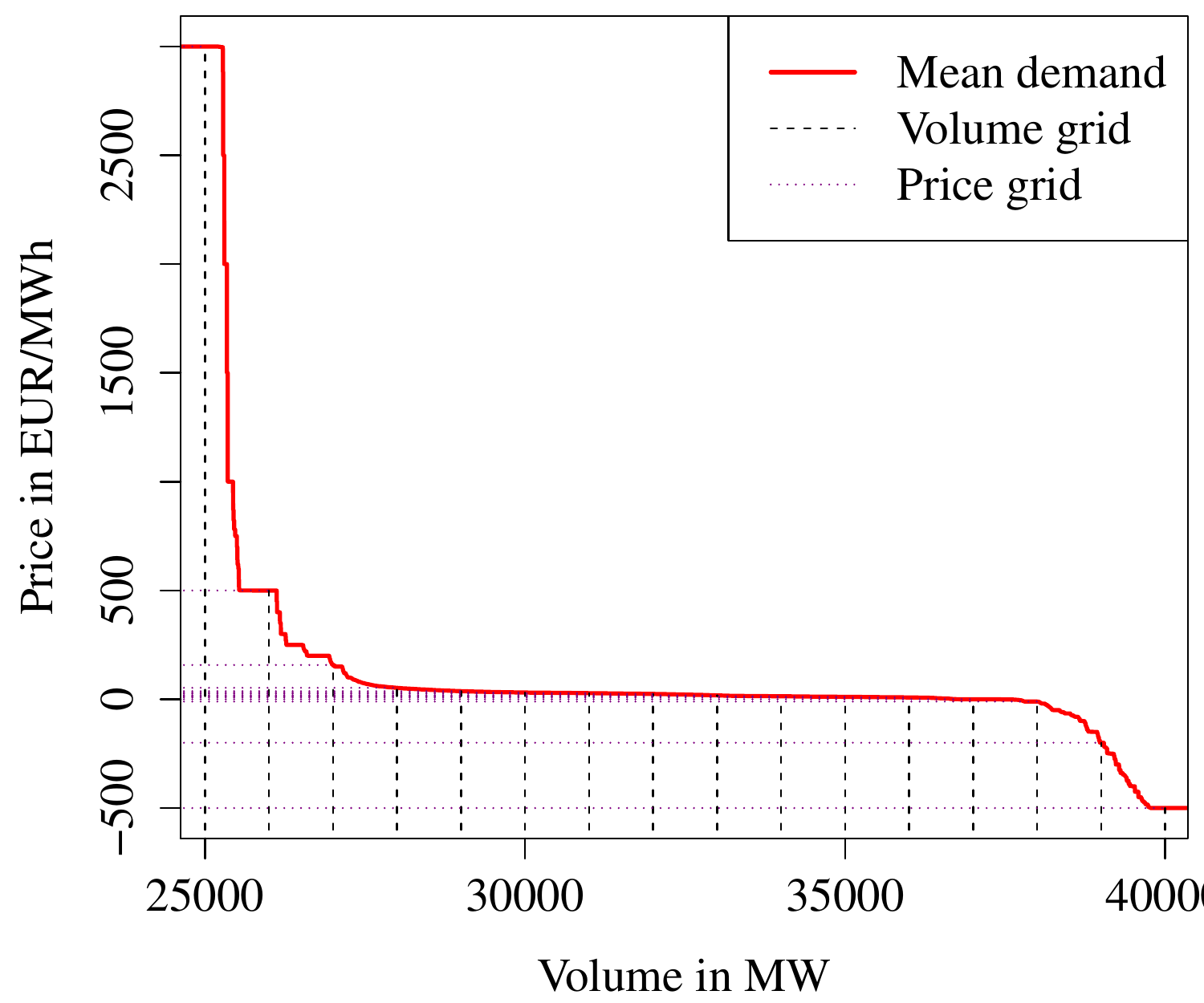} 
  \caption{Mean demand on range -500 EUR/MWh to 3000 EUR/MWh}
  \label{fig_group_dem1}
\end{subfigure}
\begin{subfigure}[b]{.49\textwidth}
 \includegraphics[width=1\textwidth]{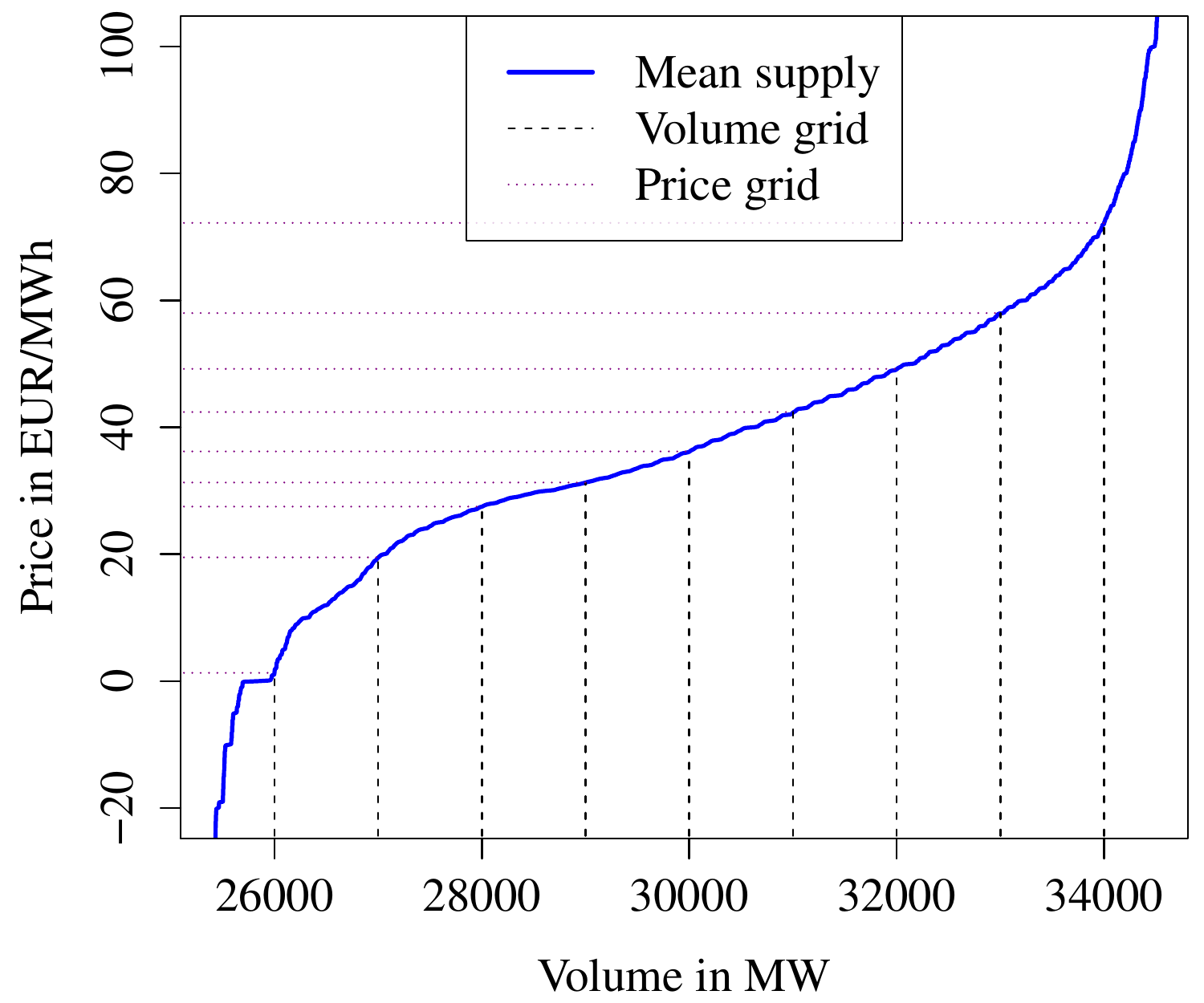} 
  \caption{Mean supply on range -20 EUR/MWh to 100 EUR/MWh}
  \label{fig_group_sup2}
\end{subfigure}
\begin{subfigure}[b]{.49\textwidth}
 \includegraphics[width=1\textwidth]{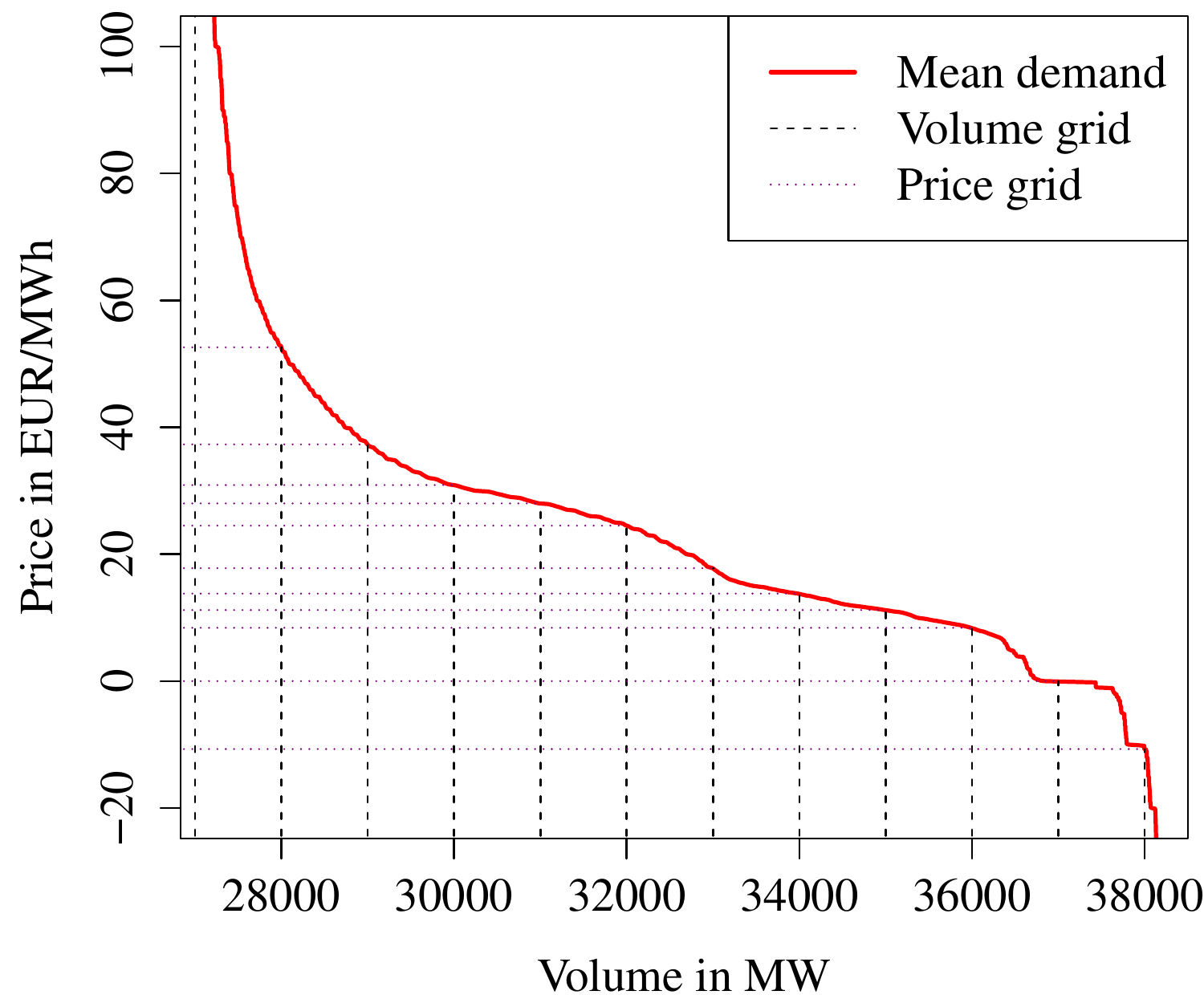} 
  \caption{Mean demand on range -20 EUR/MWh to 100 EUR/MWh}
  \label{fig_group_dem2}
\end{subfigure}
\caption{Mean supply and demand curves $\ov{S}$ and $\ov{D}$ on two selected price ranges
with volume grid $\V_*$, {price class bounds} $\C_S$ and $\C_D$ for $V^* = 1000$.}
 \label{fig_price_group}
\end{figure}

For creating the price classes we consider
{the mean bid volume $\ov{V_S}(P)$ and $\ov{V_D}(P)$ at price $P$ as defined in equation \eqref{eq_mean_volume}. We use them as} a measure of the importance for the price $P$ for the supply and 
demand side of trading. 
{
Similarly to the definiton of the price curves in equation \eqref{eq_price_curve},  we define the mean supply and demand curves $\ov{S}$ and $\ov{D}$.
They are characterized by accumulating the mean bid volumes from equation  \eqref{eq_mean_volume}
\begin{equation}
\ov{S}(P) = \sum_{ \substack{p \in \P_{S} \\ p \leq P }} \ov{V_{S}}(p)  \text{ for } P \in \P_{S} \ \ \text{and} \ \
\ov{D}(P) = \sum_{ \substack{p \in \P_{D} \\ p \geq P }} \ov{V_{D}}(p)   \text{ for } P \in \P_{D} 
\label{eq_mean_price_curve}
\end{equation}
where $\P_{S} = \bigcup_{t=1}^T \P_{S,t}$ and $\P_{D} = \bigcup_{t=1}^T \P_{D,t}$ are the sets of all bid prices for the supply and demand side.
As in \eqref{eq_price_curve} the complete mean supply and demand curve is given by the linear interpolation of the characterized points of \eqref{eq_mean_price_curve}.
The resulting mean supply and demand curve is given in Figure  \ref{fig_price_group}. 
Note that the corresponding mean supply and demand functions $\ov{S}$ and $\ov{D}$ on the price grid $\P$ are monotonically increasing.
Therefore we can use the inverse functions $\ov{S}^{-1}$ and $\ov{D}^{-1}$ for the creation of the price classes. }


For creating the price classes we additionally require an amount of volume $V_*$ 
which will give the average amount of volume that should be represented by every price class.
Then we define an equidistant volume grid $\V_* = \{iV_* | i\in \N\}$.
Using $\V_*$ and the inverse supply and demand $\ov{S}^{-1}$ and $\ov{D}^{-1}$ we define the upper and lower values of the price classes by
\begin{align}
\C_S = \ov{S}^{-1}(\V_*) = \{ \ov{S}^{-1}(iV_*) | i\in \N \} 
\ \ \text{and} \ \
\C_D = \ov{D}^{-1}(\V_*) =\{ \ov{D}^{-1}(iV_*) | i\in \N \} . 
\label{eq_price_classes} 
\end{align}


Figure \ref{fig_price_group} visualizes the classifying procedure for a volume of $V^* = 1000$. For our modeling approach later on we decided to stick with a volume size for classifying of 1000, as it provides us a manageable size of classes to estimate. However, other amounts of volume are definitely plausible. Given our data we receive a total of $M_S = 16$ and $M_D=16$ 
classes for the supply side and demand size. 
{
The collection of the price class bounds $\C_S$ and $\C_D$ which represent the price classes are given in Table \ref{tab_price_classes}. 
\begin{table}[htb!]
\small
\begin{tabular}{rl}
  &  price class bounds  \\ \hline
 $\C_S$ & -500, -103.9,  -55.1, 1.3, 19.5, 27.5,   31.3,   36.2,   42.4,   49.2, 58.0,   72.2,  225.0,  950.0, 2883.0, 3000 \\ 
  $\C_D$ & 3000,  499.9,  157.4,   52.6,   37.3,   30.9,  28.0,   24.5,   17.8,   13.8,   11.2,    8.4,    0.0,  -10.7, -200.0, -500 
\end{tabular}
\caption{Price class bounds in $\C_S$ for the supply and $\C_D$ and for the demand} 
\label{tab_price_classes}
\end{table}
Note that for supply price classes $c\in \C_S$ the price class is always represented by the upper bound $c$, {whereas for demand price classes $c\in \C_D$ the price class is always represented by the lower bound $c$.} {
The price classes $\P_{S}(c)$ for an price class upper bound $c \in \C_S$ and  $\P_{D}(c)$ for an price class lower bound $c \in \C_D$
are given by
\begin{align*} 
 \P_{S}(c) &= \left\{ P \in \P | P > \max \{p\in \C_S |  p<c\} , P \leq \min\{ p \in \C_S | p \geq c\} \right\},  \\
 \P_{D}(c) &= \left\{ P \in \P | P \geq \max \{ p \in \C_D | p\leq c\} , P < \min\{p \in \C_D| p> c\} \right\}  .
\end{align*}
Here $\P_{S}(c)$ and $\P_{D}(c)$ are all prices that belong to the same price class as $c$. 
This means for instance that $\P_{S}( -500 ) = \{ -500\}$ and $\P_{S}( -103.9 ) = \{ -499.9, -499.8, \ldots, -103.9\}$.
As $c\in \C_S$ and $c\in \C_D$ uniquely describe the price classes $\P_{S}(c)$ and $\P_{D}(c)$ we can take $c$ as the price class representative and refer to $\C_S$ and $\C_D$ 
as price classes even though they are only collections of price class bounds.

Moreover, the associated volumes at time $t$ to the prices classes $\C_S$ and $\C_D$ are given by
\begin{align*}
X^{(c)}_{S,t} &= \sum_{ \substack{P\in \P_{S}(c) }} V_{S,t}(P) \ \ \text{ for } \ \ c\in \C_S \\
X^{(c)}_{D,t} &= \sum_{ \substack{P\in \P_{D}(c) }} V_{D,t}(P) \ \ \text{ for } \ \ c\in \C_D .
\end{align*}
Hence $X^{(-500)}_{S,t}$ gives the amount of volume bid on the supply side at exactly -500 at time $t$ and
$X^{(-103.9)}_{S,t}$ the amount between inclusively -499.9 and -103.9 at time $t$.
}}


\begin{figure}[htb!]
\centering
\begin{subfigure}[b]{.49\textwidth}
 \includegraphics[width=1\textwidth]{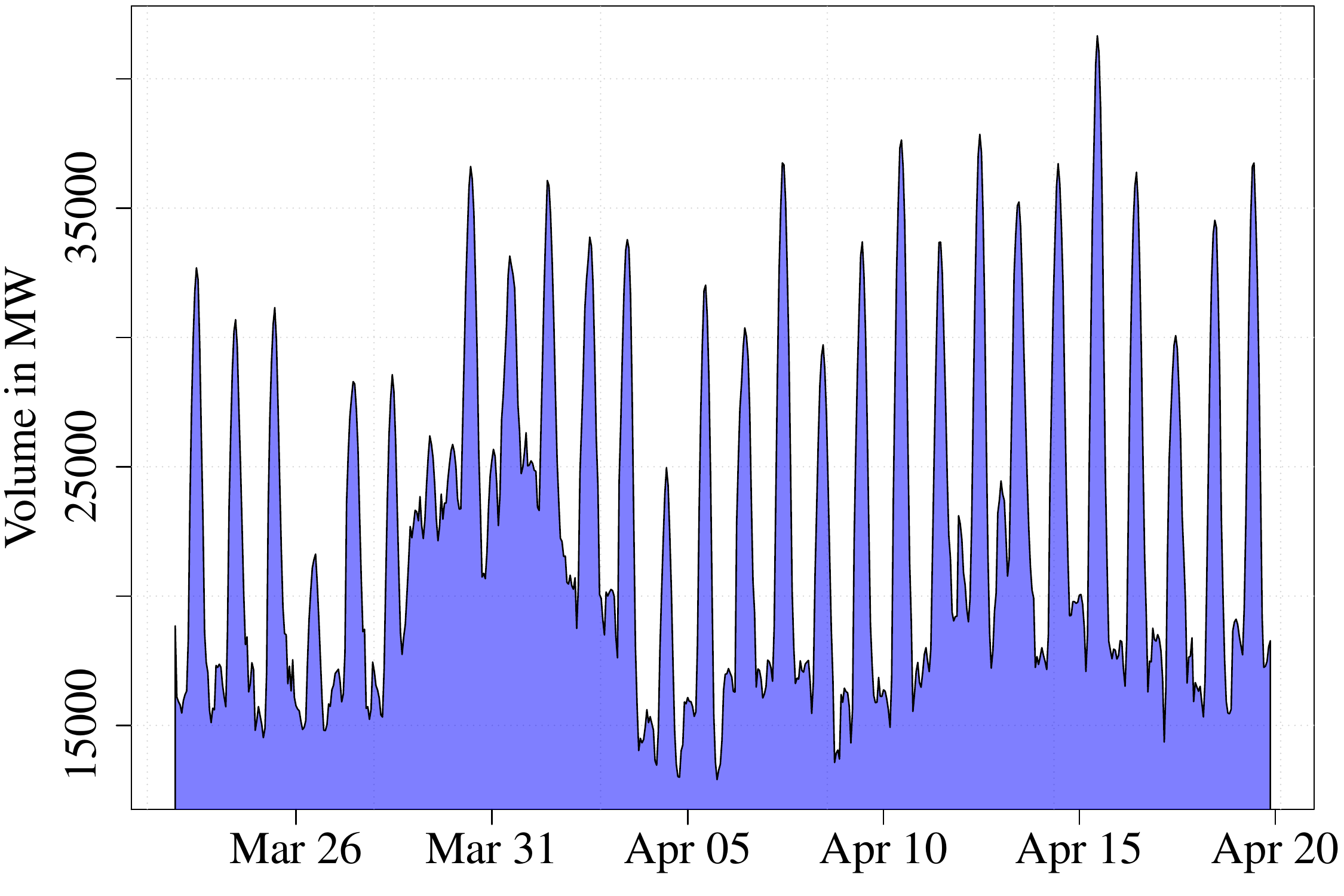} 
  \caption{$X^{(-500)}_{S,t}$ with bids at exactly  $-500$}
  \label{fig_bid1}
\end{subfigure}
\begin{subfigure}[b]{.49\textwidth}
 \includegraphics[width=1\textwidth]{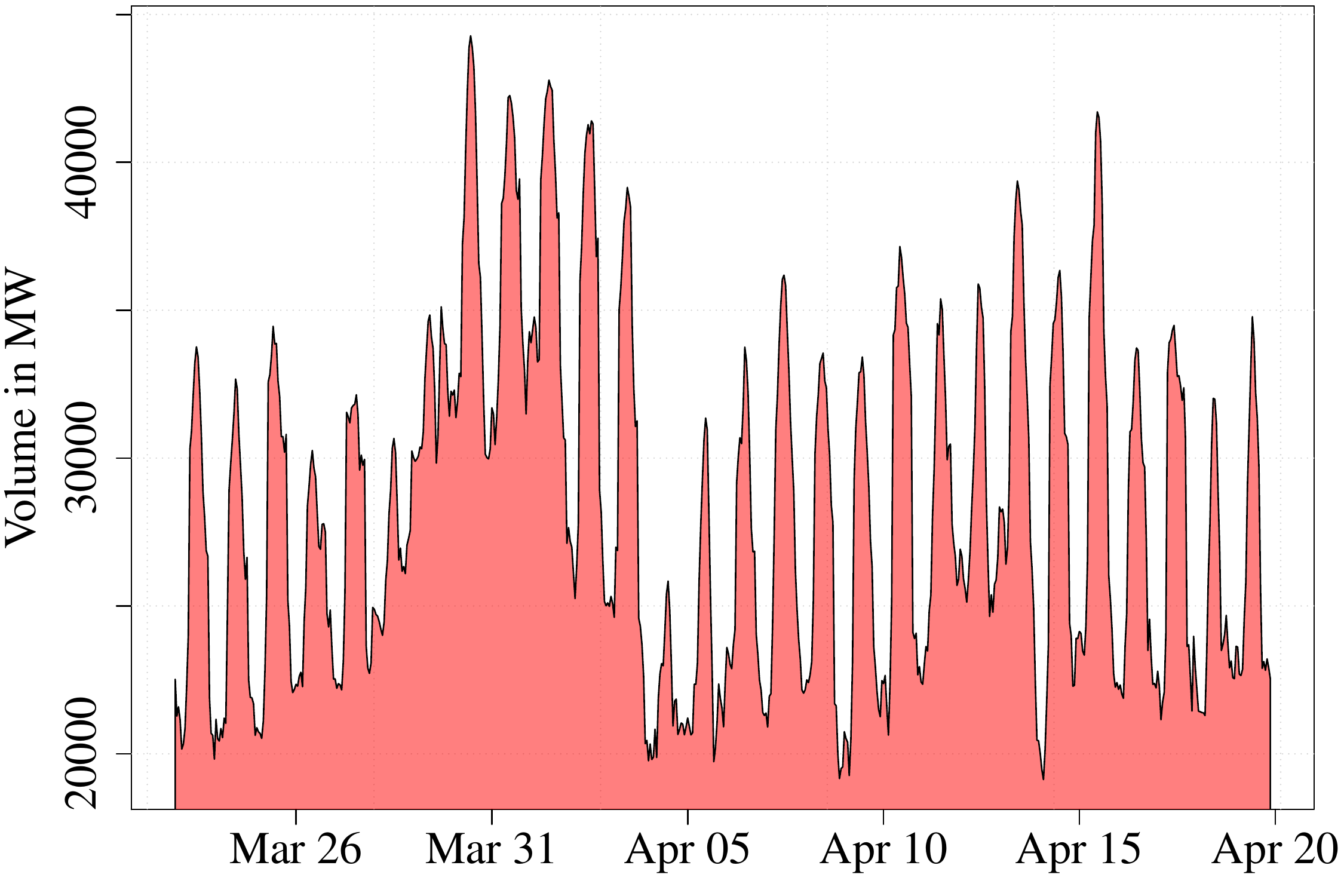} 
  \caption{ $X^{(3000)}_{D,t}$ with bids at exactly $3000$}
  \label{fig_bid2}
\end{subfigure}
\begin{subfigure}[b]{.49\textwidth}
 \includegraphics[width=1\textwidth]{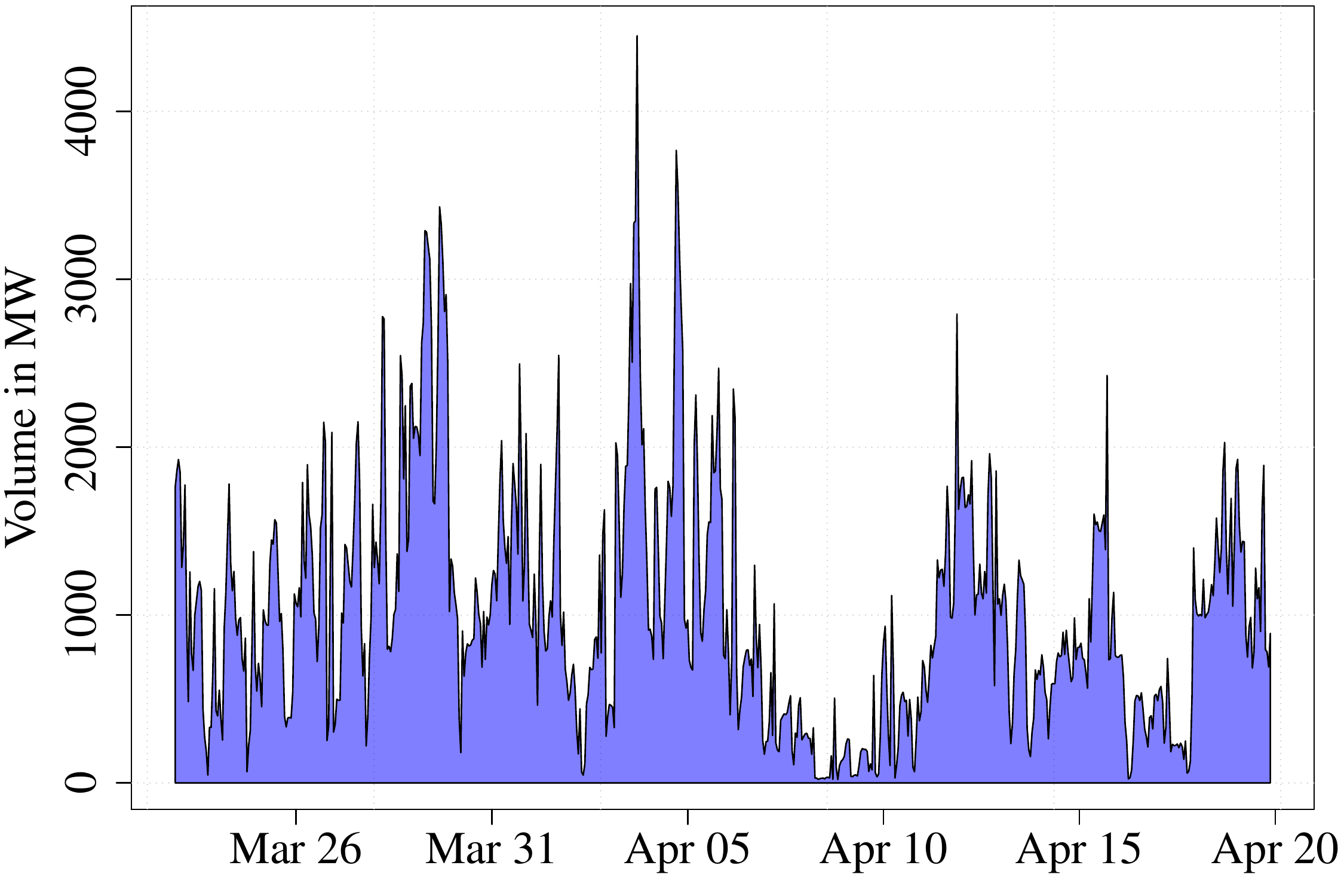} 
  \caption{$X^{(19.5)}_{S,t}$ with bids in $[1.4,19.5]$}
  \label{fig_bid3}
\end{subfigure}
\begin{subfigure}[b]{.49\textwidth}
 \includegraphics[width=1\textwidth]{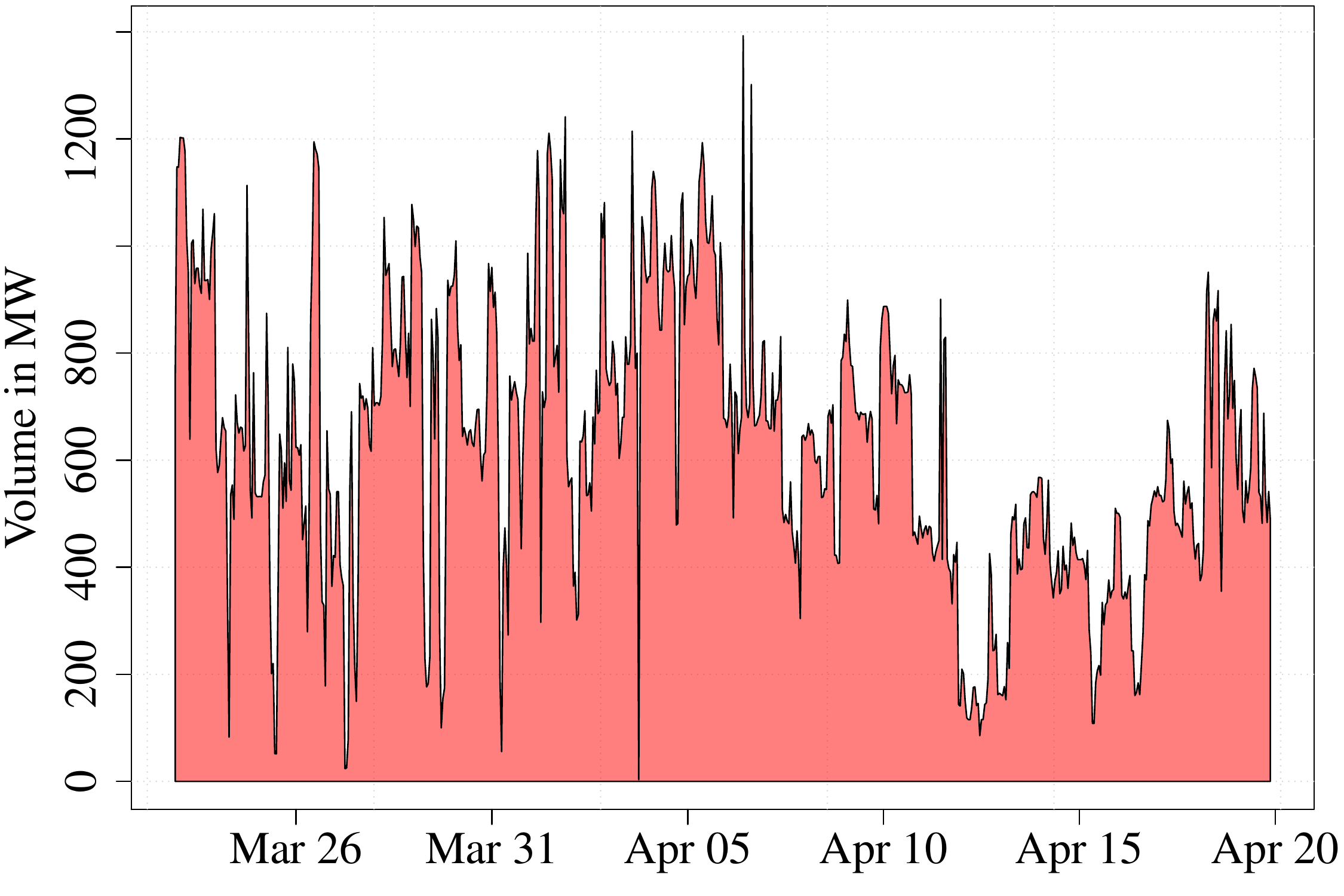} 
  \caption{$X^{(8.4)}_{D,t}$ with bids in $[8.4,11.1]$}
  \label{fig_bid4}
  \end{subfigure}
\begin{subfigure}[b]{.49\textwidth}
 \includegraphics[width=1\textwidth]{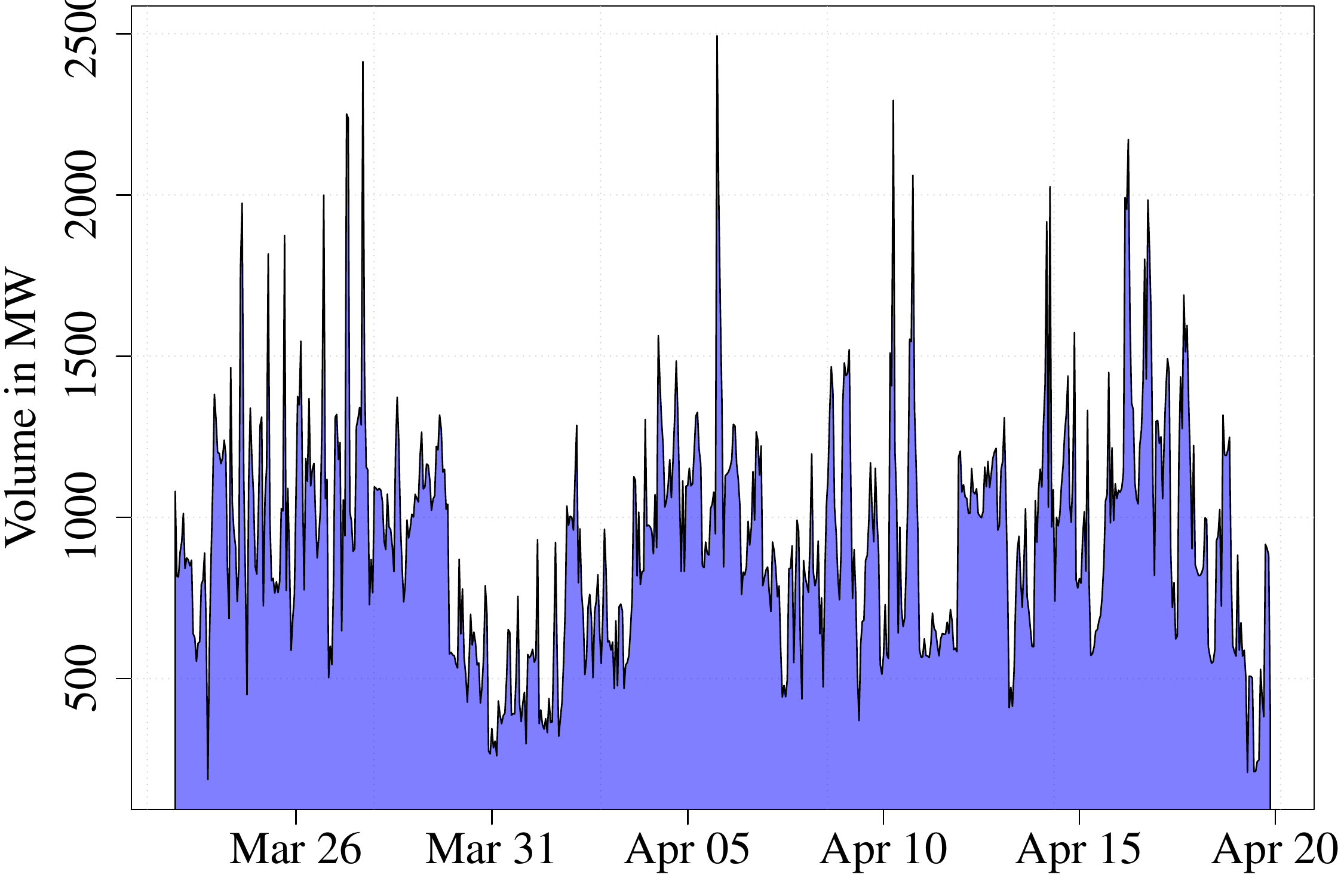} 
  \caption{ $X^{(58.0)}_{S,t}$ with bids in $[49.3,58]$ }
  \label{fig_bid5}
\end{subfigure}
\begin{subfigure}[b]{.49\textwidth}
 \includegraphics[width=1\textwidth]{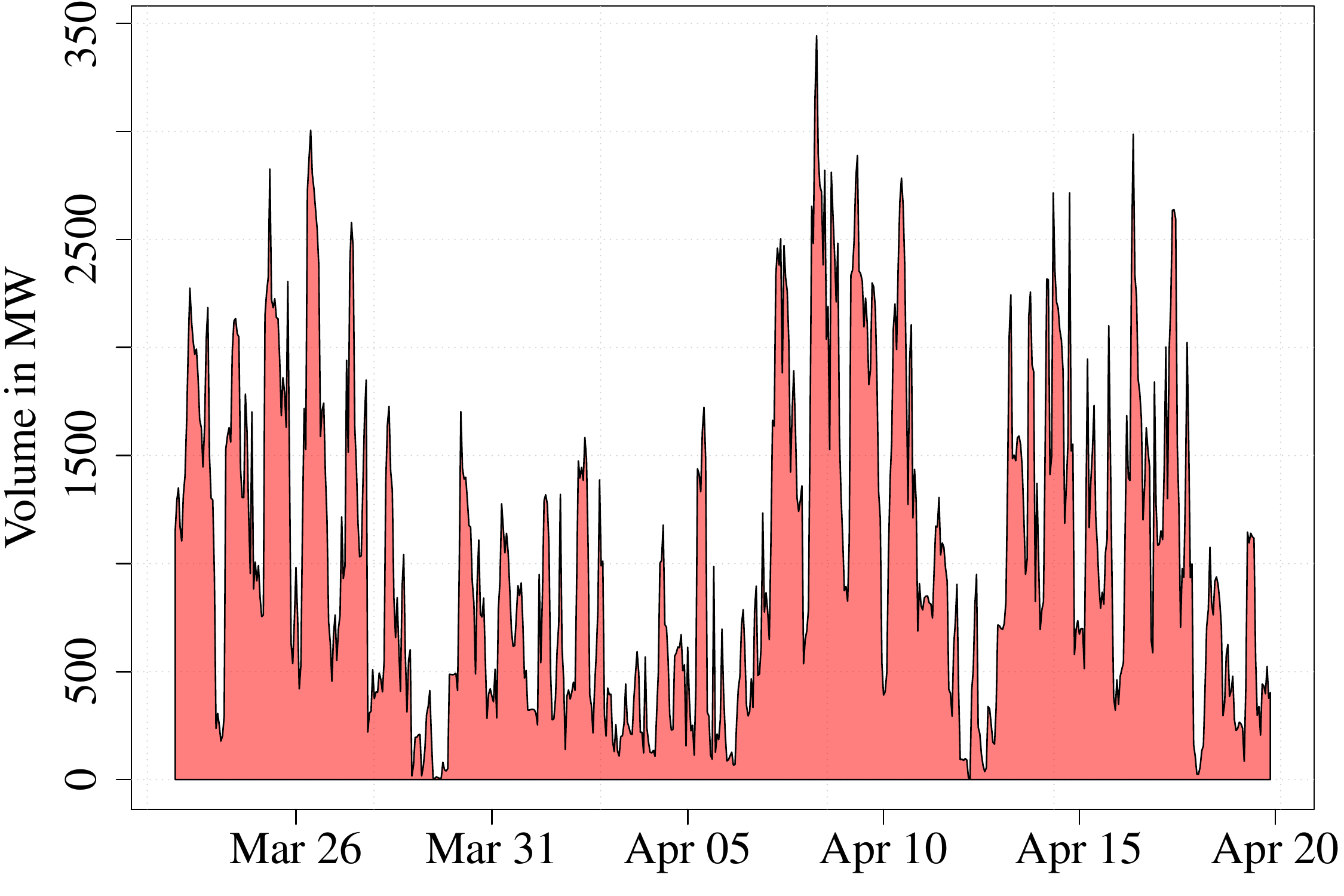} 
  \caption{$X^{(37.3)}_{D,t}$ with bids in $[37.3,52.5]$}
  \label{fig_bid6}
\end{subfigure}
\caption{{Bid volumes of certain price classes} for supply and demand 
for four observed weeks in 2015.}
 \label{fig_example_xts}
\end{figure}

As an example we show several {bid volume processes of selected price classes} in Figure \ref{fig_example_xts} in a short time perspective.
Note that the illustrated bid volumes $X^{(-500)}_{S,t}$ and $X^{(3000)}_{D,t}$ are very important in practice,
as they represent large volumes. Moreover, both bids are favored by some market participants as they will always be realized.
{The corresponding bid volumes are covering the price inelastic supply or demand and are also known as the must-run stack.
However, these volumes are not only covering the must-run bids but also the net import and export positions and 
specific block orders like limit orders.} 
We observe that they have a more distinct seasonal structure than the 
common bids. Due to the way we construct our classes, $X^{(-500)}_{S,t}$ and $X^{(3000)}_{D,t}$ are independent of the choice of volume $V^*$.

Every other bid volume processes, e.g. $X^{(19.5)}_{S,t}$, $X^{(8.4)}_{D,t}$, $X^{(58.0)}_{S,t}$ and $X^{(37.3)}_{D,t}$, in Figure \ref{fig_example_xts} has a a more complex structure. 
But many of them exhibit also a daily and weekly seasonal pattern. In the online appendix we provide time series plots as in Figure \ref{fig_example_xts} for all {bid volume processes} for the full time range.

{
As mentioned $X^{(-500)}_{S,t}$ and $X^{(3000)}_{D,t}$ represent large volumes. The other processes cover approximately a volume of 1000, but there is an exception as well.
Because of the construction of the price classes using the equidistant volume grid the last classes $X^{(3000)}_{S,t}$ and $X^{(-500)}_{D,t}$ tend to cover smaller volumes. However, as the corresponding bids are hardly realized the influence is negligible.
}

\subsection{Time series model for bid classes} \label{time_series_model}

Now we provide a model for the {bid volume process $X^{(c)}_{S,t}$ and $X^{(c)}_{D,t}$ of the price classes $\P_{S}(c)$ and $\P_{D}(c)$.}
Therefore, we introduce $X^{(c)}_{S,d,h}$ and $X^{(c)}_{S,d,h}$ as the bid supply and demand volume 
of price class $c\in \C_S$ resp. $c\in \C_D$ at day $d$ and hour $h$. For the well-known issue with the clock change due to daylight saving time we decided to interpolate the missing hour in March with the two hours around the missing hour and use the average of the double hours in October so that there are 24 observable prices each day. Thus, the volume processes $X^{(c)}_{S,d,h}$ and $X^{(c)}_{S,d,h}$ are well defined.

\begin{figure}[htb!]
\centering
 \includegraphics[width=.99\textwidth, height=.49\textwidth]{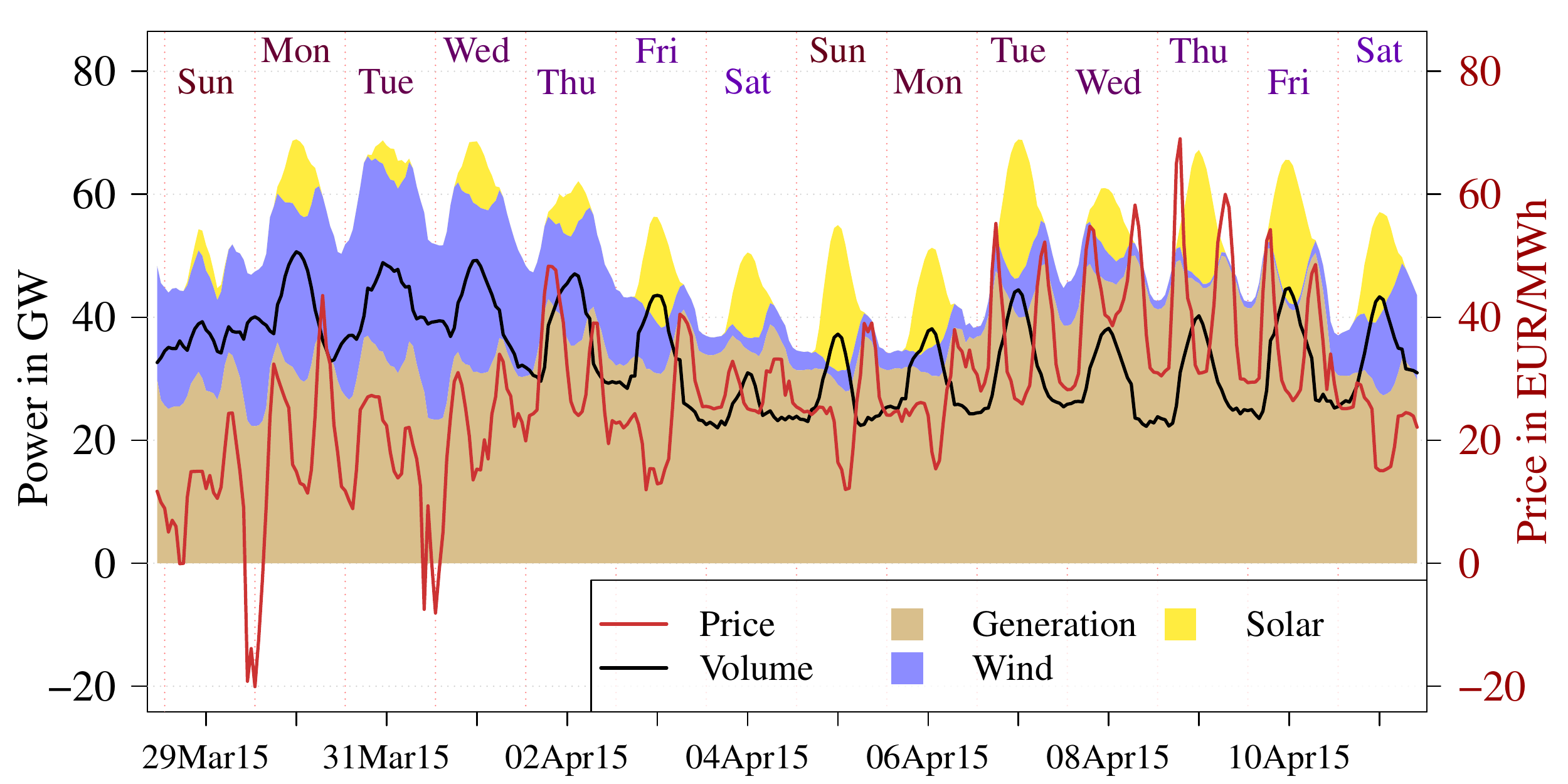} 
\caption{Sample of $X_{\text{price},t}$, $X_{\text{volume},t}$, $X_{\text{generation},t}$, $X_{\text{wind},t}$
$X_{\text{solar},t}$ from 29.03.2015 to 11.04.2015.}
 \label{fig_add_sample}
\end{figure}

As mentioned above the processes $X^{(-500)}_{S,d,h}$ and $X^{(3000)}_{D,d,h}$ play an important role.
But we also consider the impact of other possible sources that might influence the bidding behavior.
In particular we use the EPEX market clearing price and volume of Germany and Austria of previous auctions, the planned electric power generation in Germany of conventional power plants with more than 100 MW power as well as the planned wind and solar power feed-in. 
The last three processes are provided by the EEX transparency database. Hence, we assume that market participants
have access to this database or similar information and base their bids at least partially on those time series.
Especially the impact of wind and solar energy on electricity prices {due to the merit-order effect} is well known (see e.g. \cite{hirth2013market}, \cite{cludius2014merit} or \cite{ketterer2014impact}).
Therefore we introduce $M_X = 5$ additional processes 
denoted by $X_{\text{price},t}$, $X_{\text{volume},t}$, $X_{\text{generation},t}$, $X_{\text{wind},t}$
$X_{\text{solar},t}$ that represent the additional information that is available at the time where the auction will take place.
A sample of the considered processes is given in Figure \ref{fig_add_sample}.

Similarly to $X^{(c)}_{S,d,h}$ and $X^{(c)}_{D,d,h}$ we introduce the slightly transformed processes 
$X_{\text{price},d,h}$, $X_{\text{volume},d,h}$, $X_{\text{generation},d,h}$, $X_{\text{wind},d,h}$ and $X_{\text{solar},d,h}$
at day $d$ and hour $h$. 
Note that the planned generation as well as the projected wind and solar power is known for one day in advance so we can use e.g. 
$X_{\text{solar}, d+1,h}$ to predict $X^{(c)}_{S,d+1,h}$ and $X^{(c)}_{S,d+1,h}$. 


The considered model is a simple regression approach and similar to the basic autoregressive model as used in
\cite{weron2008forecasting}, \cite{maciejowska2016probabilistic} or \cite{ziel2016forecasting}
for modeling the electricity price. But we will use it in a more flexible way for the bid volume processes of the price classes.
For example, \cite{weron2008forecasting} allow for a linear dependency of $X_{\text{price},d,h}$ to 
$X_{\text{price},d-1,h}$, $X_{\text{price},d-2,h}$ and $X_{\text{price},d-7,h}$ as well as 
dummies on Sunday, Monday and Saturday.
However, the choice of lags $1$, $2$ and $7$ as well as the selection of the weekday dummies is the same for all 24 hours.
As in \cite{ziel2016forecasting}, we will allow for much more flexibility in the model, 
as the structure of the data is far more complex. 
This applies to both, the autoregressive lag structure and the weekday structure.

\begin{figure}[htb!]
\centering
\begin{subfigure}[b]{.49\textwidth}
 \includegraphics[width=1\textwidth, height=.85\textwidth]{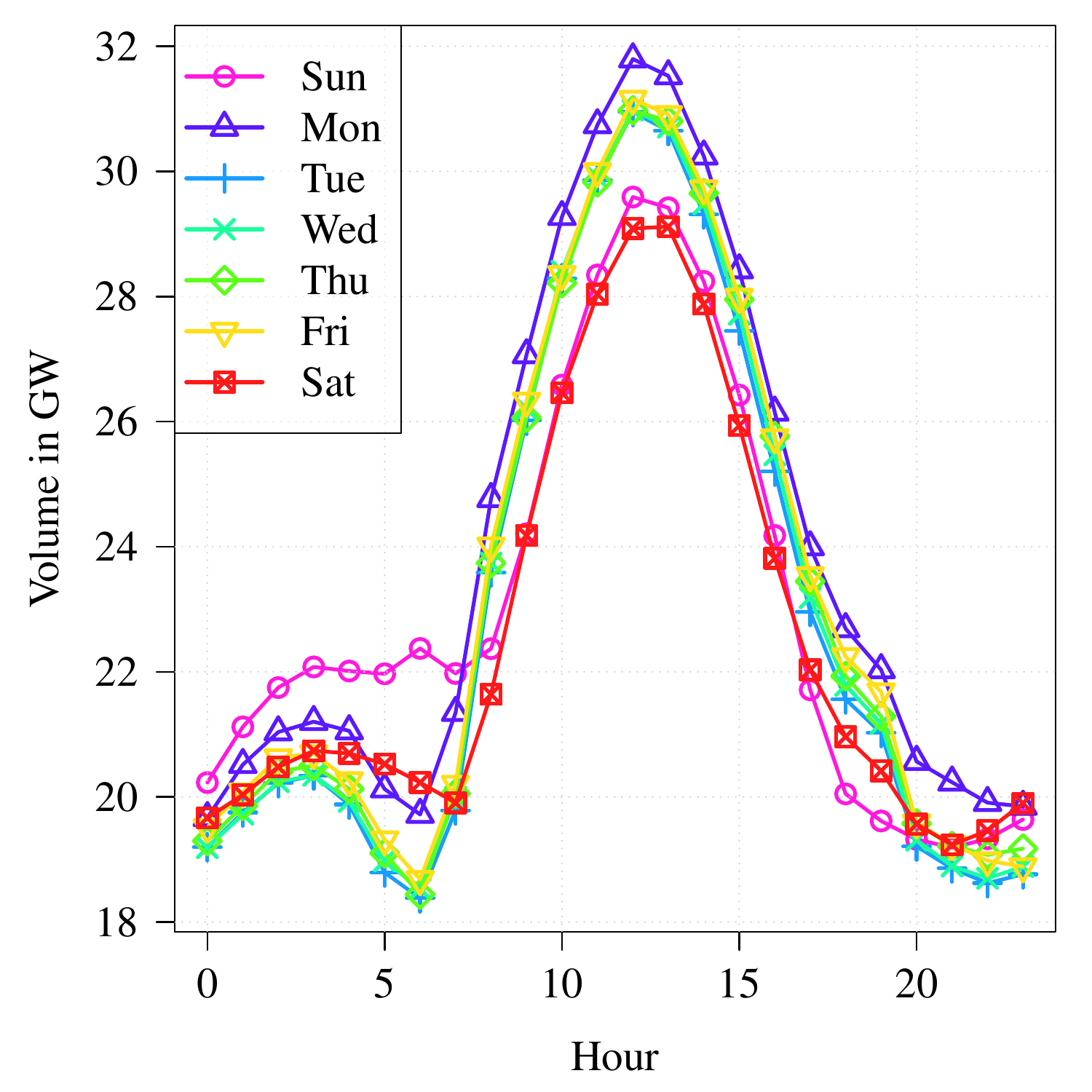} 
  \caption{Supply $X^{(-500)}_{S,d,h}$ }
  \label{fig_wds}
\end{subfigure}
\begin{subfigure}[b]{.49\textwidth}
 \includegraphics[width=1\textwidth, height=.85\textwidth]{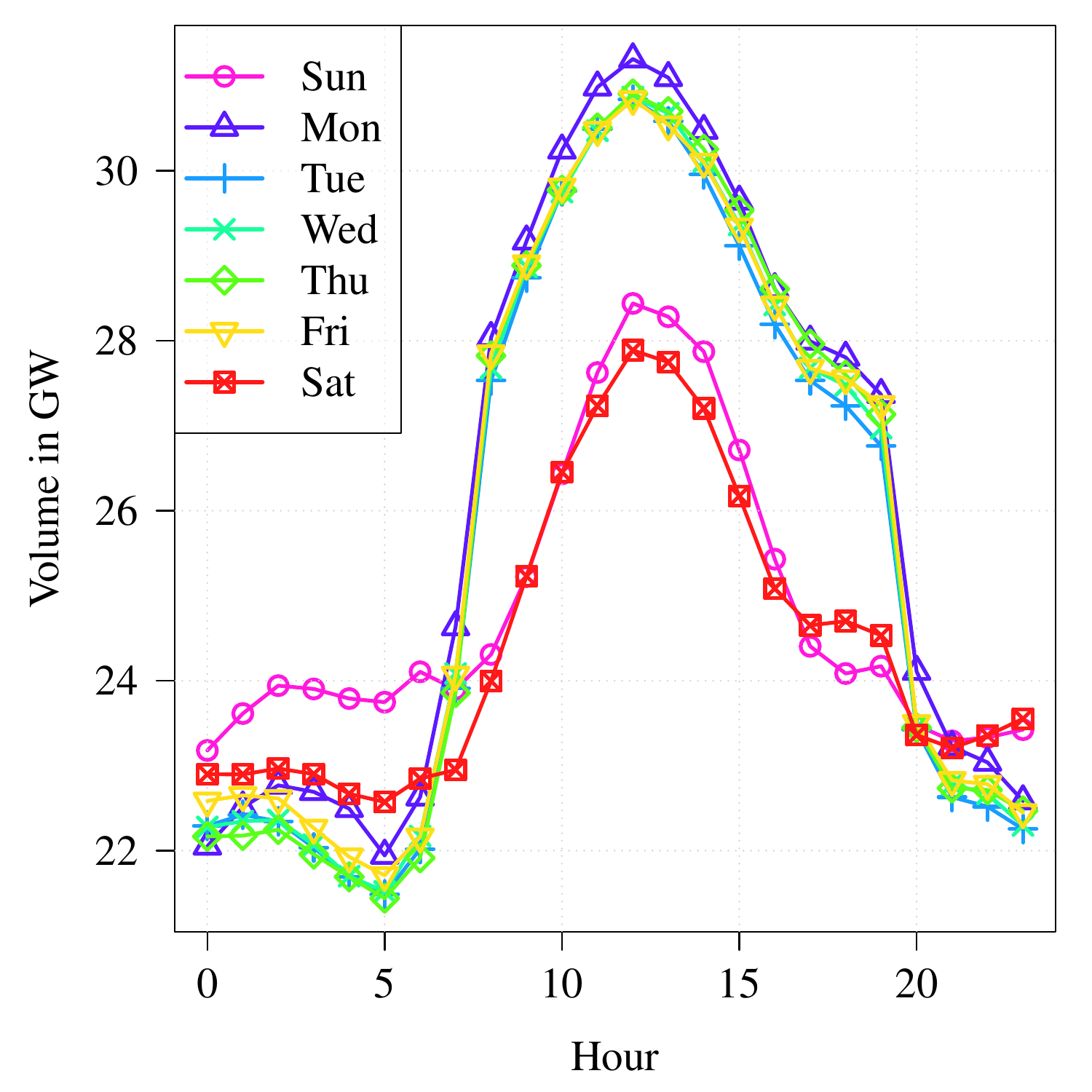} 
  \caption{Demand $X^{(3000)}_{D,d,h}$}
  \label{fig_wdd}
\end{subfigure}
\caption{Weekly mean bid volumes for each day of the week and hour of the day. }
 \label{fig_wd_sd}
\end{figure}

In Figure \ref{fig_wd_sd} the weekly sample 
mean of the bid volume processes $X^{(-500)}_{S,d,h}$ and $X^{(3000)}_{D,d,h}$ for our full sample time range is given.
There we can see that the daily seasonal structure seems to depend on the day of the week, as it is typical for the electricity market clearing price or the 
electricity load (see e.g. \cite{ziel2015efficient}). So we see that the Saturdays and Sundays have a clearly different behavior than the other weekdays.
But from 0:00 to 6:00 the Saturday seems to leave the typical pattern of a Sunday.  
Furthermore, we recognize that for the demand side the hours from 8:00 to 19:00 are clearly
on a higher level during the working days. This is interesting, as it exactly matches 
the peakload standard block order at EPEX. 

For modeling the day of the week impact we define the weekday indicators
$$W_{k}(d) = \begin{cases}
              1 &, \WW(d) < k \\
              0 &, \WW(d) \geq k \\
             \end{cases}
$$
where $\WW(d)$ is a function that gives a number that corresponds to the weekday of day $d$. We use without loss
of generality $k=1$ for a Monday, for a Tuesday $k=2$
up to $k=7$ for a Sunday.

To fully present the considered time series model, it is necessary to
introduce the object 
\begin{align*}
\bsX_{d,h} &= (X_{1,d,h}, \ldots, X_{M,d,h})' \\ &= ( (X^{(c)}_{S,d,h})_{c\in \C_S}, (X^{(c)}_{D,d,h})_{c\in \C_D} , 
X_{\text{price},d,h}, X_{\text{volume},d,h}, X_{\text{generation},d+1,h},X_{\text{wind},d+1,h},  X_{\text{solar},d+1,h})'. 
\end{align*}
As the planned processes (generation, wind and solar) are known one day in advance they are represented with the day $d+1$ in the object $\bsX_{d,h}$.
Note that the dimension of $\bsX_{d,h}$ is {$M=M_S+M_D + M_X$  ($M = 16+16+5 = 37$ given the used data).
However, we have to model and forecast only the first $M_S + M_D$ components for each hour $h$ which exactly match 
the bid volume processes of the supply and demand price classes.
Moreover,} we do not impose a time series model to $\bsX_{d,h}$ directly, but to its zero mean process 
$\bsY_{d,h} = \bsX_{d,h} - \bsmu_h$ with $\bsmu_h= \E(\bsX_{d,h})$. We estimate the mean $\bsmu_h$ 
by the corresponding sample mean.

Now for each hour $h$ the considered time series model of $\bsY_{d,h} =  (Y_{1,d,h}, \ldots, Y_{M,d,h})'$ is constructed that
it can potentially depend linearly on $\bsY_{d-k,h}$, but also on a different hour 
$\bsY_{d-k,j}$ with $j\neq h$ and the introduced weekday dummies. 
The considered time series model for $Y_{m,d,h}$ for each hour $h$ and $m \in \{1, \ldots, M_S+M_D\}$ is given by
\begin{equation}
 Y_{m,d,h} = 
  \sum_{l = 1}^{M} \sum_{j=1}^{24}  \sum_{k \in \II_{m,h}(l,j)} \phi_{m,h, l,j,k} Y_{l,d-k,j} 
 + \sum_{k=2}^7 \psi_{m,h,k} W_k(d) + \eps_{m,d,h}
\label{eq_main_ar_model}
\end{equation}
with parameters $\phi_{m,h, l,j,k}$ and $\psi_{m,h,k}$, 
 $\II_{m,h}(l,j)$ as lag sets of lags and $\eps_{m,d,h}$ as error term.
 We assume that the error process $(\eps_{m,d,h})_{d\in \Z}$ is i.i.d. 
 with constant variance $\sigma_{m,h}^2$. The introduced parameters $\phi_{m,h, l,j,k}$ 
 will model the linear autoregressive impact and $\psi_{m,h,k}$ the day of the week effect.
 
The choice of lag sets $\II_{m,h}(l,j)$ in \eqref{eq_main_ar_model} is crucial for the full model, as they 
specify the possible model structure.
In general it holds true that larger sets $\II_{m,h}(l,j)$ increase the likelihood of overfitting,
even though this likelihood is limited due to our used regularized estimation technique.
However, if the lag sets were chosen too small, we might miss important features in the data.
Thus, we should always choose $\II_{m,h}(l,j)$ of reasonable size. This size is determined by the user and can be chosen freely or be backed up by fundamental data analysis, e.g. the correlation structure. Please note that this procedure only determines the possible lag structure and not the final lag structure as it only defines the set of lags which our estimation algorithm will consider. 
The coefficients that correspond to these lags can have zero impact because of the estimation procedure. 
For this paper we decided on 
$$\II_{m,h}(l,j) = \begin{cases}
\{1, 2, \ldots, 36\} &, m= l \ \text{ and } \ h=j \\
\{1, 2, \ldots, 8\} &, (m= l \ \text{ and } \ h\neq j) \text{ or } ( m\neq l \ \text{ and } \ h=j)   \\
\{1\} &,  m\neq l \ \text{ and } \ h\neq j
             \end{cases}$$
{for every bid volume process of price class $m$.}
Thus, the process $Y_{m,d,h}$ of price class $m$ at day $d$ and hour $h$ can 
depend on the values of the past 36 days of price class $m$ at hour $h$. 
In contrast, $Y_{m,d,h}$ for a specific price class $m$ and a specific hour $h$ is only allowed to depend on the value of another process at another hour one with a maximum lag of 1. In all other cases a maximum lag of eight is possible. To illustrate this setting Figure \ref{fig_illustration_dep} shows the possible dependency structure 
of $Y_{m,d,h}$ for an {exemplary price class $m$ for an hour $h=2$}. The left hand side of the figure shows a specific price class {$m$, called target price class}. The blue rectangle symbolizes the hour which was modeled, e.g. the hour 2:00. Every green rectangle gives information, if this lag is considered for modeling that hour. Red rectangles indicate that the lag is not considered as it lays outside of our lag definition, gray rectangles indicate that this data is not available as it is future information. Therefore, the target price class for one hour can be dependent on every other hour for that same price class up to eight days back in time, and on the same hour for the same price class up 
to 36 days back in time. 
The possible dependencies on other price classes is provided in the illustration in the middle. The allowed dependencies on planned regressors (generation, wind and solar) is depicted on the right hand side of the figure. 
The color scheme 
applies as well to those classes. It is worth mentioning that the model for hour 2:00 of a specific day and price class can be dependent on other hours of the regressors for planned generation of that same day, as those information is indeed available before the auction starts. Besides that it can only depend on their lagged values for exactly hour 2:00 of the previous seven days. The shift in dependence on historical values is due to our definition of the regressors.

\definecolor{white}{rgb}{1,1,1}
\definecolor{tikcol1}{rgb}{1,.8,.8}
\definecolor{tikcol2}{rgb}{0.7,1,.7}
\definecolor{tikcol3}{rgb}{0.5,.8,1}
\definecolor{tikcol4}{rgb}{0.8,.8,.8}

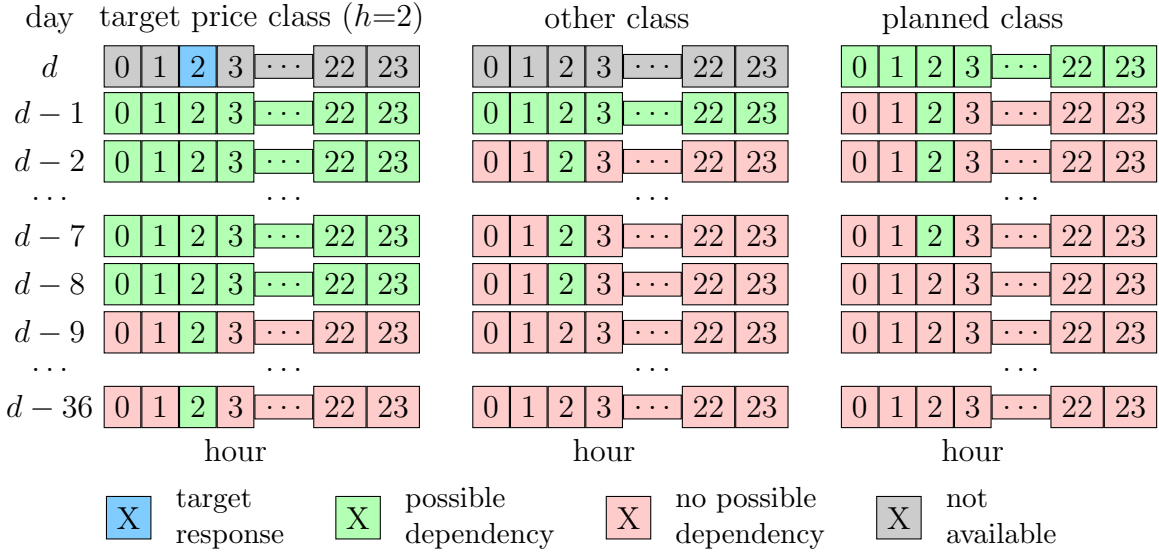
\begin{figure}
\begin{tikzpicture}
 
 \tikzstyle{ann} = [draw=none, fill=none, right]
    \matrix[nodes={draw, fill=tikcol1},
        row sep=1pt,column sep=0pt] {
        \node[draw=none,fill=none] (day_node) {$d  $}; &
    \node[rectangle, fill=tikcol4] {0}; &
    \node[rectangle, fill=tikcol4] {1}; &
    \node[rectangle, fill=tikcol3] {2}; &
    \node[rectangle, fill=tikcol4] (main_d0_h3) {3}; &
    \node[rectangle, fill=tikcol4] {$\ldots$}; &
    \node[rectangle, fill=tikcol4] {22}; &
    \node[rectangle, fill=tikcol4] {23};&
        \node[draw=none,fill=none] {$  \ \ \ $}; &
    \node[rectangle, fill=tikcol4] {0}; &
    \node[rectangle, fill=tikcol4] {1}; &
    \node[rectangle, fill=tikcol4] {2}; &
    \node[rectangle, fill=tikcol4] (other_d0_h3) {3}; &
    \node[rectangle, fill=tikcol4] {$\ldots$}; &
    \node[rectangle, fill=tikcol4] {22}; &
    \node[rectangle, fill=tikcol4] {23}; &
        \node[draw=none,fill=none] {$  \ \ \ $}; &
    \node[rectangle, fill=tikcol2] {0}; &
    \node[rectangle, fill=tikcol2] {1}; &
    \node[rectangle, fill=tikcol2] {2}; &
    \node[rectangle, fill=tikcol2] (solar_d0_h3) {3}; &
    \node[rectangle, fill=tikcol2] {$\ldots$}; &
    \node[rectangle, fill=tikcol2] {22}; &
    \node[rectangle, fill=tikcol2] {23};    \\
        \node[draw=none,fill=none] {$d-1 $}; &
    \node[rectangle, fill=tikcol2] {0}; &
    \node[rectangle, fill=tikcol2] {1}; &
    \node[rectangle, fill=tikcol2] {2}; &
    \node[rectangle, fill=tikcol2] {3}; &
    \node[rectangle, fill=tikcol2] {$\ldots$}; &
    \node[rectangle, fill=tikcol2] {22}; &
    \node[rectangle, fill=tikcol2] {23};&
        \node[draw=none,fill=none] {$  \ \ \ $}; &
    \node[rectangle, fill=tikcol2] {0}; &
    \node[rectangle, fill=tikcol2] {1}; &
    \node[rectangle, fill=tikcol2] {2}; &
    \node[rectangle, fill=tikcol2] {3}; &
    \node[rectangle, fill=tikcol2] {$\ldots$}; &
    \node[rectangle, fill=tikcol2] {22}; &
    \node[rectangle, fill=tikcol2] {23}; &
        \node[draw=none,fill=none] {$  \ \ \ $}; &
    \node[rectangle] {0}; &
    \node[rectangle] {1}; &
    \node[rectangle, fill=tikcol2] {2}; &
    \node[rectangle] {3}; &
    \node[rectangle] {$\ldots$}; &
    \node[rectangle] {22}; &
    \node[rectangle] {23};
    \\
    \node[draw=none,fill=none] {$d-2  $}; &
    \node[rectangle, fill=tikcol2] {0}; &
    \node[rectangle, fill=tikcol2] {1}; &
    \node[rectangle, fill=tikcol2] {2}; &
    \node[rectangle, fill=tikcol2] {3}; &
    \node[rectangle, fill=tikcol2] {$\ldots$}; &
    \node[rectangle, fill=tikcol2] {22}; &
    \node[rectangle, fill=tikcol2] {23};&
        \node[draw=none,fill=none] {$  \ \ \ $}; &
    \node[rectangle] {0}; &
    \node[rectangle] {1}; &
    \node[rectangle, fill=tikcol2] {2}; &
    \node[rectangle] {3}; &
    \node[rectangle] {$\ldots$}; &
    \node[rectangle] {22}; &
    \node[rectangle] {23}; &
        \node[draw=none,fill=none] {$  \ \ \ $}; &
    \node[rectangle] {0}; &
    \node[rectangle] {1}; &
    \node[rectangle, fill=tikcol2] {2}; &
    \node[rectangle] {3}; &
    \node[rectangle] {$\ldots$}; &
    \node[rectangle] {22}; &
    \node[rectangle] {23};
    \\
        \node[draw=none,fill=none] {$\ldots  $}; &
         & & & & \node[draw=none,fill=none] {$\ldots $};  &&&& 
         & & & & \node[draw=none,fill=none] {$\ldots $}; &&&& 
         & & & & \node[draw=none,fill=none] {$\ldots $};
        \\
         \node[draw=none,fill=none] {$d-7  $}; &
    \node[rectangle, fill=tikcol2] {0}; &
    \node[rectangle, fill=tikcol2] {1}; &
    \node[rectangle, fill=tikcol2] {2}; &
    \node[rectangle, fill=tikcol2] {3}; &
    \node[rectangle, fill=tikcol2] {$\ldots$}; &
    \node[rectangle, fill=tikcol2] {22}; &
    \node[rectangle, fill=tikcol2] {23};&
        \node[draw=none,fill=none] {$  \ \ \ $}; &
    \node[rectangle] {0}; &
    \node[rectangle] {1}; &
    \node[rectangle, fill=tikcol2] {2}; &
    \node[rectangle] {3}; &
    \node[rectangle] {$\ldots$}; &
    \node[rectangle] {22}; &
    \node[rectangle] {23}; &
        \node[draw=none,fill=none] {$  \ \ \ $}; &
    \node[rectangle] {0}; &
    \node[rectangle] {1}; &
    \node[rectangle, fill=tikcol2] {2}; &
    \node[rectangle] {3}; &
    \node[rectangle] {$\ldots$}; &
    \node[rectangle] {22}; &
    \node[rectangle] {23};    \\  
         \node[draw=none,fill=none] {$d-8  $}; &
    \node[rectangle, fill=tikcol2] {0}; &
    \node[rectangle, fill=tikcol2] {1}; &
    \node[rectangle, fill=tikcol2] {2}; &
    \node[rectangle, fill=tikcol2] {3}; &
    \node[rectangle, fill=tikcol2] {$\ldots$}; &
    \node[rectangle, fill=tikcol2] {22}; &
    \node[rectangle, fill=tikcol2] {23};&
        \node[draw=none,fill=none] {$  \ \ \ $}; &
    \node[rectangle] {0}; &
    \node[rectangle] {1}; &
    \node[rectangle, fill=tikcol2] {2}; &
    \node[rectangle] {3}; &
    \node[rectangle] {$\ldots$}; &
    \node[rectangle] {22}; &
    \node[rectangle] {23}; &
        \node[draw=none,fill=none] {$  \ \ \ $}; &
    \node[rectangle] {0}; &
    \node[rectangle] {1}; &
    \node[rectangle] {2}; &
    \node[rectangle] {3}; &
    \node[rectangle] {$\ldots$}; &
    \node[rectangle] {22}; &
    \node[rectangle] {23};    \\  
         \node[draw=none,fill=none] {$d-9  $}; &
    \node[rectangle] {0}; &
    \node[rectangle] {1}; &
    \node[rectangle, fill=tikcol2] {2}; &
    \node[rectangle] {3}; &
    \node[rectangle] {$\ldots$}; &
    \node[rectangle] {22}; &
    \node[rectangle] {23};&
        \node[draw=none,fill=none] {$  \ \ \ $}; &
    \node[rectangle] {0}; &
    \node[rectangle] {1}; &
    \node[rectangle] {2}; &
    \node[rectangle] {3}; &
    \node[rectangle] {$\ldots$}; &
    \node[rectangle] {22}; &
    \node[rectangle] {23}; &
        \node[draw=none,fill=none] {$  \ \ \ $}; &
    \node[rectangle] {0}; &
    \node[rectangle] {1}; &
    \node[rectangle] {2}; &
    \node[rectangle] {3}; &
    \node[rectangle] {$\ldots$}; &
    \node[rectangle] {22}; &
    \node[rectangle] {23};    \\  
        \node[draw=none,fill=none] {$\ldots  $}; &
         & & & & \node[draw=none,fill=none] {$\ldots $};  &&&& 
         & & & & \node[draw=none,fill=none] {$\ldots $}; &&&& 
         & & & & \node[draw=none,fill=none] {$\ldots $};
        \\
         \node[draw=none,fill=none] {$d-36  $}; &
    \node[rectangle] {0}; &
    \node[rectangle] {1}; &
    \node[rectangle, fill=tikcol2] {2}; &
    \node[rectangle] (main_sub_d0_h3) {3}; &
    \node[rectangle] {$\ldots$}; &
    \node[rectangle] {22}; &
    \node[rectangle] {23};&
        \node[draw=none,fill=none] {$  \ \ \ $}; &
    \node[rectangle] {0}; &
    \node[rectangle] {1}; &
    \node[rectangle] {2}; &
    \node[rectangle] (other_sub_d0_h3) {3}; &
    \node[rectangle] {$\ldots$}; &
    \node[rectangle] {22}; &
    \node[rectangle] {23}; &
        \node[draw=none,fill=none] {$  \ \ \ $}; &
    \node[rectangle] {0}; &
    \node[rectangle] {1}; &
    \node[rectangle] {2}; &
    \node[rectangle] (solar_sub_d0_h3) {3}; &
    \node[rectangle] {$\ldots$}; &
    \node[rectangle] {22}; &
    \node[rectangle] {23};    \\      
    };

     \node[draw=none,fill=none, anchor=south] at (day_node.north){day};

     \node[draw=none,fill=none, anchor=south] at (main_d0_h3.north){$\ \ \ \ $ target price class ($h$=2)};
     \node[draw=none,fill=none, anchor=south] at (other_d0_h3.north){\color{white} g \color{black} other class };
     \node[draw=none,fill=none, anchor=south] at (solar_d0_h3.north){planned class};

          \node[draw=none,fill=none, anchor=north] (leg1) at (main_sub_d0_h3.south){hour};
          \node[draw=none,fill=none, anchor=north] at (other_sub_d0_h3.south){hour};
          \node[draw=none,fill=none, anchor=north] at (solar_sub_d0_h3.south){hour};

   \node[rectangle, draw,fill=tikcol3, anchor=north] (leg1a) at ([yshift=-9pt, xshift=-40pt]leg1.south){X};
   \node[draw=none,fill=none, anchor=west] (leg2) at (leg1a.east){ 
   \small
$   \begin{array}{l}
 \text{target} \\ \text{response}     
   \end{array}$};

   \node[rectangle, draw,fill=tikcol2, anchor=west] (leg2a) at ([xshift=10pt]leg2.east){X};
   \node[draw=none,fill=none, anchor=west] (leg3) at (leg2a.east){ \small
$   \begin{array}{l}
 \text{possible} \\ \text{dependency}     
   \end{array}$};

      \node[rectangle, draw,fill=tikcol1, anchor=west] (leg3a) at ([xshift=10pt]leg3.east){X};
   \node[draw=none,fill=none, anchor=west] (leg4) at (leg3a.east){ \small
$   \begin{array}{l}
 \text{no possible} \\ \text{dependency}     
   \end{array}$};
   
      \node[rectangle, draw,fill=tikcol4, anchor=west] (leg4a) at ([xshift=10pt]leg4.east){X};
   \node[draw=none,fill=none, anchor=west] (leg5) at (leg4a.east){ \small
$   \begin{array}{l}
 \text{not} \\ \text{available}     
   \end{array}$};
   
\end{tikzpicture}
\caption{Illustration of the dependency structure for a target {bid volume process of a price class} at hour 2.}
\label{fig_illustration_dep}
\end{figure}

For the estimation of the parameters in \eqref{eq_main_ar_model} we use a method of high-dimensional statistics, namely the lasso 
estimation procedure, introduced by \cite{tibshirani1996regression}. Recently it was also used in a context of electricity price 
forecasting by \cite{ziel2015efficient}, \cite{ludwig2015putting}, \cite{ziel2016forecasting} and \cite{gaillard2016additive}.

The lasso estimator is a penalized least square estimator, thus we require the regression
representation of model \eqref{eq_main_ar_model}.
Therefore, let the multivariate ordinary least squares representation of \eqref{eq_main_ar_model}
be: 
\begin{align}
Y_{m,d,h} &=  \X_{m,d,h} \bsbeta_{m,h} + \eps_{m,d,h}, 
\label{eq_model_ols}
\end{align}
with the $p_{m,h}$-dimensional vector of regressors $\X_{m,d,h} = (\X_{m,d,h,1}, \ldots, \X_{m,d,h,p_{m,h}})'$ and $\bsbeta_{m,h} = (\bsbeta_{m,h,1}, \ldots, \bsbeta_{m,h,p_{m,h}})'$ as $p_{m,h}$-dimensional parameter vector.
For the considered lasso estimation procedure it is also important that the regressors in 
\eqref{eq_model_ols} are standardized. 
Thus, we introduce 
with
\begin{align}
\wtilde{Y}_{m,d,h} &=  \wtilde{\X}_{m,d,h} \wtilde{\bsbeta}_{m,h} + \wtilde{\eps}_{m,d,h}, 
\label{eq_model_ols_scale}
\end{align}
the standardized version of equation \eqref{eq_model_ols}. Here 
$\wtilde{Y}_{m,d,h}$ and the elements of $\wtilde{\X}_{m,d,h}$ are scaled so that they have a variance of 1.
Given the scaled parameter vector $\wtilde{\bsbeta}_{m,h}$ we can easily reproduce $\bsbeta_{m,h}$ by rescaling.
We estimate the scaled parameter vectors $\wtilde{\bsbeta}_{m,h}$ given $n$ observable days by using the lasso estimator $\what{\wtilde{\bsbeta}}_{m,h}$:
   \begin{align}
   \what{\wtilde{\bsbeta}}_{m,h} &= \argmin_{\bsbeta \in \R^{p_{m,h}}} 
\sum_{d=1}^n (\wtilde{Y}_{m,d,h} -  \wtilde{\X}_{m,d,h} \bsbeta)^2
+ \lambda_{m,h} \sum_{j=1}^{p_{m,h}} |\bsbeta_{j} | 
\label{eq_lasso}
   \end{align}
where $\lambda_{m,h}\geq 0$ is a penalty parameter. 
Note that for $\lambda_{m,h}= 0$ we receive the common ordinary least square estimator 
and for sufficiently large $\lambda_{m,h}$ we get $\what{\wtilde{\bsbeta}}_{m,h} = \bsnull$.
{In general, the lasso estimator is a biased estimator. 
However, under certain regularity conditions the lasso estimator is consistent and asymptotic normal for the non-zero parameter components.
For example, if we impose stationarity to the underlying process $\wtilde{Y}_{m,d,h}$ {we arrive at the mentioned asymptotic properties.}
Still, even if the process is heteroscedastic or only periodically stationary we can achieve the same asymptotic results, see e.g. \cite{ziel2016iteratively}.
But in general, it holds roughly spoken, the more the stationarity assumption of process is violated and the stronger the correlation structure in the process
the worse the convergence behavior of the lasso estimator. For more theoretical and applied details on the lasso estimator we suggest \cite{hastie2015statistical}.
}

As estimation algorithm we use the
coordinate descent approach of \cite{friedman2007pathwise} which 
is a fast estimation procedure. { For implementation we use the \texttt{R} package \texttt{glmnet}, see \cite{friedman2010regularization}. }
The algorithm solves the lasso problem 
on a given grid $\Lambda_{m,h}$ of $\lambda_{m,h}$ values. This grid $\Lambda_{m,h}$ is usually
chosen to be exponential decaying.
Given a grid $\Lambda_{m,h}$, we select our optimal tuning parameter $\lambda_{m,h}$ by minimizing the
popular Baysian information criteria (BIC) which performs conservative model selection. 
{However, the tuning parameter could be chosen by another information criteria. 
Cross-validation techniques or test based approaches as introduced in \cite{lockhart2014significance} might be plausible as well. }

Given the estimated parameters $\what{\wtilde{\bsbeta}}_{m,h}$ for $\wtilde{\bsbeta}_{m,h}$, 
we calculate the lasso estimator  $\what{\bsbeta}_{m,h}$ for $\bsbeta_{m,h}$ by rescaling.
With $\what{\bsbeta}_{m,h}$ which contains the estimates for $\what{\phi}_{m,h, l,j,k}$ and $\what{\psi}_{m,h,k}$ we can compute a day-ahead point forecast by 
$$\what{Y}_{m,n+1,h} = 
\sum_{l = 1}^{M} \sum_{j=1}^{24}  \sum_{k \in \II_{m,h}(l,j)} \what{\phi}_{m,h, l,j,k} Y_{l,n+1-k,j} 
 + \sum_{k=2}^7 \what{\psi}_{m,h,k} W_k(n+1) .$$
 If we have the predicted values $\what{Y}_{1,n+1,h}, \ldots, \what{Y}_{M_S+M_D,n+1,h}$  we obtain the bid volume forecasts $\what{X}_{1,n+1,h}, \ldots, \what{X}_{M_S+M_D,n+1,h}$ by adding the sample means.
{
Using residual based bootstrap as in \cite{ziel2016lasso} we can compute $B$ bootstrap samples
$\what{X}^b_{m,n+1,h} $ for $b\in \{1, \ldots, B\}$. 
To capture the correlation structure of the residuals adequately we sample from the residual vector 
$\what{\bseps}_{d,h} = ( \what{\eps}_{1,d,h}, \ldots, \what{\eps}_{M_s+M_D,d,h} )'$ only over the days $d$.
So, if $\what{\EE}_d = (\what{\bseps}_{d,0}, \ldots, \what{\bseps}_{d,23})'$ denotes the daily residual vector for a day
we sample from $(\what{\EE}_1, \ldots, \what{\EE}_n)$. This guarantees that the residual correlation structure within the 24 single auctions 
is preserved. The $B$ bootstrap samples together with the reconstruction scheme described in the next subsection
are used to receive probabilistic forecasts for $X_{m,n+1,h}$. }

\subsection{Reconstructing bids and price curves} \label{reconstruction}

After computing the forecast $\what{X}_{m,n+1,h}$ for each class $m\in \{1,\ldots, M_S+M_D \}$ and hour $h$
we model the apportionment of the forecasted bid volume $\what{X}_{m,n+1,h}$ for each price class.
This will be useful for computing both point and probabilistic forecasts of the price curves.
Especially, for the probabilistic forecast it is important to understand the 
bidding structure within a price class as we can use it for simulation methods. 
However, for forecasting 
the overall behavior, e.g. if we just want to see if there is a large probability for high prices, the reconstruction of the bids is not that relevant.
We show that the reconstruction of the bids is relevant for the local price behavior, especially to explain price clustering. 

\begin{figure}[htb!]
\centering
\begin{subfigure}[b]{.49\textwidth}
\includegraphics[width=1\textwidth, height= .6\textwidth]{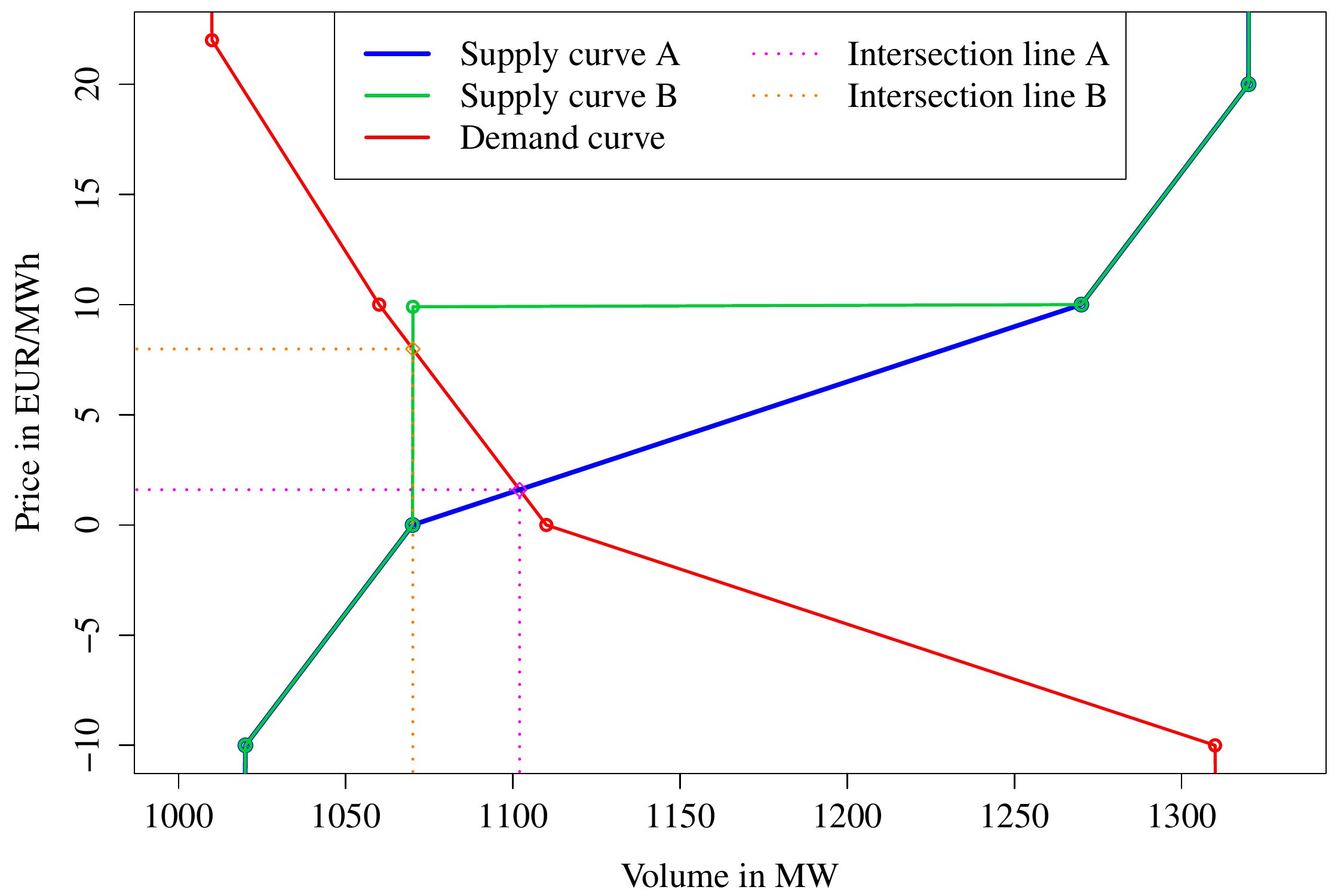}
\caption{Supply and Demand curves with the corresponding market clearing prices.}
  \label{fig_toy_plot}
\end{subfigure}
\begin{subfigure}[b]{.49\textwidth}
\small
\centering Supply, Scenario A \\
\begin{tabular}{rrrrrrr}
  \hline
 Price & -500 & -10 & 0  & 10.0 & 20 & 3000 \\ 
  Volume & 1000 & 20 & 50  & 200.0 & 50 & 70 \\ 
   \hline
\end{tabular}

\vspace{4mm}

\centering Supply, Scenario B \\
\begin{tabular}{rrrrrrrr}
  \hline
 Price & -500 & -10 & 0 & 9.9 & 10.0 & 20 & 3000 \\ 
  Volume & 1000 & 20 & 50 & 0.1 & 199.9 & 50 & 70 \\ 
   \hline
\end{tabular}

\vspace{4mm}

\centering Demand \\
\begin{tabular}{rrrrrrrr}
  \hline
Price & 3000 & 22 & 10 & 0 & -10 & -500 \\ 
  Volume & 1000 & 10 & 50 & 50 & 200 & 20 \\ 
   \hline
\end{tabular}
  \caption{Bidding structure of the supply and demand curves.}
  \label{fig_toy_table}
\end{subfigure}
 \caption{Toy example for two supply scenarios A and B}
 \label{fig_toy}
 \end{figure}

 For example, if we forecast the sale volume of the price class ranging from -55.0 EUR/MWh to 1.3 EUR/MWh 
 to be 1000 MW, we have to redistribute this volume over the different price levels within that class, e.g. -55.0, -54.9, $\ldots$, 1.3, so that 
 the real bidding behavior is captured well. In this example {due to price clustering} it is very likely  that a significant amount of the 1000 MW is bid at 0.0 EUR/MWh, as already explained in the previous section. Furthermore we have to take into account that
 many prices are not bid at all. This is important because of the considered 
 linear interpolation method for creating the price curves. 
 So even a tiny bid of 0.1 MW can have a relatively big impact on the electricity price. 
 This holds for both, bids on the supply and demand side. 
 As this procedure is crucial for our analysis, we briefly discuss this issue in a toy example for a minor change in the supply bidding structure.
 Therefore we consider two scenarios, A and B, for the supply curve and keep the demand curve constant.
The scenario A differs only marginal from the scenario B in the bidding structure. 
In A there are 100 MW offered at 10 EUR/MWh, whereas in B 99.9 MW were bid at 10 EUR/MWh and 0.1 MW are bid 
at 9.9 EUR/MWh. The detailed assumed bids and the corresponding price curves are given in 
Figure  \ref{fig_toy}. 
There we can observe that the supply curve in scenario B looks more rectangular. Indeed, also judging by our dataset market participants on the sale and purchase side seem to aim for a price curve that is close to a step function.
In our short example, scenario A results in a market clearing price of 1.60 at a volume of 1102.0,
whereas the market clearing price in B is 7.98 at a volume of 1070.1. The market clearing price is 6.38 EUR/MWh higher.
This shows exemplarily, that minor change in the bidding structure can cause a severe price change, especially in price areas with only some bids, e.g. very large or very small (negative) prices.
Even though this is a toy example it is surprisingly real. We can assume that the described behavior is known by at least some market participants, as we can observe that some agents try to strategically chose their bids to achieve the rectangular shape of the function.

To take into account whether a certain price is traded or not, we have to model the probability of that event. We will refer to this approach as ``reconstructing'' throughout our paper. 
For reconstructed objects, we will use the accent $ \breve{} $.
 { Remember that $V_{S,t}(P)$ and $V_{S,t}(P)$ denote the bid volume for the supply and demand at price $P\in \P$ at time $t$.
 Similarly as for the bid classes $X^{(c)}_{S,d,h}$ and $X^{(c)}_{D,d,h}$, we introduce the hour-day transformation 
 $V_{S,d,h}(P)$ and $V_{S,d,h}(P)$ of $V_{S,t}(P)$ and $V_{S,t}(P)$ that handles the clock change. }
 We can express the bid volumes $X^{(c)}_{S,d,h}$ and $X^{(c)}_{D,d,h}$ of the price classes by 
 $$X{(c)}_{S,d,h} = \sum_{P \in \P_S(c)} V_{S,d,h}(P) \ \ \text{ and } \ \
 X^{(c)}_{D,d,h} = \sum_{P \in \P_D(c)} V_{D,d,h}(P),$$
the sum of the bid volumes $V_{S,d,h}$ and $V_{D,d,h}$ of the prices within the price classes. 
However, after the price class forecasting we only have bid volumes $X^{(c)}_{S,d,h}$ and $X^{(c)}_{D,d,h}$ available to derive the price bids $V_{S,d,h}(P)$ and $V_{D,d,h}(P)$ for all prices $P\in \P$.
Therefore, we introduce the reconstructed bid volumes $\breve{V}_{S,d,h}(P)$ and $\breve{V}_{D,d,h}(P)$ at price $P\in \P$ for the supply and demand side.
The reconstructed volumes $\breve{V}_{S,d,h}(P)$ and $\breve{V}_{D,d,h}(P)$ should be as close as possible to the true 
bids $V_{S,d,h}(P)$ and $V_{D,d,h}(P)$ for all $P\in \P$.

Let $\pi_{S,d,h}(P)$ and $\pi_{D,d,h}(P)$ be the probabilities that $V_{S,d,h}(P)$ and $V_{D,d,h}(P)$ respectively is greater 
than zero, so there is actually a bid at this price.
We assume these probabilities for the bids 
are constant over time.
We simply estimate $\pi_{S,d,h}(P)$ and $\pi_{D,d,h}(P)$ by the relative frequencies $\what{\pi}_{S,d,h}(P)$ and $\what{\pi}_{D,d,h}(P)$
in the given sample. 

Furthermore, we assume proportionality within the bid prices in the price classes with respect to the mean volume 
$\ov{V}_S(P)$ and $\ov{V}_D(P)$.
Then we can express the reconstructed volumed $\breve{V}_{S,d,h}(P)$ and $\breve{V}_{D,d,h}(P)$ by
\begin{align}
\breve{V}_{S,d,h}(P) &= \frac{R_S(P) \ov{V}_S(P)}{ \sum_{Q\in \P_S(c)} R_{S}(Q)\ov{V}_S(Q)  } X^{(c)}_{S,d,h}, 
\label{eq_ungroup_vol_S}
\\
\breve{V}_{D,d,h}(P) &= \frac{R_D(P) \ov{V}_D(P)}{ \sum_{Q\in \P_D(c)} R_{D}(Q)\ov{V}_D(Q)  } X^{(c)}_{D,d,h} 
\label{eq_ungroup_vol_D}
\end{align}
where 
$c$ is the price class of $\C_S$ or $\C_D$ associated with price $P\in \P$
and $R_S(P) \stackrel{\text{}}{\sim} \text{Ber}(\pi_{S,d,h}(P))$ as well as $R_D(P) \stackrel{\text{}}{\sim}\text{Ber}(\pi_{D,d,h}(P)) $
are Bernoulli random variables with probabilities $\pi_{S,d,h}(P)$ and $\pi_{D,d,h}(P)$. 
We assume that the Bernoulli random variables $R_S(P)$ and $R_D(P)$ are independent from each other 
over the full price grid. { Furthermore, we assume that they are independent from the error term $\bseps_{d,h}$ of the time series model in \eqref{eq_main_ar_model} as well. }


As we have estimates for the probabilities of the Bernoulli random variables $\pi_{S,d,h}$ and $\pi_{D,d,h}$ and 
the mean bid volumes $\ov{V}_S$ and $\ov{V}_D$ we can easily simulate 
$\breve{V}_{S,n+1,h}(P)$ and $\breve{V}_{D,n+1,h}(P)$ by equations \eqref{eq_ungroup_vol_S} and \eqref{eq_ungroup_vol_D}
given {the volume forecast} $X^{(c)}_{S,n+1,h}$ and $X^{(c)}_{D,n+1,h}$ of price classes from the time series model.
These simulations can be utilized to construct forecasts. 
If we only want to receive point estimates for $\breve{V}_{S,n+1,h}(P)$ and $\breve{V}_{D,n+1,h}(P)$
we recommend to set $R_S(P)$ and $R_D(P)$ to one, if 
$\what{\pi}_{S,n+1,h}(P)$ and $\what{\pi}_{D,n+1,h}(P)$ are greater than a certain threshold and to zero otherwise.
For our purpose we will consider the probability threshold of $1/12$. So in our point forecasts a price is active
if it occurs in average at least twice a day.
{ For probabilistic forecasts we utilize the bootstrap samples $\what{X}^b_{m,n+1,h}$. For any bootstrap sample $\what{X}^b_{m,n+1,h}$
we can reconstruct the prices bidding structure using \eqref{eq_ungroup_vol_S} and \eqref{eq_ungroup_vol_D} as well.
As we assume independence between the Bernoulli random variables and the error term, we simply draw from the underlying Bernoulli distributions independently for each bootstrap sample.

Similarly to equation \eqref{eq_price_curve} and \eqref{eq_mean_price_curve} we can calculate the supply and demand volumes $\breve{S}_{d,h}(P)$ and $\breve{D}_{d,h}(P)$ 
associated with the price curves given the volumes $\breve{V}_{S,d,h}(P)$ and $\breve{V}_{D,d,h}(P)$ for the full price grid $\P$ by aggregating
\begin{equation}
\breve{S}_{d,h}(P) = \sum_{ \substack{p \in \breve{\P}_{S,d,h} \\ p\leq P}} \breve{V}_{S,d,h}(p)  \text{ for } P \in \breve{\P}_{S,d,h} \ \ \text{ and } \ \
\breve{D}_{d,h}(P) = \sum_{\substack{p \in \breve{\P}_{D,d,h} \\ p\geq P}} \breve{V}_{D,d,h}(p) \text{ for } P \in \breve{\P}_{D,d,h} ,
\label{eq_reconstr_curve_pairs} 
\end{equation}
where $\breve{\P}_{S,d,h} = \{ P \in \P | R_S(P) =1\}$ and $\breve{\P}_{D,d,h} = \{ P \in \P | R_D(P) =1\}$ 
are the sets of reconstructed bid prices.
As for the sale and purchase curves in \eqref{eq_price_curve} the reconstructed points of the curve
\eqref{eq_reconstr_curve_pairs} must be interpolated linearly to receive the fully reconstructed supply and demand curve.
The intersection of the reconstructed sale and purchase curves $\breve{S}_{d,h}$ and $\breve{D}_{d,h}$ provides the required market clearing volume and price.

}

 \begin{figure}[htb!]
\centering
\begin{subfigure}[b]{.49\textwidth}
\includegraphics[width=1\textwidth]{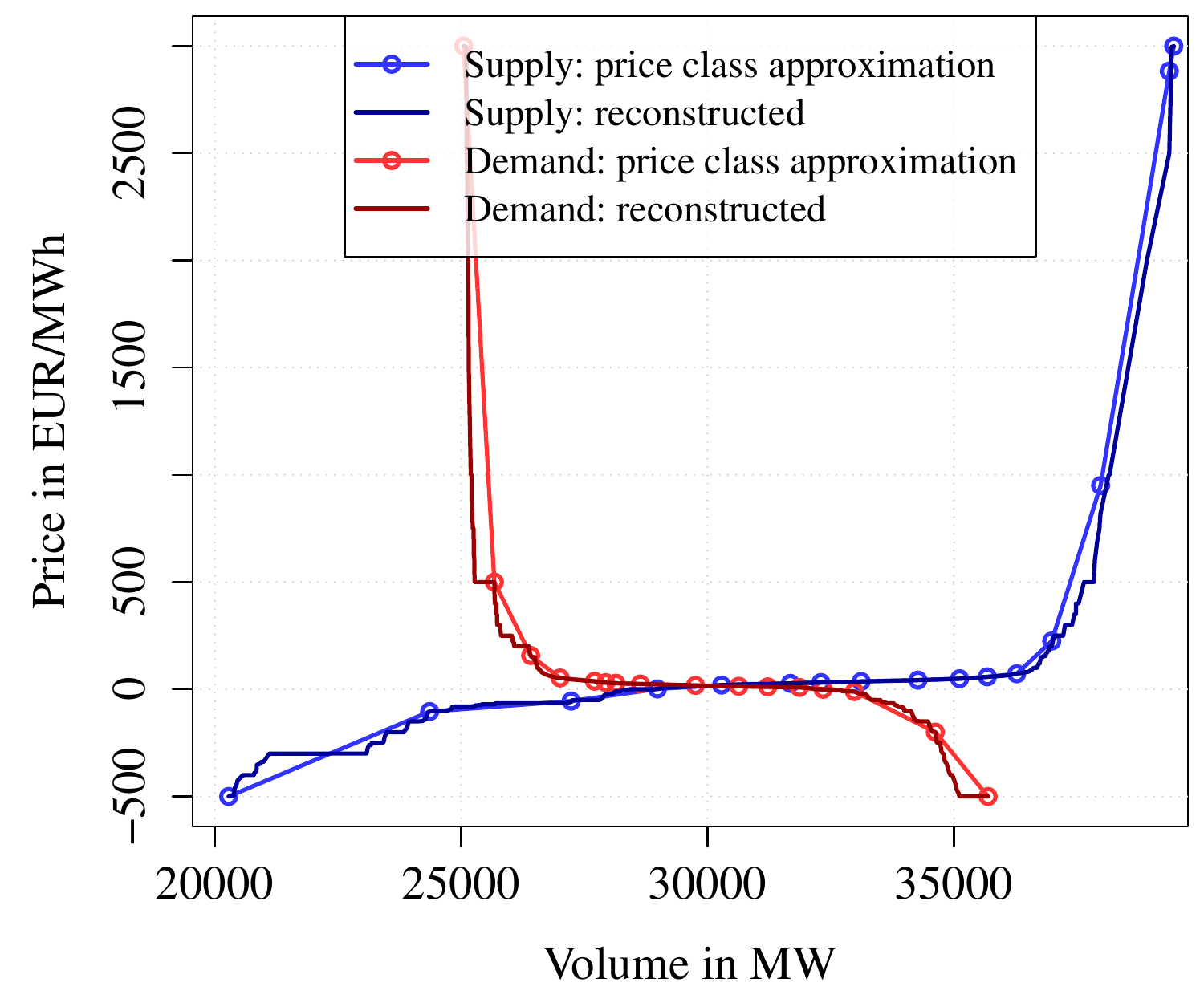}
 \caption{-500 to 3000}
  \label{fig_ungroup1}
\end{subfigure}
\begin{subfigure}[b]{.49\textwidth}
\includegraphics[width=1\textwidth]{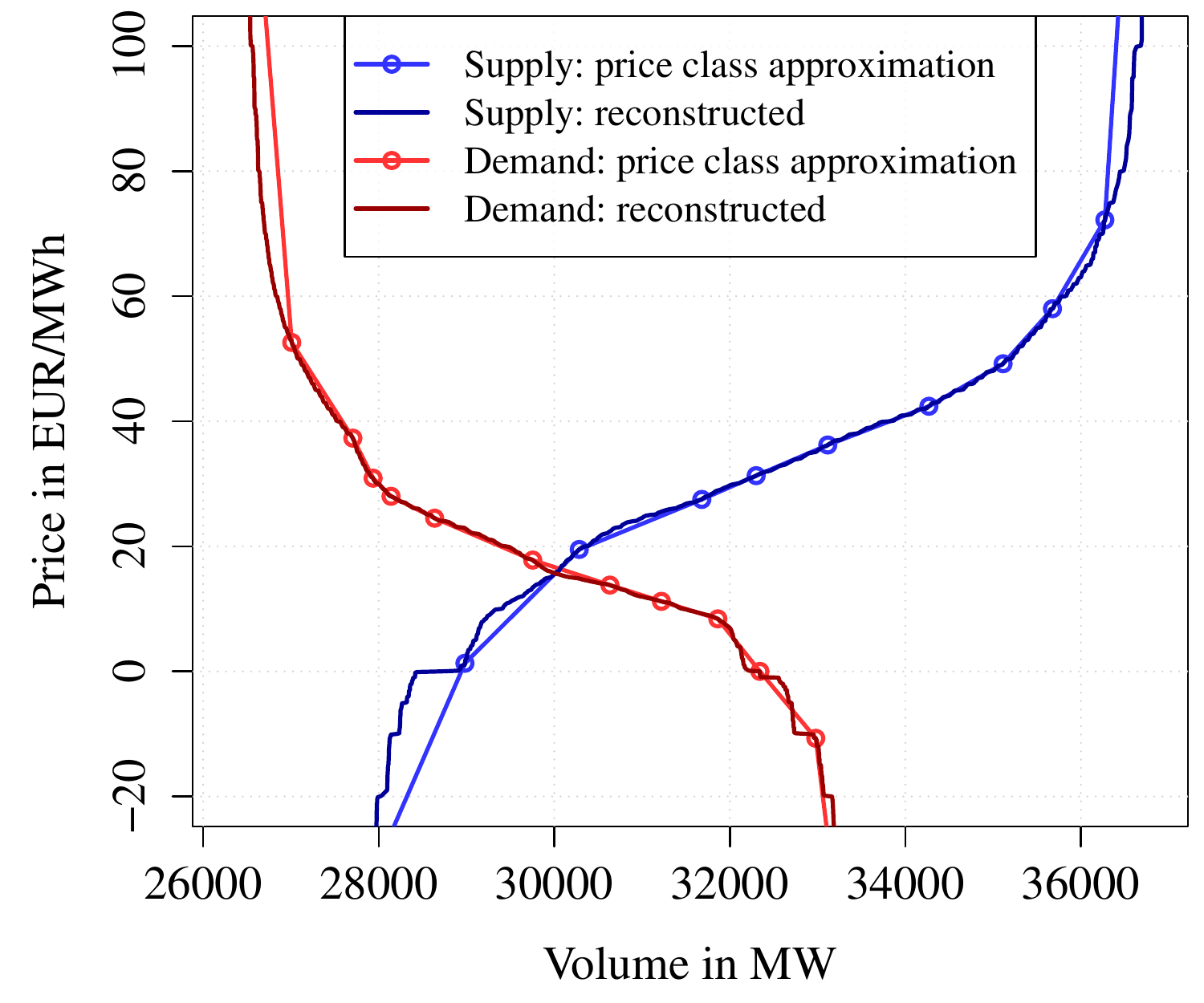}
  \caption{-20 to 100}
  \label{fig_ungroup2}
\end{subfigure}
\caption{Example of price class approximation and reconstructed price curves $\breve{S}_{d,h}$ and $\breve{D}_{d,h}$ for selected price ranges.}
 \label{fig_ungroup}
 \end{figure}
In Figure \ref{fig_ungroup}
illustrates the resulting reconstructed supply and demand curves $\breve{S}_{d,h}$ and $\breve{D}_{d,h}$ with its
plain price class approximation counterpart for a selected example. As mentioned, in the main price region around 20 to 50 with many bids 
the difference is marginal. But in uncommon price regions, e.g. around 0, the impact is larger.
In general the reconstructed price curves look much more realistic than the grouped versions.

\section{Empirical results}

In order to show the results of our X-Model under real world conditions, we performed an rolling window out-of-sample study for the time period from 01.11.2014 to 19.04.2015. 
To evaluate our results, we compare our model with the results of standard models and models used frequently in the literature.
 Additionally, we show a detailed forecasting analysis for three days namely the 19.12.2014, 24.03.2015 and 12.04.2015.
 We chose those days for the following reasons. The first day is suitable to show how price clusters can be predicted.  The second day and third day are these days in the selected out-of-sample data range 
 with the largest positive and negative price spike respectively. 
 All in all, these days are also suitable to show all important features of the model,
 even though they are far from having the best point forecast performance. 
 The detailed forecasting results of all considered days can be found in the appendix. 

 For estimation and forecasting we use for all days in the previous $730=2\times 365$ days (2 years) of data.
 {Note that as we consider a rolling window forecasting study with re-estimation,
 all objects like the estimated price classes $\C_S$ and $\C_D$ as given in Table \ref{tab_price_classes} vary in the out-of-sample period.}
 We forecast the supply and demand curves and compute the corresponding market clearing price and volume
 as described in the previous section. 
 For receiving probabilistic forecasts we perform residual based bootstrap
 with  a bootstrap sample size of $B=10000$. 
 First, we will discuss the forecasted results for the market clearing price and volume of the beforementioned three selected days. This is followed up by the results for the forecasted price curves of some hours of the 12.04.2015. Finally we will show an out-of-sample forecasting study for market clearing price over the whole forecasting period

 For comparing our results regarding the probabilistic forecast, we consider two benchmarks.
As a simple benchmark we take the weekly persistent model, sometimes called naive model,
given by
\begin{align}
X_{\text{price}, d, h} = X_{\text{price}, d-7, h} + \eps_{d,h} \ \text{ with } \ 
\eps_{d,h} \stackrel{\text{iid}}{\sim} \NN(0, \sigma_h^2) .
\label{eq_bench_naiv}
 \end{align}
Furthermore, we take a more advanced regime switching model that is in principle able to cover price spikes. The model,
is very close to the one used in \cite{karakatsani2008forecasting}.
It is a Markov switching model and is given by
\begin{align}
 X_{\text{price},d, h} = \X_{d,h} \bsb_{s(d,h)} + \eps_{s(d,h),h} \ \text{ with } \ \eps_{s(d,h),h} \stackrel{\text{iid}}{\sim}  \NN(0, \sigma^2_{s(d,h),h}) 
\label{eq_bench_rsm}
 \end{align}
with $\X_{d,h}=(1, X_{\text{price},d-1, h}, X_{\text{price},d-7, h}, \ov{X}_{\text{price},d-1}, 
X_{\text{generation},d, h}, X_{\text{wind},d, h}, X_{\text{solar},d, h})$, parameter vector $\bsb_{s(d,h)}$, 
transition probabilities $p_{i,j} = P(s(d,h)=i| s(d-1,h)=j)$ and $s(d,h)$ as the
latent regime at day $d$ and hour $h$ with $ s_{\max}$ possible states.
Here $\ov{X}_{\text{price},d-1}$ is the mean price of the last day. 
Note that the solar component is not included for the hours from 0:00, 1:00, 2:00, 3:00 and 23:00 as there is no solar energy produced during night.
We estimate the regime switching model with $s_{\max} = 2$ regimes by maximizing the likelihood with the EM-algorithm. 
{For all benchmarks we consider the same amount of data as used for the X-Model estimation and forecasting, namely always two years.}



 



\begin{figure}[htb!]
\centering
\begin{subfigure}[b]{.49\textwidth}
\includegraphics[width=1\textwidth, height= .8\textwidth]{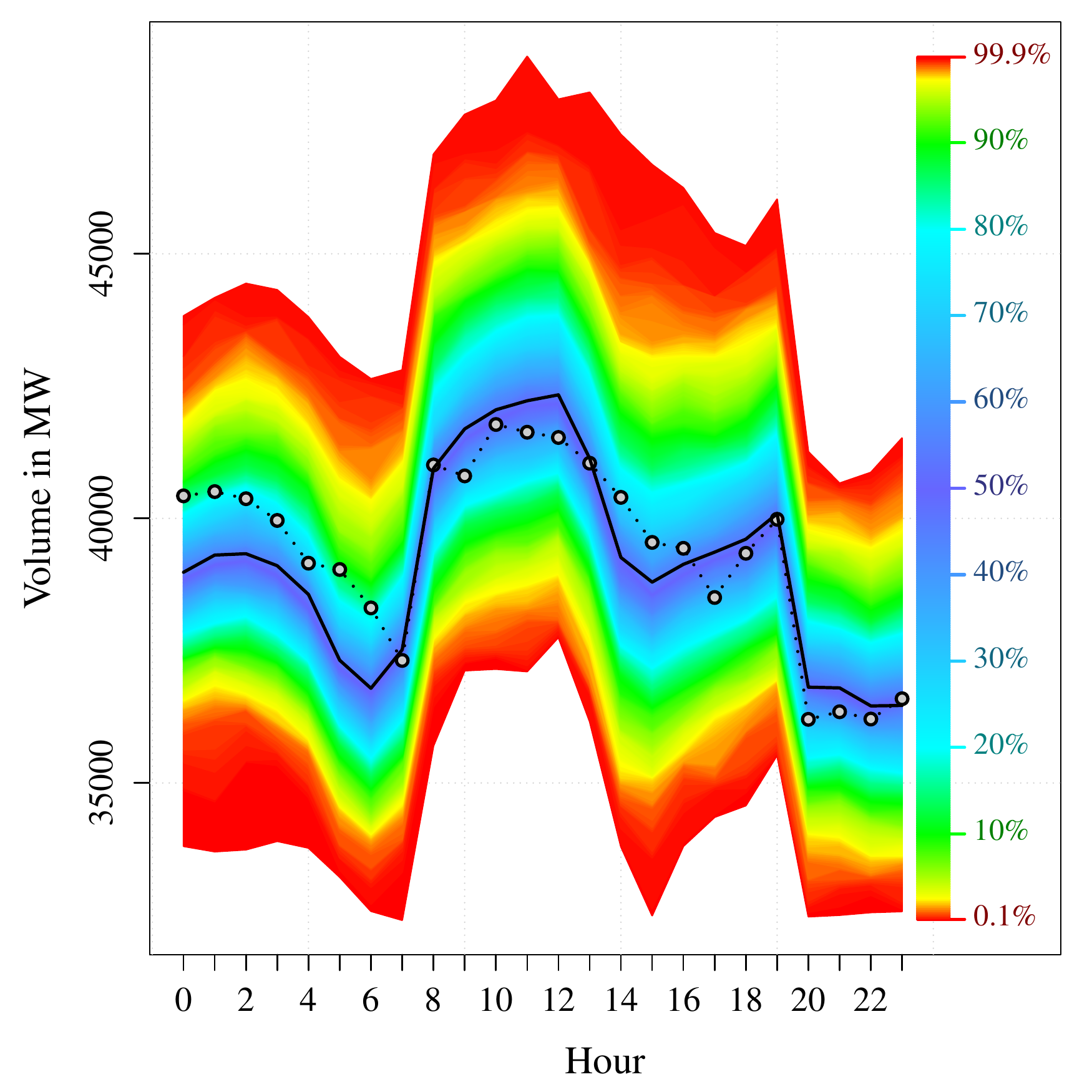}
\caption{Volume forecast for 19.12.2014}
  \label{fig_dmmV_1}
\end{subfigure}
\begin{subfigure}[b]{.49\textwidth}
 \includegraphics[width=1\textwidth, height= .8\textwidth]{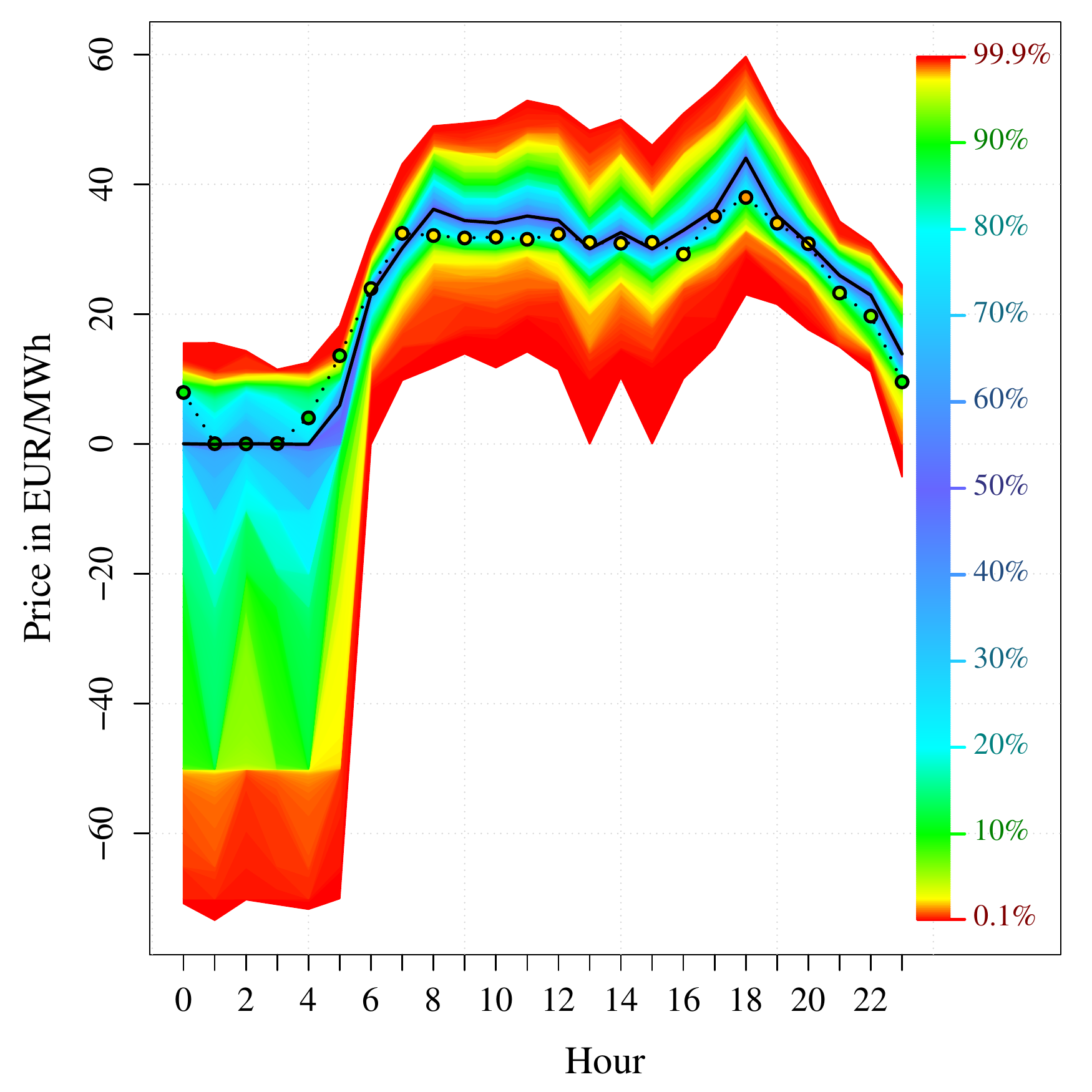} 
  \caption{Price forecast for 19.12.2014}
  \label{fig_dmmP_1}
\end{subfigure}
\begin{subfigure}[b]{.49\textwidth}
\includegraphics[width=1\textwidth, height= .8\textwidth]{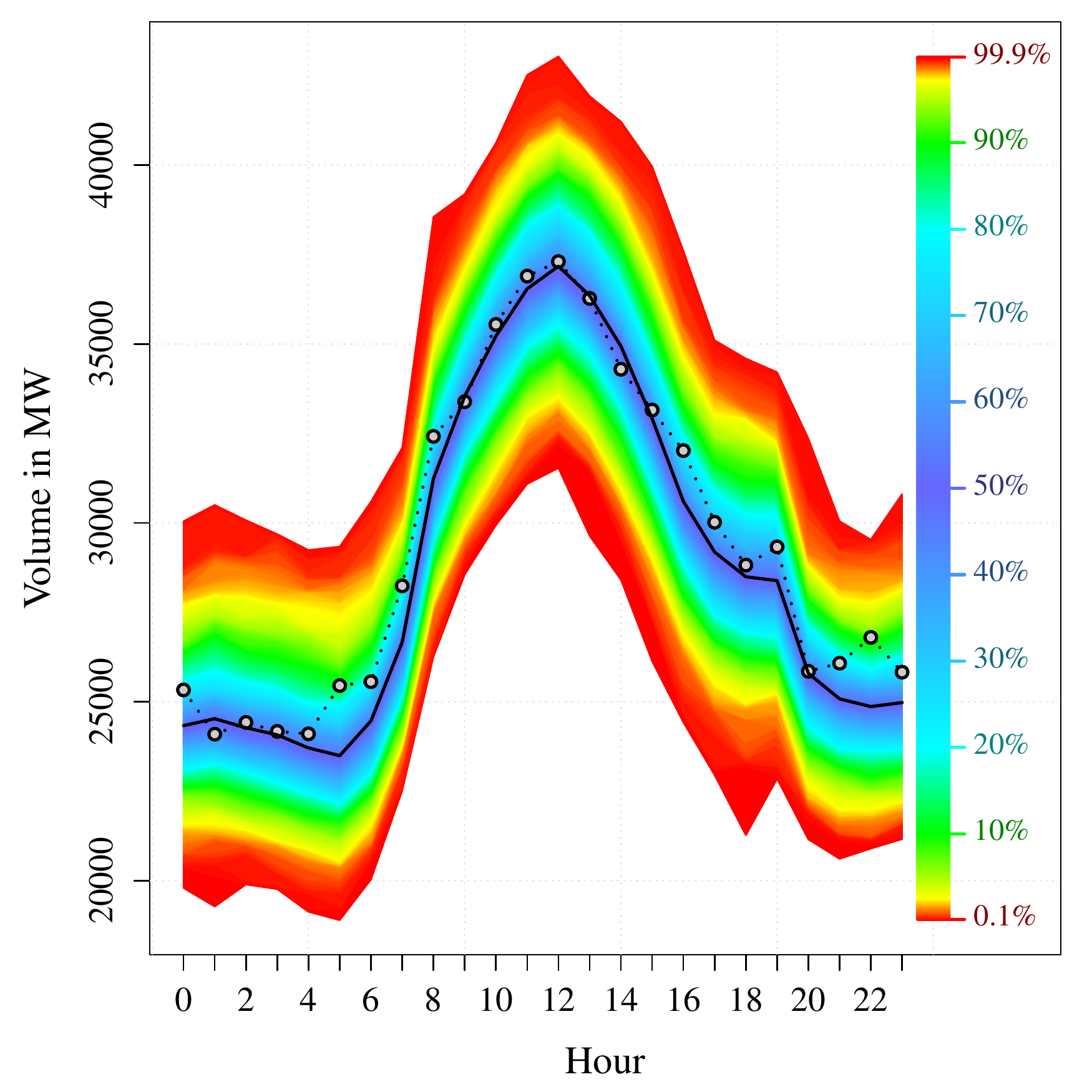}
\caption{Volume forecast for 24.03.2015}
  \label{fig_dmmV_2}
\end{subfigure}
\begin{subfigure}[b]{.49\textwidth}
 \includegraphics[width=1\textwidth, height= .8\textwidth]{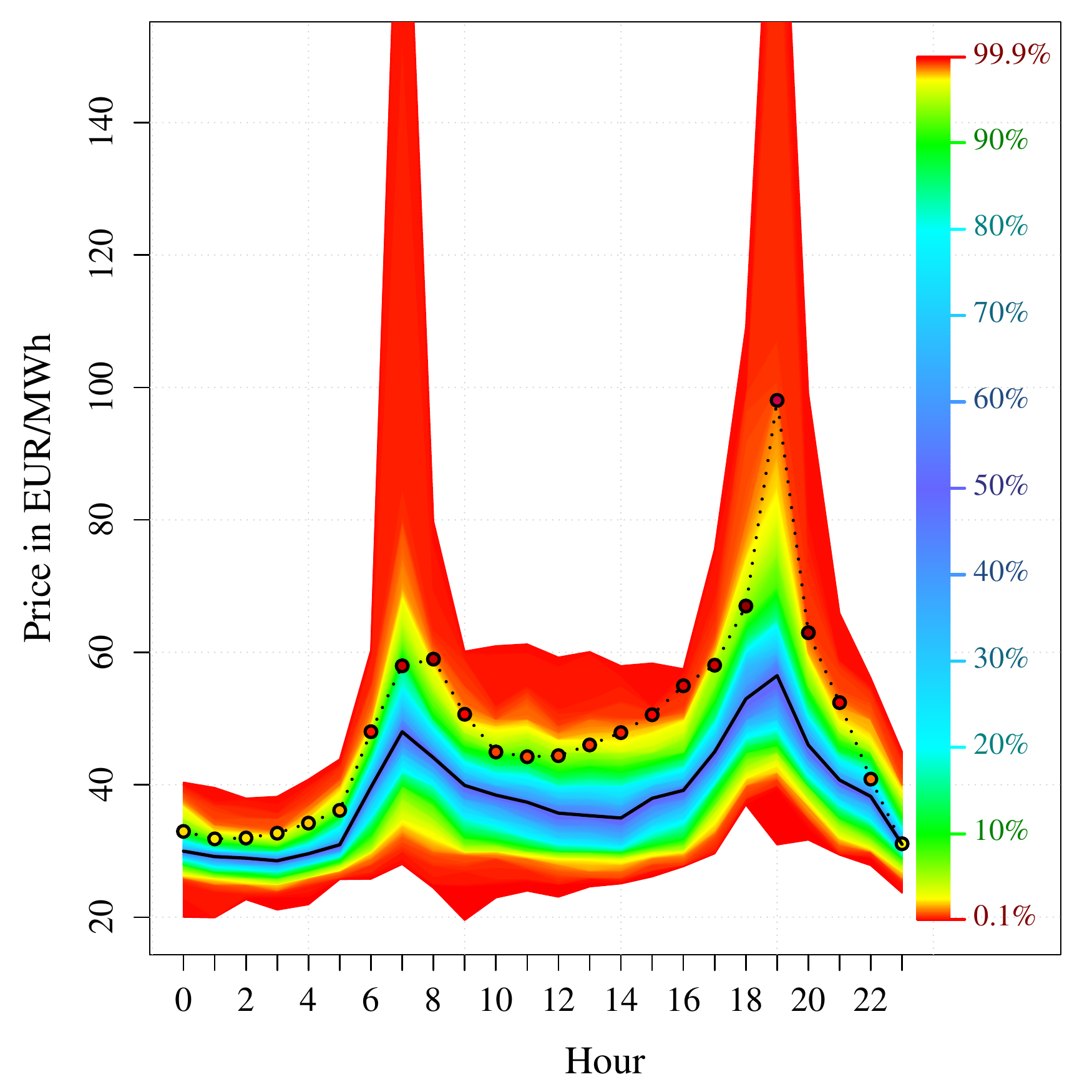} 
  \caption{Price forecast for 24.03.2015}
  \label{fig_dmmP_2}
\end{subfigure}
\begin{subfigure}[b]{.49\textwidth}
\includegraphics[width=1\textwidth, height= .8\textwidth]{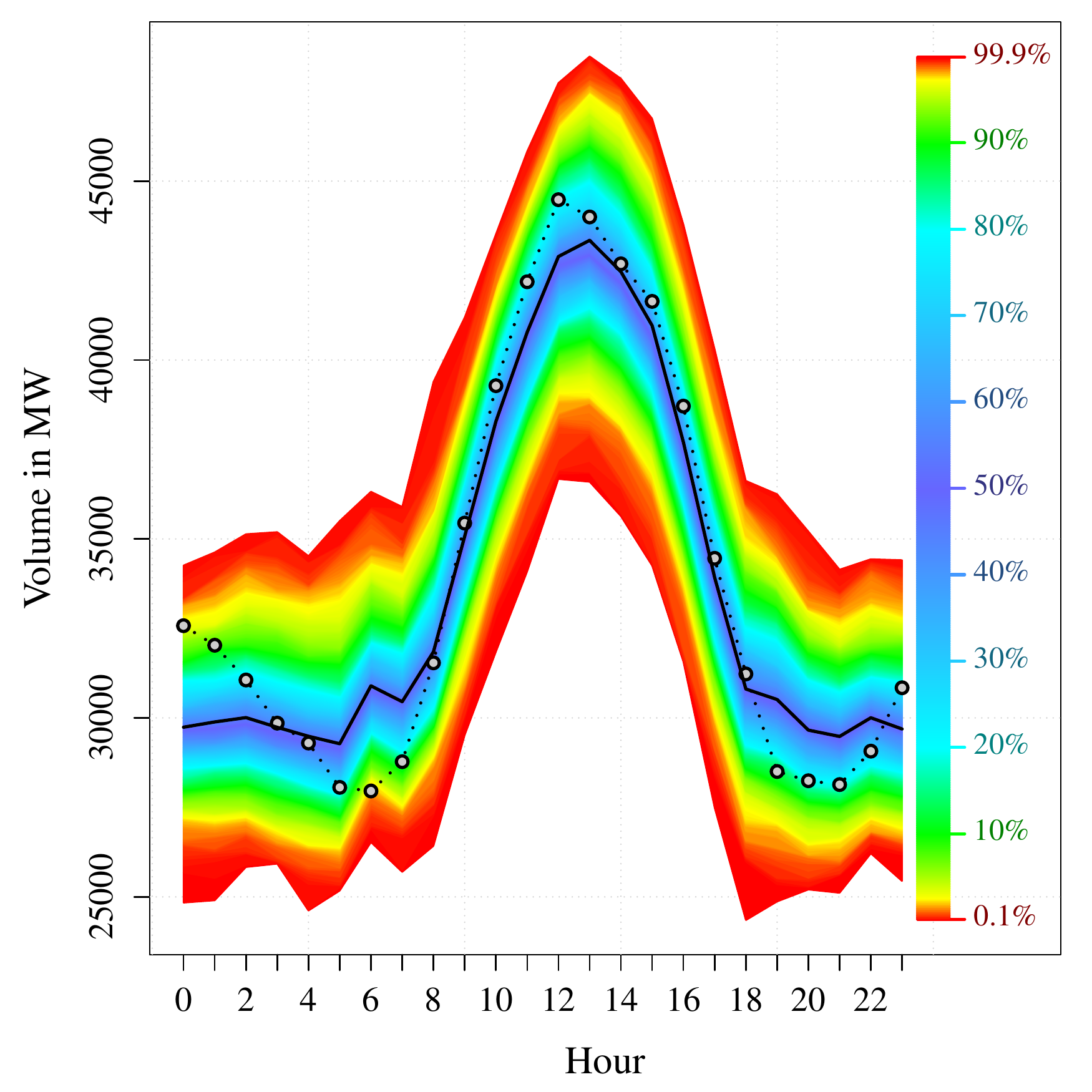}
\caption{Volume forecast for 12.04.2015}
  \label{fig_dmmV_3}
\end{subfigure}
\begin{subfigure}[b]{.49\textwidth}
 \includegraphics[width=1\textwidth, height= .8\textwidth]{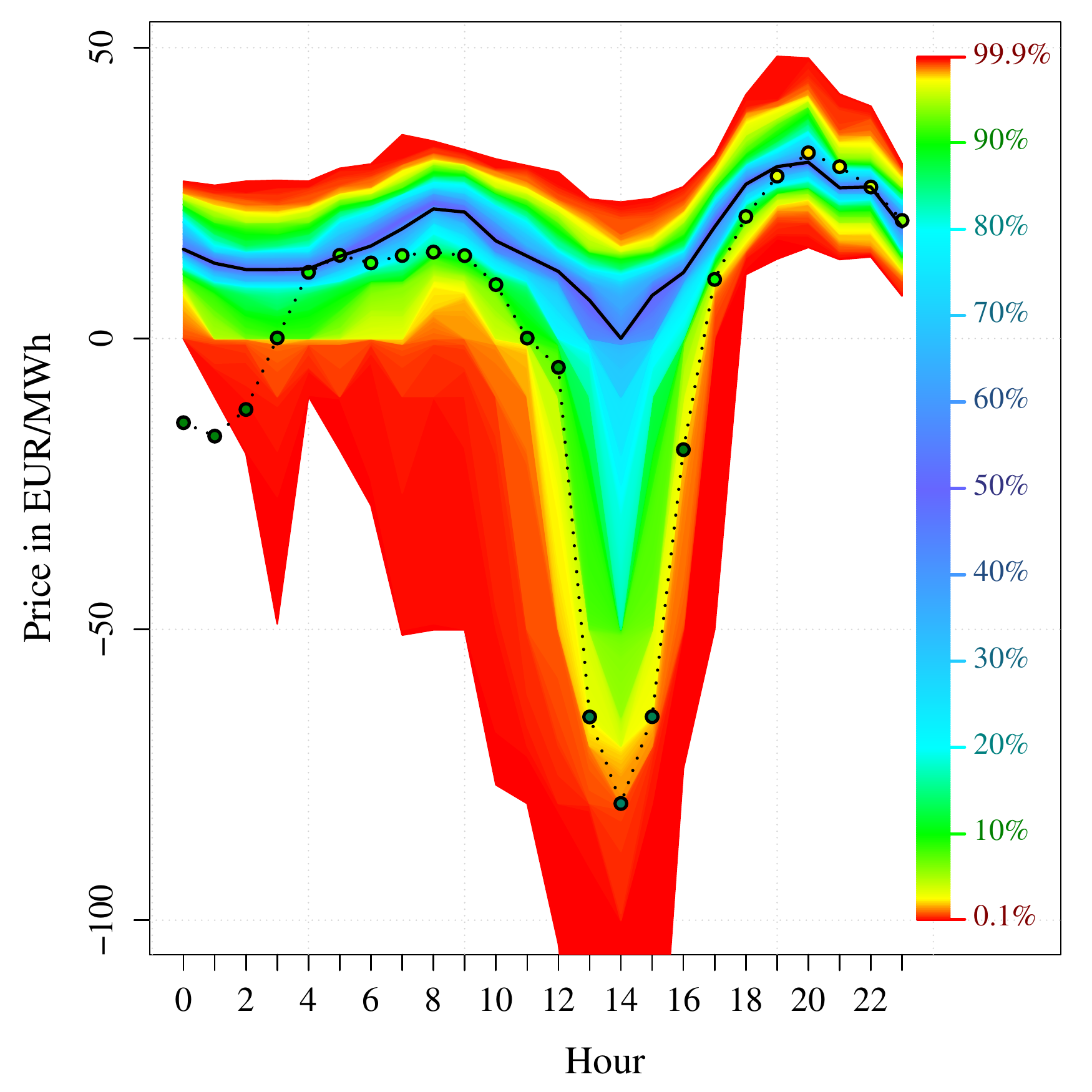} 
  \caption{Price forecast for 12.04.2015}
  \label{fig_dmmP_3}
\end{subfigure}\caption{Probabilistic volume and price forecast of the X-Model with point estimate (black line) and observed values (colored dots) with legend for the 
19.12.2014, 24.03.2015 and 12.04.2015. The observed prices are colored as in Figure \ref{fig_3d-sd-curve-time-example}.}
 \label{fig_price_vol_forecast}
\end{figure}

\begin{figure}[htb!]
\centering
\begin{subfigure}[b]{.49\textwidth}
 \includegraphics[width=1\textwidth, height= .8\textwidth]{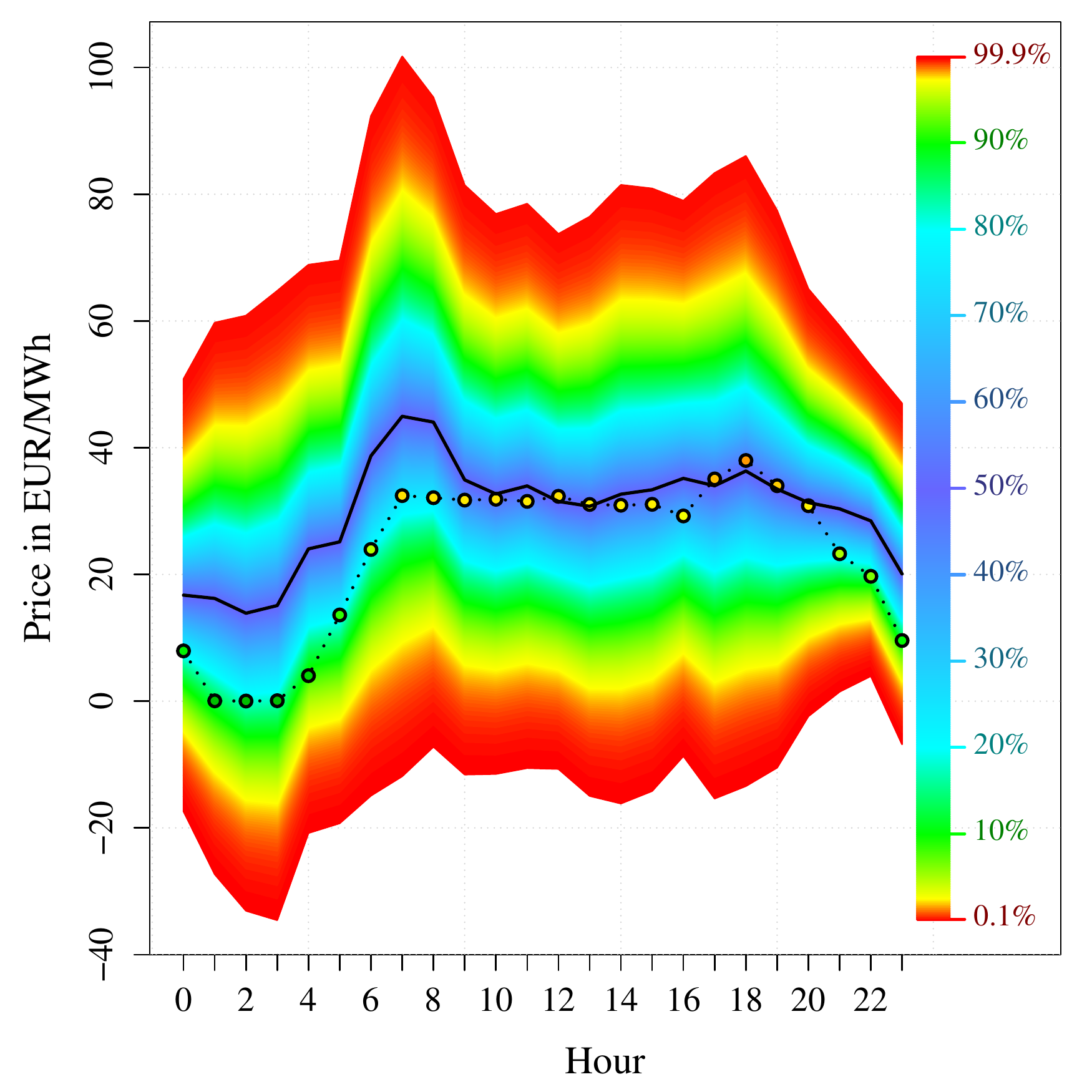} 
  \caption{Persistent model for 19.12.2014}
  \label{fig_B_1_1}
\end{subfigure}
\begin{subfigure}[b]{.49\textwidth}
 \includegraphics[width=1\textwidth, height= .8\textwidth]{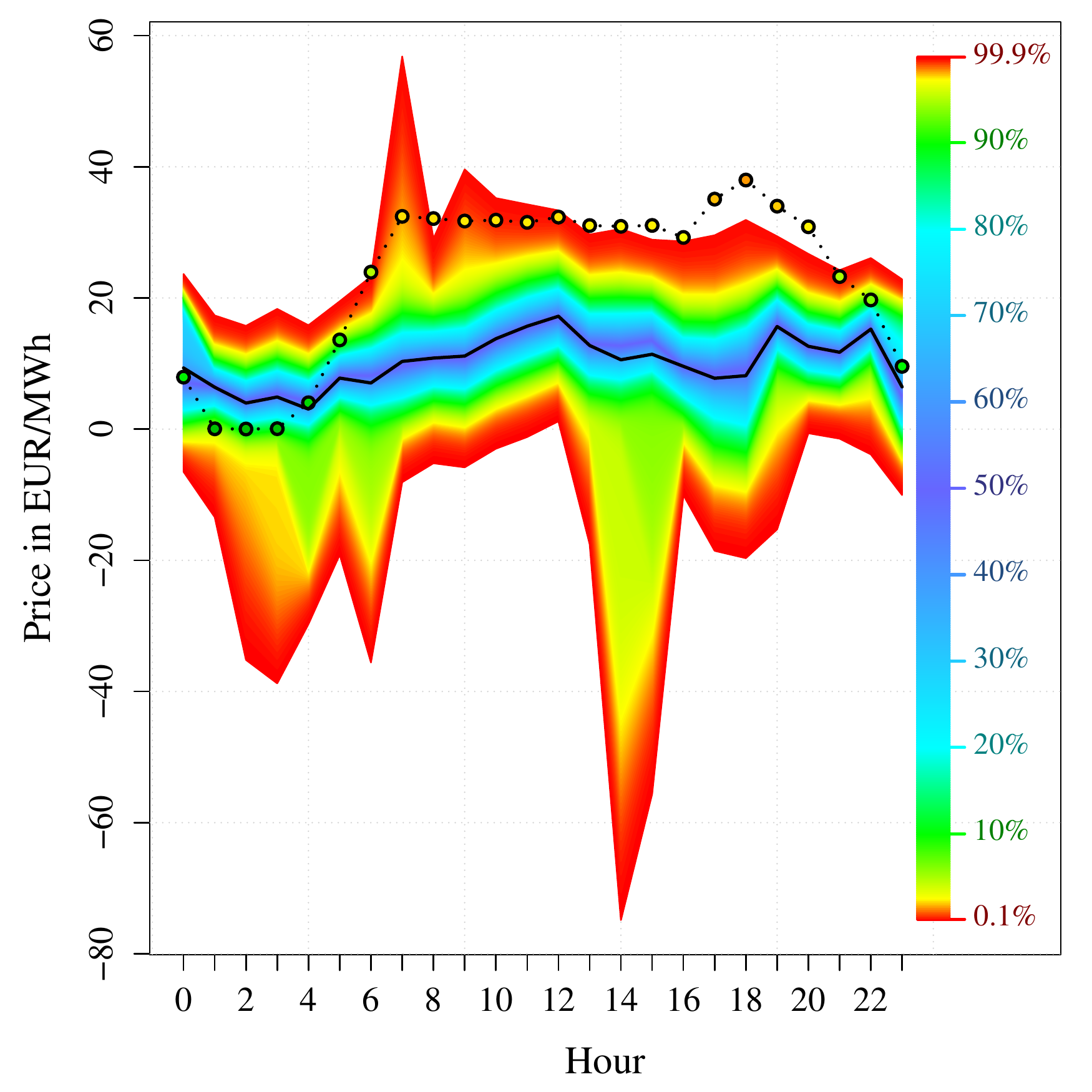} 
  \caption{Markov-Chain-Switching for 19.12.2014}
  \label{fig_B_1_2}
\end{subfigure}
\begin{subfigure}[b]{.49\textwidth}
 \includegraphics[width=1\textwidth, height= .8\textwidth]{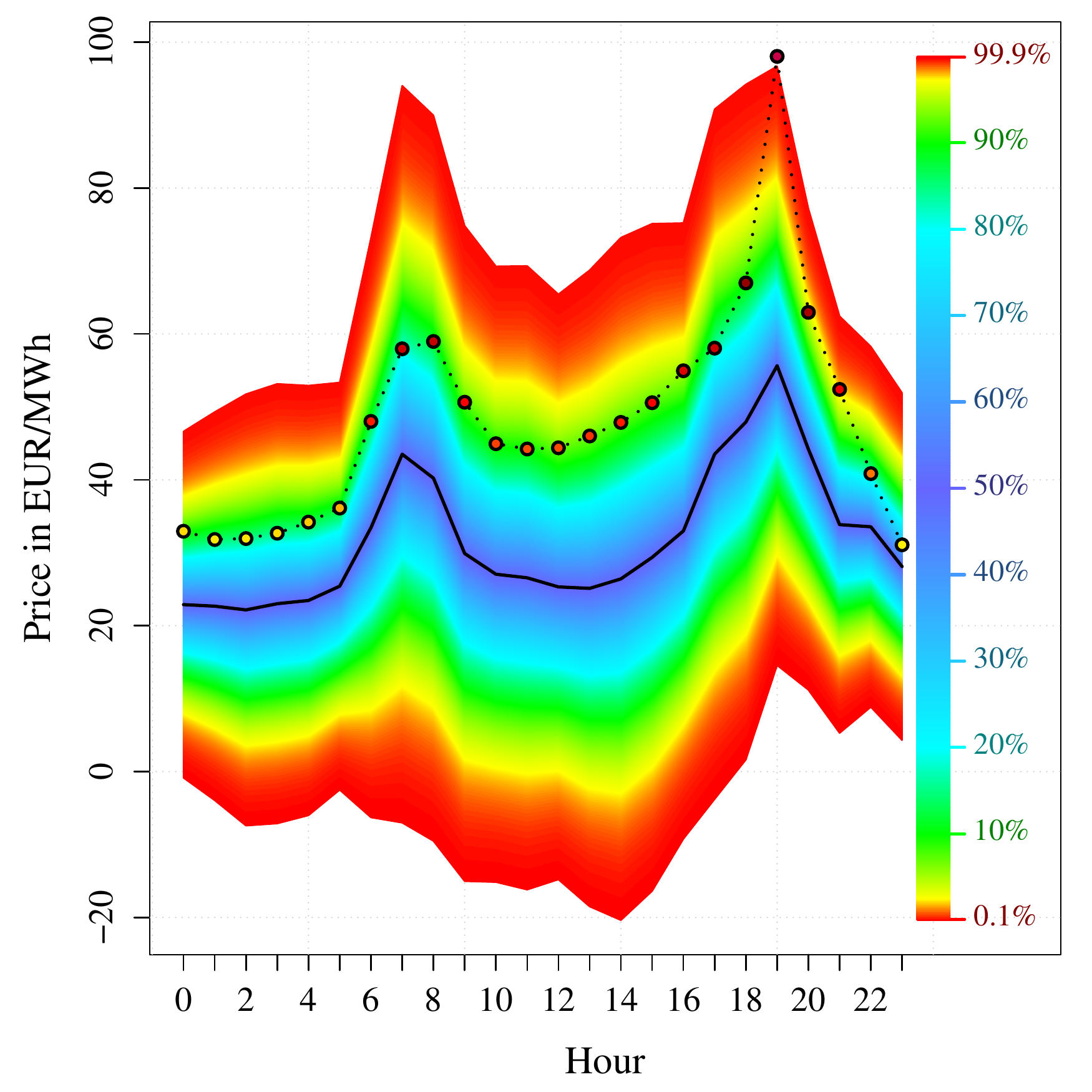} 
  \caption{Persistent model  for 24.03.2015}
  \label{fig_B_2_1}
\end{subfigure}
\begin{subfigure}[b]{.49\textwidth}
 \includegraphics[width=1\textwidth, height= .8\textwidth]{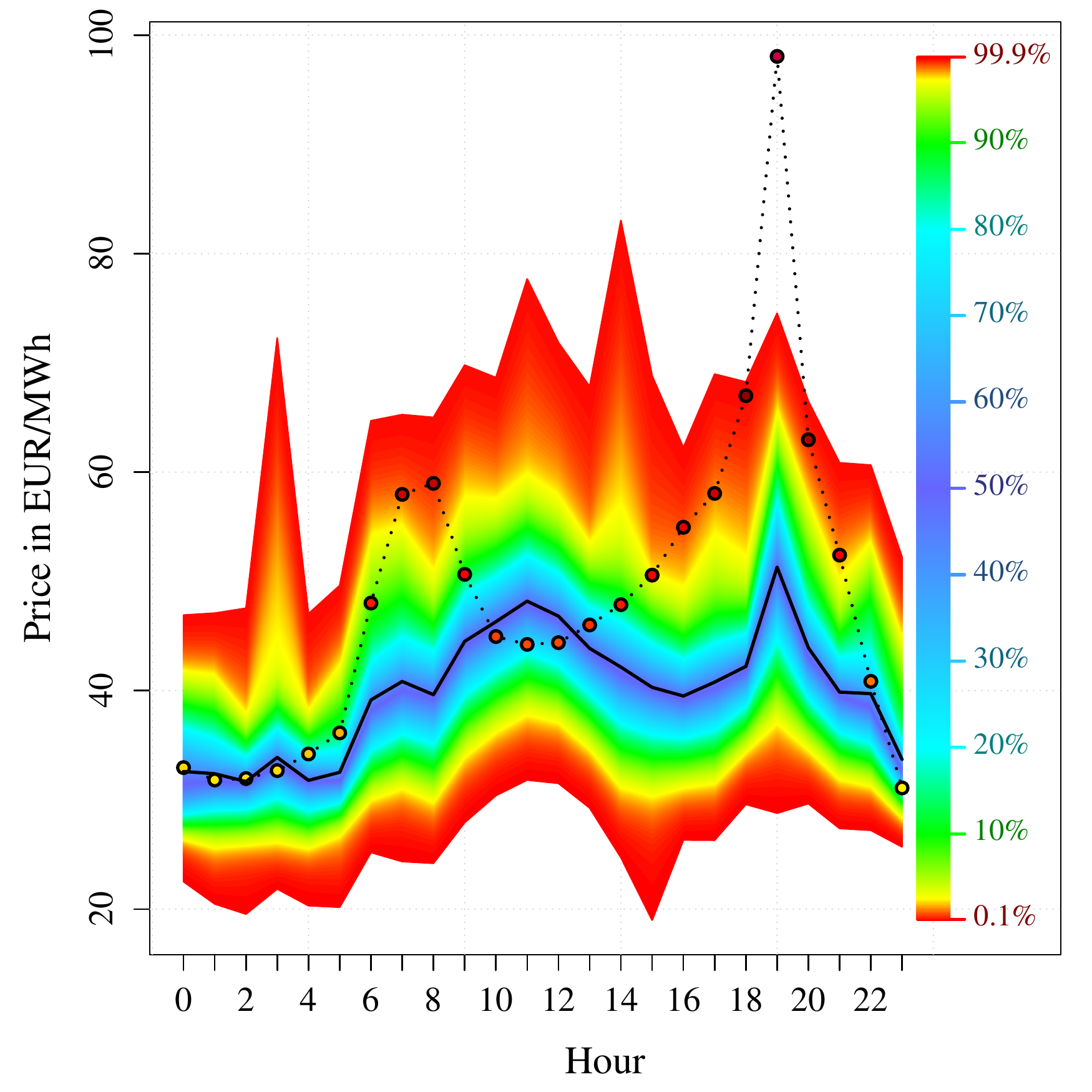} 
  \caption{Markov-Chain-Switching  for 24.03.2015}
  \label{fig_B_2_2}
\end{subfigure}
\begin{subfigure}[b]{.49\textwidth}
 \includegraphics[width=1\textwidth, height= .8\textwidth]{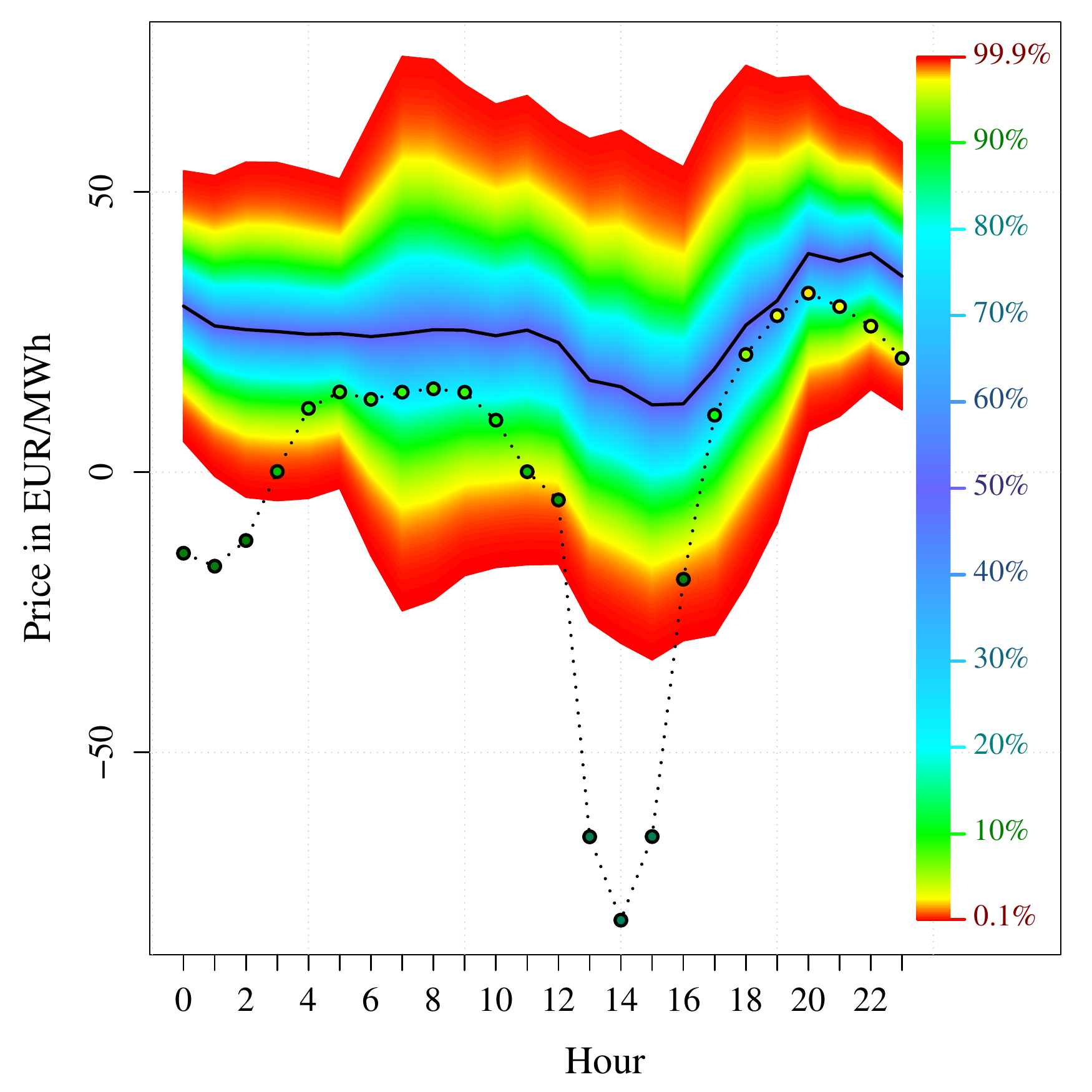} 
  \caption{Persistent model for 12.04.2015}
  \label{fig_B_3_1}
\end{subfigure}
\begin{subfigure}[b]{.49\textwidth}
 \includegraphics[width=1\textwidth, height= .8\textwidth]{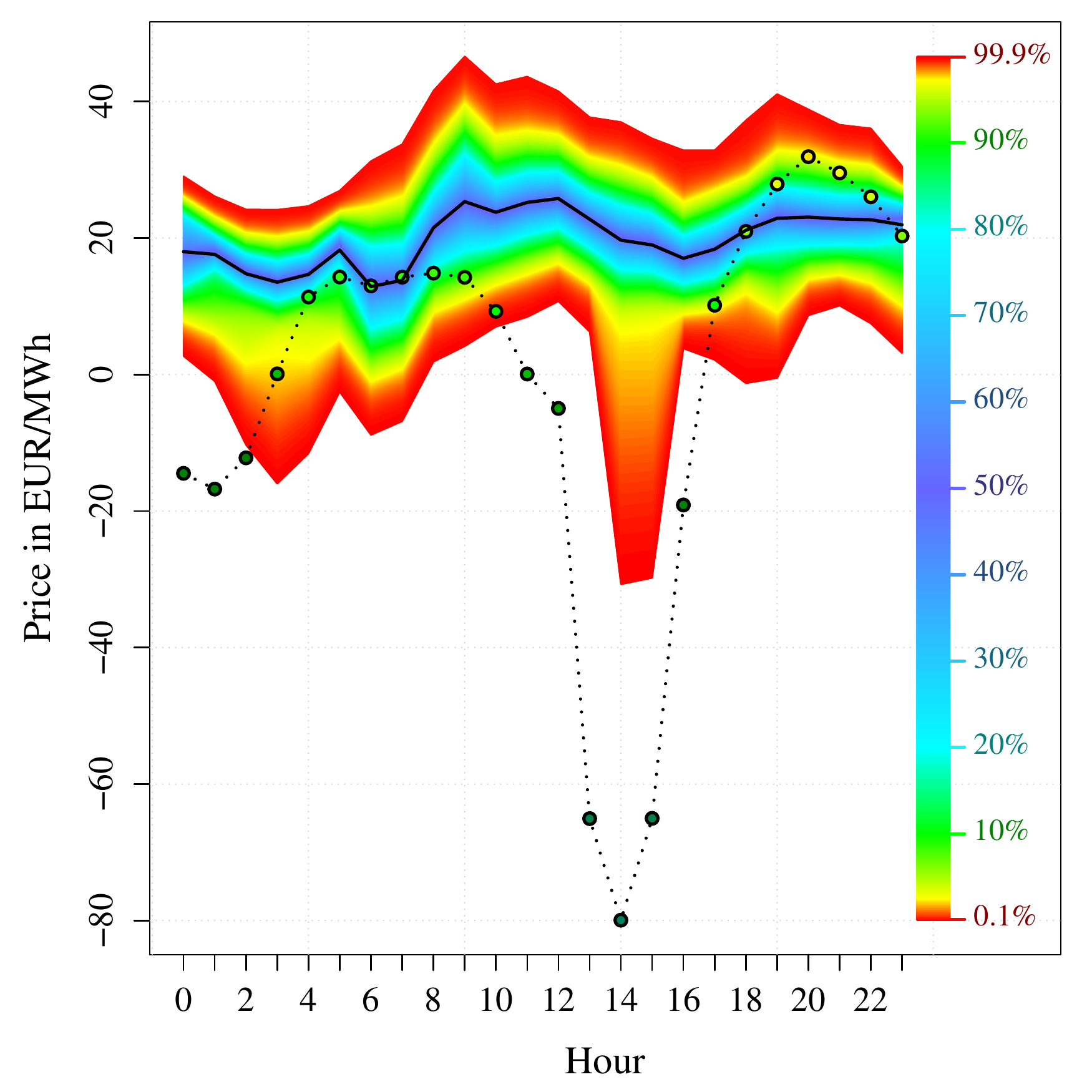} 
  \caption{Markov-Chain-Switching for 12.04.2015}
  \label{fig_B_3_2}
\end{subfigure}
\caption{
Probabilistic price forecast of considered benchmarks with point estimate (black line) and observed values (colored dots) with legend for the 
19.12.2014, 24.03.2015 and 12.04.2015. The observed prices are colored as in Figure \ref{fig_3d-sd-curve-time-example}.}
 \label{fig_price_forecast_bench}
\end{figure}

The results of the X-Model for the market clearing price and the volume of the three chosen days are given in Figure \ref{fig_price_vol_forecast}.
The price forecast of the two benchmarks are given in Figure \ref{fig_price_forecast_bench}.
{Both figures provide probabilistic forecasts for the quantiles ranging from 0.1\% to 99.9\% which can be regarded as prediction intervals.}
 For the volume forecasts in \ref{fig_price_vol_forecast} we see no special behavior, the prediction intervals
 seem to map the daily pattern well.
 The observations lie all clearly within the 95\% prediction bands. 
 Thus, it seems that the method provides reliable forecasting results for the volume of these days.

More interesting are the results of the X-Model for the market clearing prices, in Figures \ref{fig_dmmP_1}, \ref{fig_dmmP_2} and \ref{fig_dmmP_3}. 
There we observe the distinct non-linear behavior of the prices. 
For small (especially negative) prices in \ref{fig_dmmP_1} and \ref{fig_dmmP_3} 
we see clearly left skewed prediction densities. 
Similarly we have noticeable right-skewed prediction densities for large prices as in \ref{fig_dmmP_2}. Therefore, the information of previous auction data seems to capture the increased likelihood of extreme price events very well.

In Figure \ref{fig_dmmP_1} we observe that the beforementioned price clustering at different integer price levels, e.g. at 0, can be modeled by this forecasting method.
For the first four hours of the day the three point forecast for the electricity price were extremely close to zero.
So it was relatively likely to receive values at the price cluster around 0.
And indeed, the three market clearing prices were in this price cluster, namely 0.05 at 1:00, 0.02 at 2:00 and 0.07 at 3:00. 
{In general, we can observe possible price clusters in Figure \ref{fig_price_vol_forecast}. They are at those spots, where the transition between the colors of the legend changes abruptly.
For example in Figure \ref{fig_dmmP_1} at the price cluster at 0 at 2:00 there is an abrupt color change from  cyan to ultramarine and another cluster at -50 with color change from red to yellow. 
}

The forecast plot \ref{fig_dmmP_3} for the 12.04.2015 is also suitable to highlight 
  the difference between 
 common statistical 
 outliers, i.e. random events that can happen, but are extremely rare, and price spikes that are predictable in the sense that the probability for such an event is relatively large given the available information.
 The 12.04.2015 was a Sunday, one week after the Easter holidays. 
 But the 12th April opened with a clearly negative price of -14.47 at 0:00 and reached values between -79.94  to 31.93 during the day.
 The prices of the past week were all in the range of 12.00 to 69.03 with the last observation 
 on 11.04.2015 23:00 at 22.11. Thus, it is usually very complicated to forecast a realistic likelihood for such negative prices with an autoregressive approach. However, our X-Model, which focuses on auction data, seems to have recognized the pattern within the data and provided a realistic confidence interval nonetheless.
 Regarding the prediction bands in \ref{fig_dmmP_3} we see clear changes over the day. 
 It starts quite narrow at 0:00, becomes significantly wider and more left-skewed at around 5:00.
 This peaks at 14:00 where the observed price also reaches its daily minimum.
 Afterwards the prediction intervals become smaller and more symmetric as the 
 forecasts moves closer to common price levels.
 However, for the first hours of the day the negative prices are not predicted by the X-Model. Thus, we have classical outliers. The benchmark models in \ref{fig_B_3_1} and \ref{fig_B_3_2} suffer from the same problem. 
However, it is remarkable that the X-Model predicts a quite large probability for negative prices for the morning and afternoon hours,
 especially from 13:00 to 15:00. For instance, the clear negative price at 13:00 with -65.06 lies clearly within the 99\% prediction intervals. 
 Both benchmark models in Figure \ref{fig_B_3_1} and \ref{fig_B_3_2} were not able to predict these price spikes well.
  Many standard electricity price models only allow for errors where the shape of the density does not depend on 
 the predicted value. As in the persistent model the shape of the prediction density is kept
 constant and simply gets shifted and scaled over time. Thus they are definitely not suitable to capture the real underlying behavior.
 In contrast, the regime switching model in \ref{fig_B_3_2} is in general able to cover price spikes, as 
 the forecast density is a mixture density. We see that at 14:00 and 15:00 the prediction density becomes left skewed, which provides clear indication
 for a price spike. However, the magnitude is not well predicted. 
 
 The largest weakness of all models known so far that are designed for modeling such price spikes
 is that they use only the information of the observed past market clearing prices and related processes like wind and solar energy. The amount of historical extreme prices, which are considered by most common models, is typically very low as they occur only rarely. Hence, such models often simply have too little data points to learn from the behavior at these price levels.  
 
 The X-Model on the other hand uses the bidding information from all time points in all price regions.
 Thus, it can learn a lot about the price behavior in every price region, even for market clearing prices, which were never realized so far.

In general, Figure \ref{fig_price_vol_forecast} shows that the X-Model adopts the non-linear shape of the price curves and hands it down to the forecasts. This automatically adjusts the shape of the prediction densities.

\begin{figure}[htb!]
\centering
 \includegraphics[width=.99\textwidth]{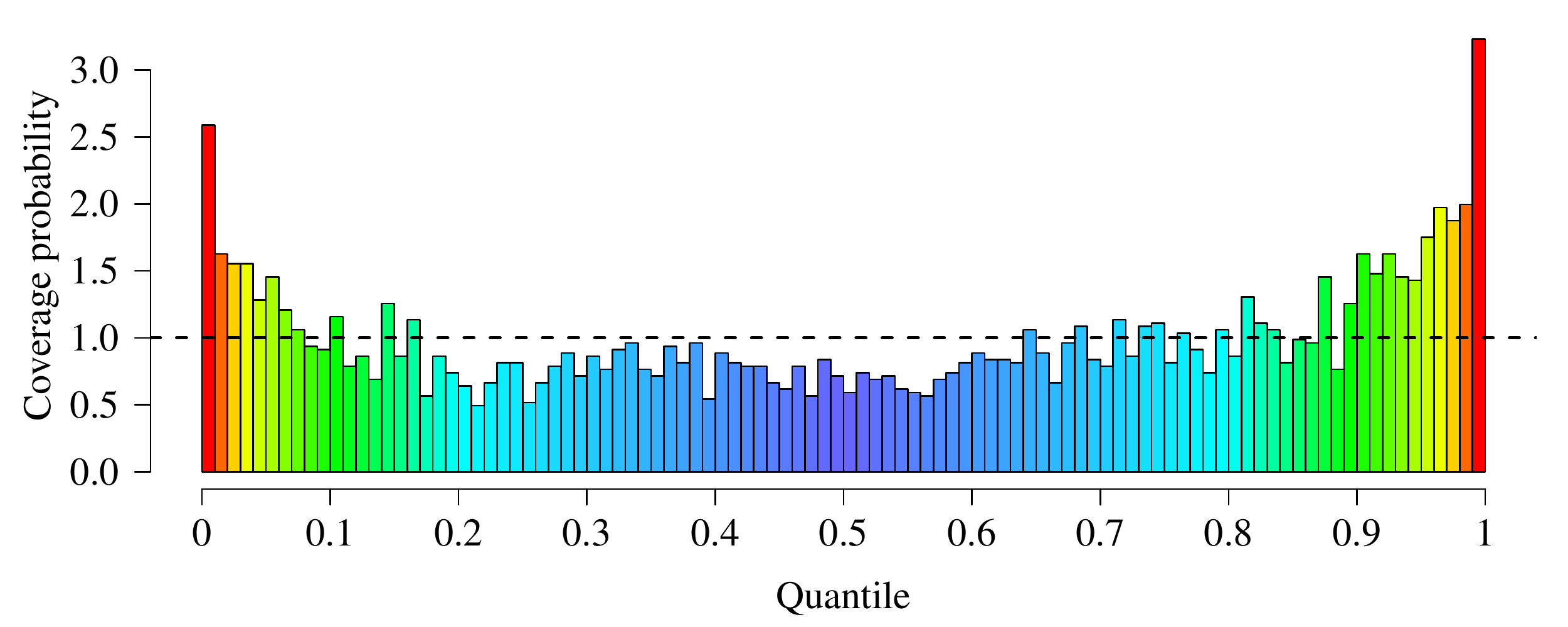} 
\caption{Histogram of the empirical coverage of the X-Model with uniform distribution (dashed line).
The colors for the quantile match these in Figures \eqref{fig_price_vol_forecast} and \eqref{fig_price_forecast_bench}. }
 \label{fig_coverage}
\end{figure}

In Figure  \ref{fig_coverage} the {coverage probability} 
 of all out-of-sample results is visualized. { Each bar represents a different 1\% quantile, whereas the color of the bar matches the specific quantile as shown in e.g. Figure \ref{fig_price_vol_forecast}. The ordinate represents the observed amount of values which fell into a specific estimated quantile divided by the theoretical amount of values of that quantile. If the values for the quantiles were all estimated perfectly, the bars would in our case all have a value of 1.
Nevertheless, we observe that the low and high probability regions around 0 and 1 (especially the yellow to red colored regions) are clearly overrepresented, indicated by their values of greater than 1.5. 
This suggests that the X-Model may estimate too conservative, as it forecasts extreme events with a too small probability.
For the quantile areas around 20\% to 60\% we observe an underrepresentation. }

\begin{figure}[htb!]
\centering
\begin{subfigure}[b]{.49\textwidth}
\includegraphics[width=1\textwidth, height= .6\textwidth]{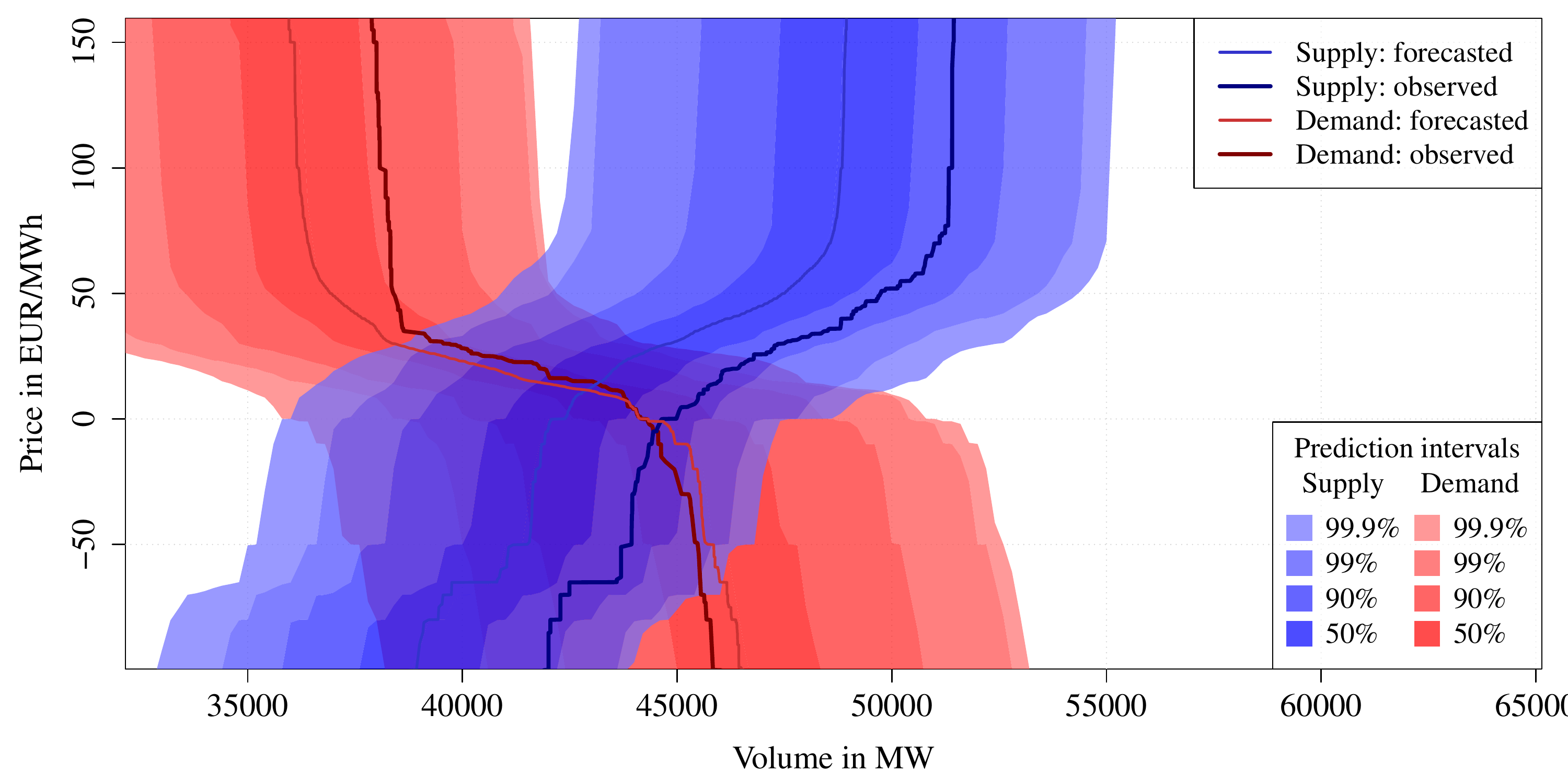}
\caption{12:00}
  \label{fig_dc_1}
\end{subfigure}
\begin{subfigure}[b]{.49\textwidth}
 \includegraphics[width=1\textwidth, height= .6\textwidth]{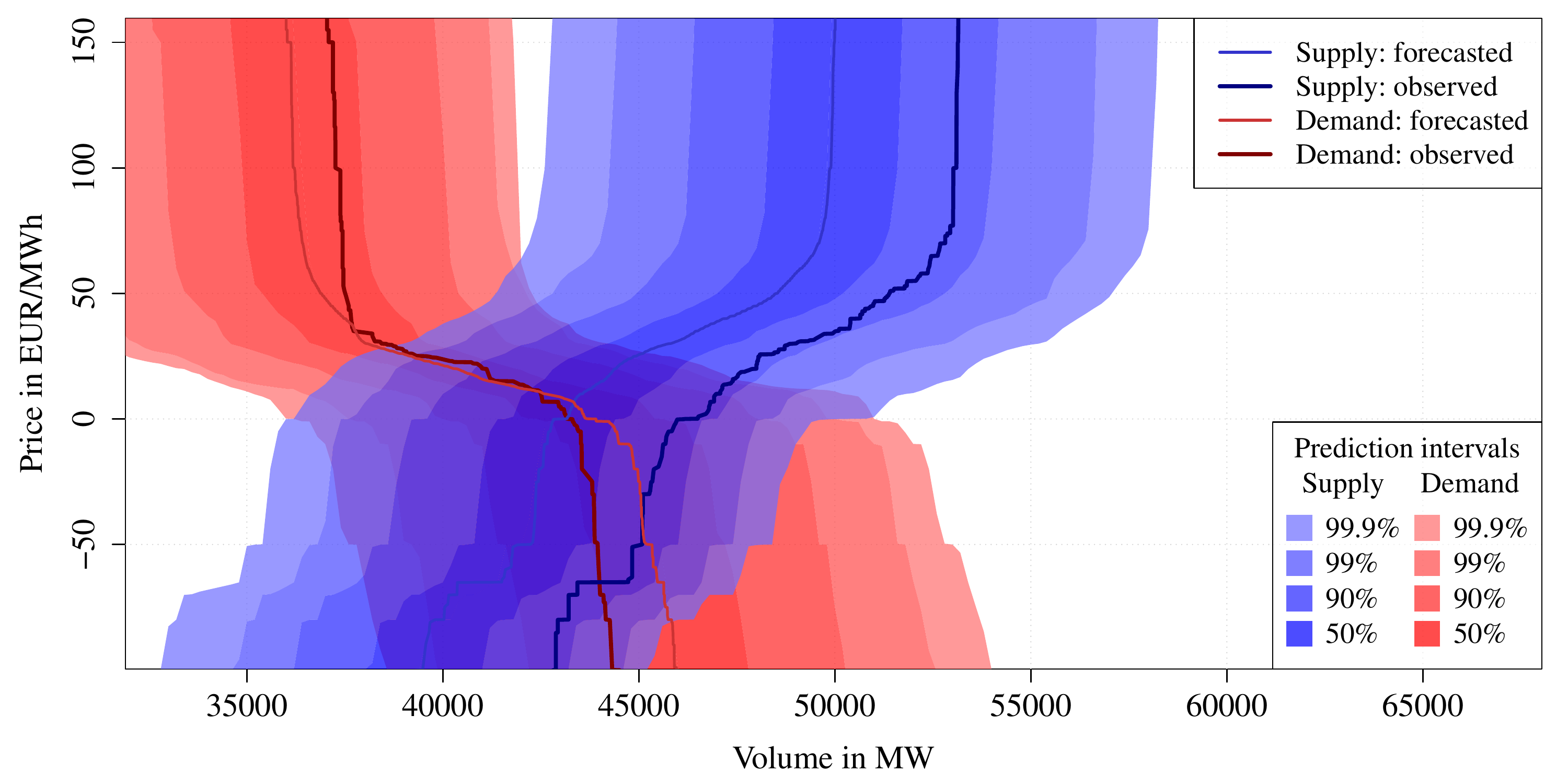} 
  \caption{13:00}
  \label{fig_dc_2}
\end{subfigure}
\begin{subfigure}[b]{.49\textwidth}
\includegraphics[width=1\textwidth, height= .6\textwidth]{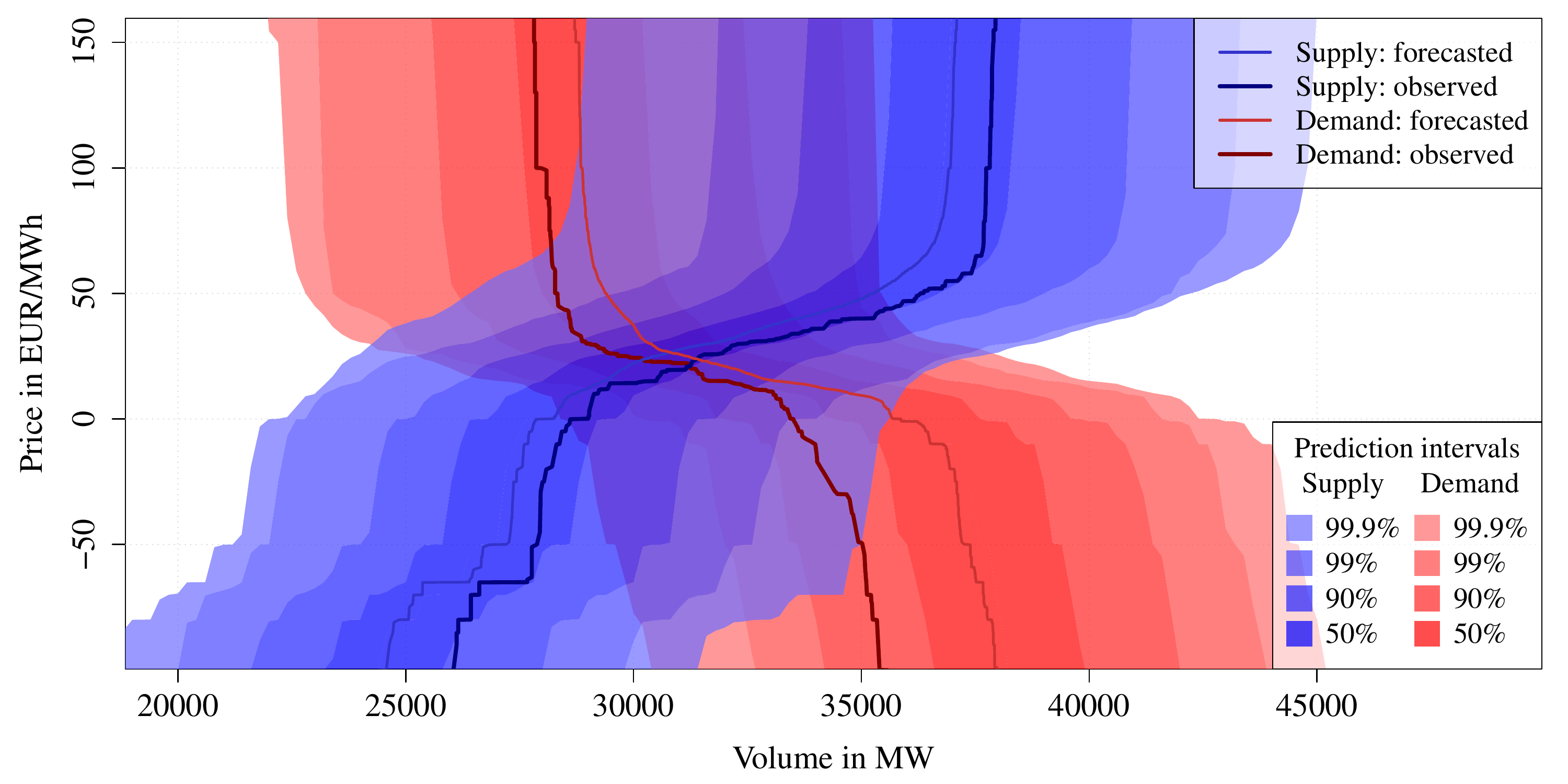}
\caption{18:00}
  \label{fig_dc_3}
\end{subfigure}
\begin{subfigure}[b]{.49\textwidth}
 \includegraphics[width=1\textwidth, height= .6\textwidth]{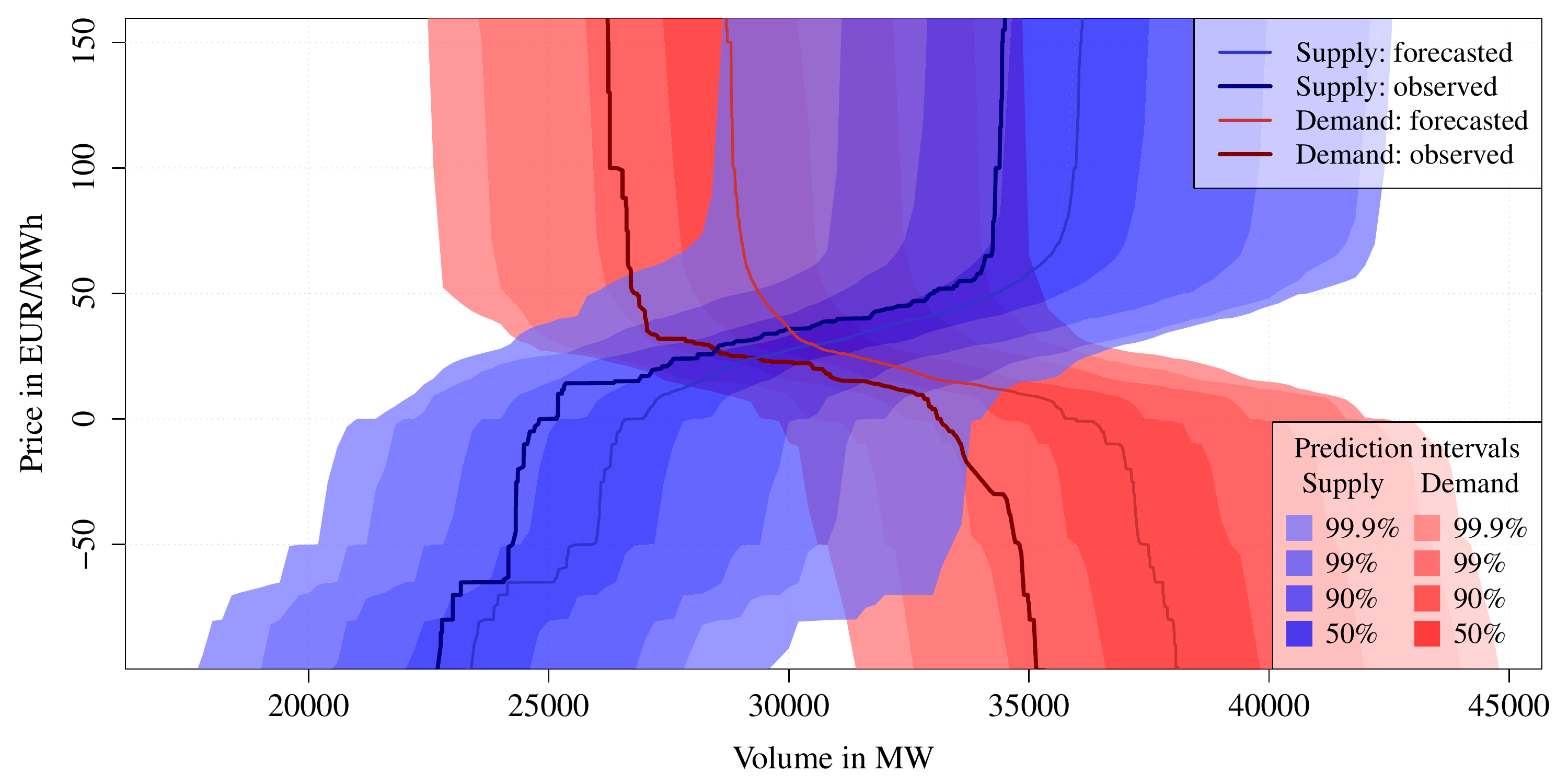} 
  \caption{19:00}
  \label{fig_dc_4}
\end{subfigure}
\caption{Supply and demand curves forecasts with prediction bands for the 12.04.2015 and selected hours.}
 \label{fig_forecast_curves}
\end{figure}

 Moreover, we are able to perform forecasts and compute prediction intervals for the full price curves.
 In Figure \ref{fig_forecast_curves} we exemplarily plot the forecast for the four selected hours that we discussed in the introduction.
Figures \ref{fig_dc_1} and \ref{fig_dc_2} show a forecast at 12:00 and 13:00 where the realized price 
 dropped from -4.96 to -65.06. 
 In Figures \ref{fig_dc_3} and \ref{fig_dc_4} we have the price curves at 19:00 and 20:00 where the market clearing price increased from 27.92 to the hightest value of that day 31.93.
 Remember that in the 12:00 and 13:00 case in \ref{fig_dmmP_3} the prediction densities of the market clearing price were highly left skewed
  and in the 19:00 and 20:00 case relatively symmetric.
Both graphs of \ref{fig_forecast_curves} show additionally to the forecasted price curves with its prediction intervals
the realized supply and demand curves of the actual auction. Note that we only show the most relevant price region between -100 and 150.

For the 13:00 case with the clear negative price, the observed demand and supply curves lie
within the relatively narrow {$90\%$ prediction bands of both curves}. 
But this does not mean that the market clearing price lies in the $90\%$ prediction interval as well. 
The reason is that both confidence intervals have a complex dependence structure. 
In fact, the observed price in \ref{fig_dc_2} lies {only in the $99\%$ prediction interval in Figure \ref{fig_dmmP_3}}.
However, we can see quite well that the predicted intersection is at a region where both, the 
supply and demand curve, have a relatively large absolute slope. The magnitude of the slope even increases for more negative values.
This is the reason for the clearly left-skewed prediction density, because a relatively moderate 
increase in the supply curve or decrease in the demand curve causes relatively large price movements. In economic terms this coincides with a situation where
both the supply and the demand side is relatively inelastic in the negative price region. This induces high price volatility.

At 19:00, where the market clearing price turned out to be relatively high, the prediction intervals look similar in general, but also have important differences in the detail.
Here the demand side is still elastic at the region close to the market clearing price. However, for a price level of around 30 to 40 the demand side also seems to loose elasticity quite dramatically. The supply side is still quite elastic up to a price region of 50 to 60. 
Small volume shocks in the bidding structure are therefore likely to be compensated by the elastic supply resulting in a high likelihood of small price changes to occur. However, for medium to large volume shocks the intersection might be shifted out of the quite elastic area of the demand curve. This is detected by our model and indicated by a large volatility with a clear right-skewness of the confidence interval, as can be also obtained from Figure \ref{fig_dmmP_1}.

Now we want to compare the point forecast of the proposed X-Model
in the out-of-sample region from 01.11.2014 to 19.04.2015 with several common benchmarks.
Even though the model is primarily designed to detect and model extreme price events with the corresponding prediction densities, it is interesting to see the performance purely based on standard error measures in comparison to other established electricity models.

Denote $\what{X}_{\text{price},d, h}$ the predicted point forecast of a electricity price model at day $d$ and hour $h$
that corresponds to $X_{\text{price},d, h}$.
Further we denote by $\DD$ the set of all days from 01.11.2014 to 19.04.2015, except for 
the 29.03.2015 which we ignore here as it is the day where the time was switched due to daylight saving time. So $\DD$ contains
in total $\#(\DD) = 169$ days.
We define the common error measures, e.g. the absolute mean absolute error ($\text{MAE}_h$) at hour $h$ and the root mean square error  
($\text{RMSE}_h$) at hour $h$ by
\begin{align*}
 \text{MAE}_h &= \frac{1}{\#(\DD)} \sum_{d\in \DD} |X_{\text{price},d, h} - \what{X}_{\text{price},d, h}|,  \\ 
 \text{RMSE}_h &= \sqrt{ \frac{1}{\#(\DD)} \sum_{d\in \DD} |X_{\text{price},d, h} - \what{X}_{\text{price},d, h}|^2  } .
\end{align*}
Both measures are suitable to compare point-forecasts of different models at a certain $h$. 
Similarly to the $\text{MAE}_h$ and $\text{RMSE}_h$, we define the overall MAE and RMSE by
\begin{align*}
 \text{MAE} &= \frac{1}{24\#(\DD)} \sum_{d\in \DD} \sum_{h=0}^{23} |X_{\text{price},d, h} - \what{X}_{\text{price},d, h}|  , \\
 \text{RMSE} &= \sqrt{ \frac{1}{24\#(\DD)} \sum_{d\in \DD} \sum_{h=0}^{23} |X_{\text{price},d, h} - \what{X}_{\text{price},d, h}|^2  } .
\end{align*}
In general the MAE is more robust than the RMSE, as the latter is by far more sensitive to outliers.

The first two benchmarks we consider are the persistent model (Persistent) given in equation \eqref{eq_bench_naiv} and the 
regime switching model as presented in equation \eqref{eq_bench_rsm} with $s_{\max} =2$ (Regime).
The next simple benchmark that we consider is a very powerful one in terms of MAE. It uses different information as our model, namely the electricity 
price from the Energy Exchange Austria (EXAA). This is an electricity price for Germany and Austria with the same zones for physical 
settlement as the German and Austrian EPEX spot price. It is traded 
everyday at 10:12 and the prices are known at 10:20 for market participants, which means they are especially known in advance to the 
EPEX auction at 12:00. 
\cite{ziel2015forecasting} show that 
the very simple naive estimator $\what{X}_{d,h}^{\text{price}} = X_{d,h}^{\text{EXAA,price}}$ with 
$X_{d,h}^{\text{EXAA,price}}$ as EXAA electricity price at day $d$ and hour $h$ is very competitive.
However, the EXAA benchmark model (EXAA) is basically beyond the competition, as it uses information which we did not explicitly include in our X-Model. But still, it can help to gain insights about possible improvements.
Furthermore, we introduce two AR($p$) based models, namely a univariate on $X_{\text{price},t}$ (AR(p))
and a 24-dimensional model with 24 simple univariate AR models on $X^{\text{price}}_{d, h}$ for each hour $h$ (24-dim. AR).
They are formally defined by
\begin{align*}
 X_{\text{price},t} &= \phi_0 + \sum_{k=1}^p \phi_k X_{\text{price},t-k} + \eps_t  \ \text{ with } \ \eps_{t} \stackrel{\text{iid}}{\sim} \NN(0, \sigma^2) ,\\ 
 X_{\text{price},d, h} &= \phi_{h,0} + \sum_{k=1}^{p_h} \phi_{h,k} X_{\text{price},d-k, h} + \eps_{d,h} 
 \ \text{ with } \ \eps_{d,h} \stackrel{\text{iid}}{\sim} \NN(0, \sigma_h^2) .
\end{align*}
We estimate the AR models by solving the Yule-Walker equations. The optimal orders $p$ and $p_h$ are determined by
minimizing the Akaike Information Criterion (AIC) on a grid of possible orders. 
For the univariate AR model we search the optimal $p$ on $\{1,2,\ldots, 700\}$ 
which allows for dependencies of more than four weeks. For the 24-dimensional model
the optimal order $p_h$ is searched on $\{1,2, \ldots, 50\}$, which allows for a memory of up to seven weeks and one day. 

Furthermore, we consider two more models from the literature, 
a wavelet based model 
and a more advanced time series approach.
The wavelet based approached is basically the popular wavelet-ARIMA model introduced by \cite{conejo2005day}.
We use Daubechies 4 wavelet decomposition and model the coefficients of the wavelet decomposition by an ARIMA(12,1,1).
The second benchmark model is a time series based approach that is analyzed by \cite{keles2012comparison}.
We select the ARMA(5,1) model with a trend component as well as their sophisticated annual, weekly and daily seasonal components. 
The model is suggested as one  of the best models in the comparison study by \cite{keles2012comparison}.
We refer to the two models as Conejo et al. and Keles et al. respectively.

\begin{figure}[htb!]
\centering
\begin{subfigure}[b]{.49\textwidth}
\includegraphics[width=1\textwidth]{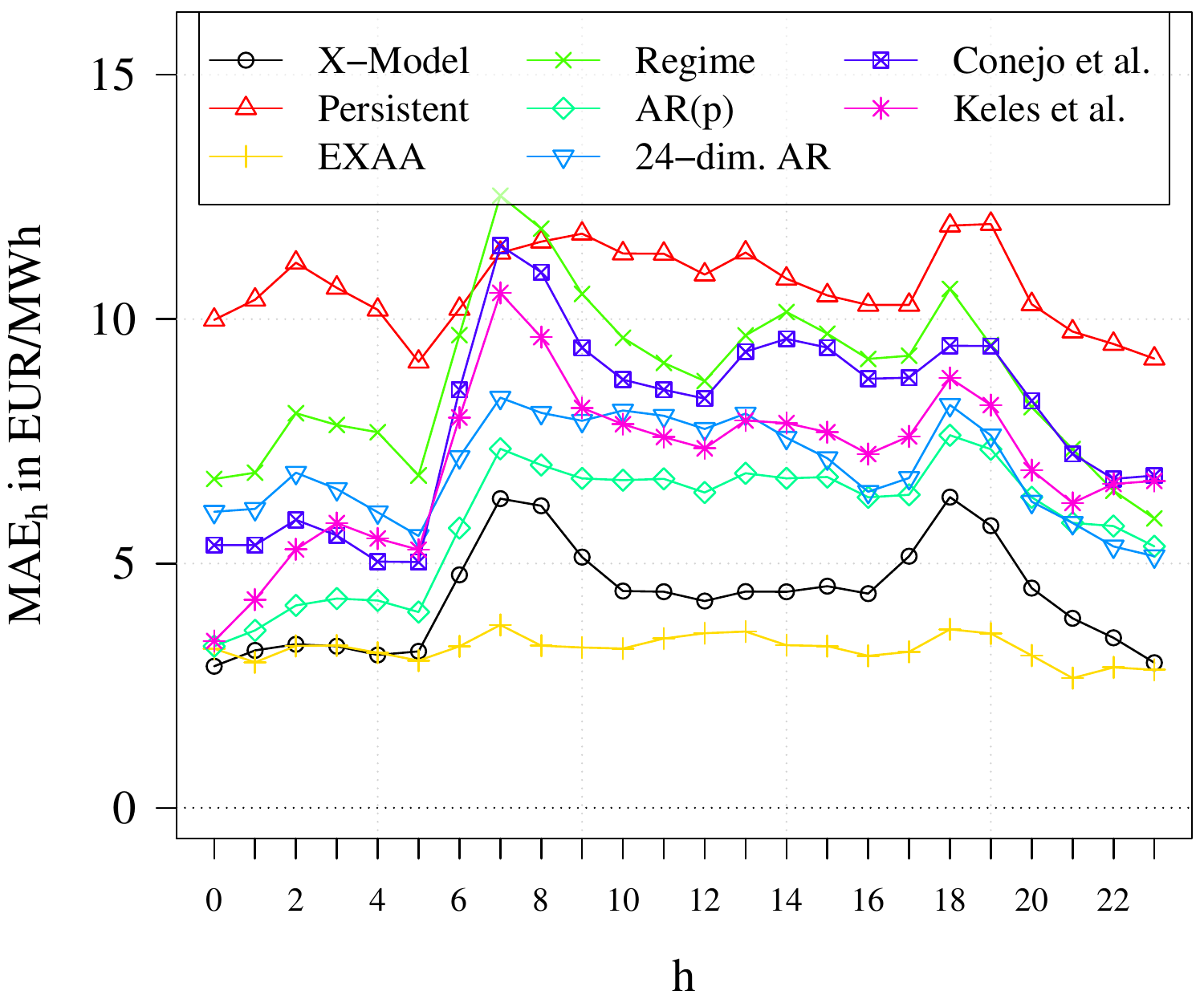}
\caption{$\text{MAE}_h$}
  \label{fig_mae}
\end{subfigure}
\begin{subfigure}[b]{.49\textwidth}
 \includegraphics[width=1\textwidth]{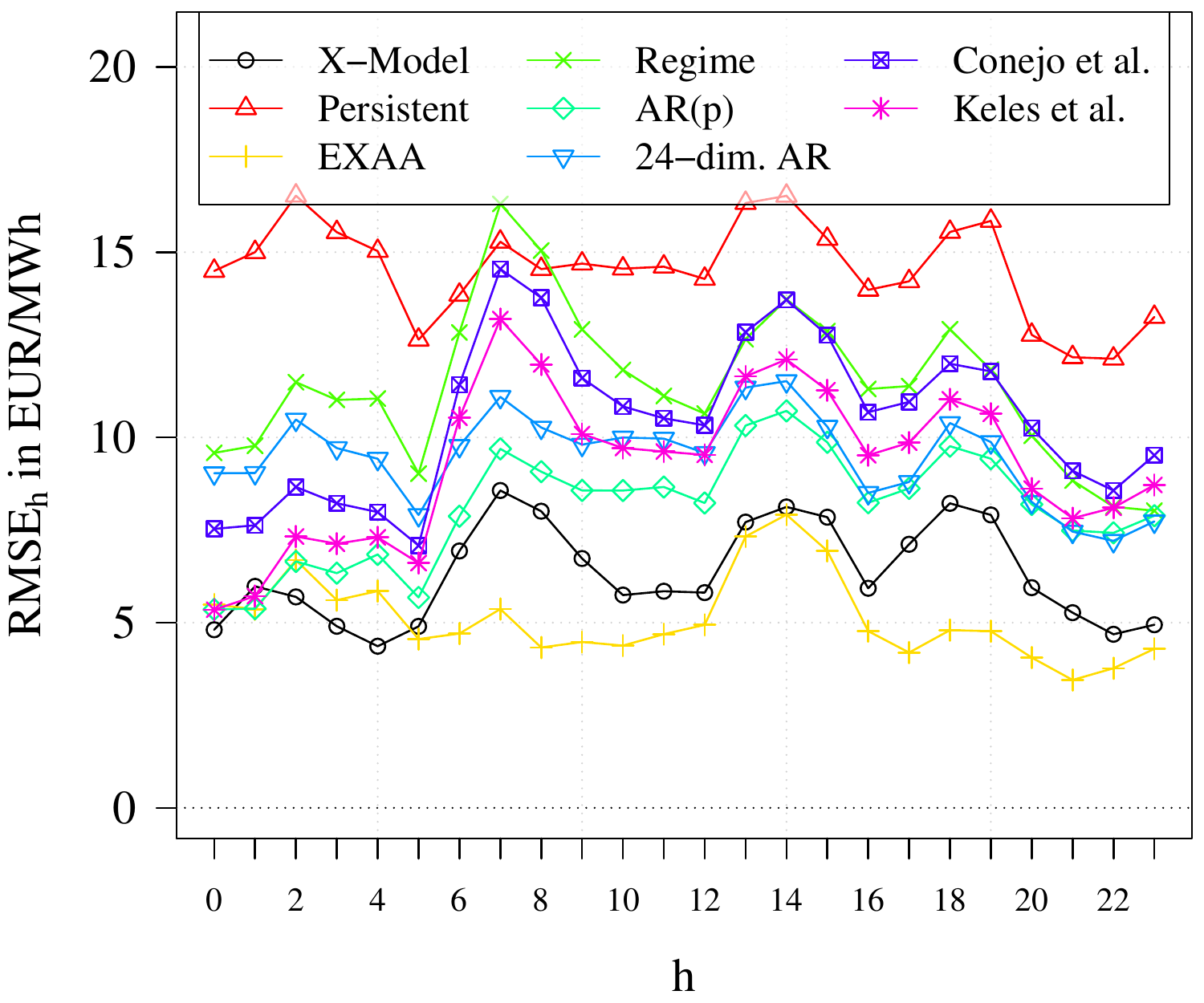} 
  \caption{$\text{RMSE}_h$}
  \label{fig_rmse}
\end{subfigure}
\caption{$\text{MAE}_h$ and $\text{RMSE}_h$ for $h\in \{0, \ldots, 23\}$ for the considered models.}
 \label{fig_mae_rmse}
\end{figure}

The estimated MAE and RMSE values of all considered models with their estimated standard deviations are given in Table \ref{tab_mae_rmse}.
The hourly $\text{MAE}_h$ and $\text{RMSE}_h$ for all models are visualized in Figure \ref{fig_mae_rmse}.

\begin{table}[ht]
\centering
\begin{tabular}{r|rr |r r}
 Models & MAE (std.dev.) & \% of persistent & RMSE (std.dev.) & \% of persistent\\ 
  \hline
X-Model & 4.35 (0.076) & 40.8 & 6.46 (0.217) & 44.3 \\ 
  Persistent & 10.66 (0.159) & 100.0 & 14.60 (0.240) & 100.0 \\ 
  Regime & 8.83 (0.117) & 82.9 & 11.60 (0.197) & 79.5 \\ 
  EXAA & 3.26 (0.065) & 30.6 & 5.23 (0.303) & 35.8 \\ 
 AR(p) & 5.91 (0.090) & 55.4 & 8.25 (0.222) & 56.5 \\ 
  24-dim. AR & 6.96 (0.103) & 65.3 & 9.55 (0.219) & 65.4 \\ 
  Conejo et al. & 8.02 (0.112) & 75.3 & 10.72 (0.213) & 73.4 \\ 
  Keles et al. & 7.11 (0.099) & 66.7 & 9.53 (0.219) & 65.3 \\ 
\end{tabular}
\caption{MAE and RMSE in EUR/MWh of the X-Model and several benchmark models}
\label{tab_mae_rmse}
\end{table}

There we observe that the proposed X-Model performs surprisingly well,
even though does not directly model the electricity price. 
With an MAE of 4.35 and RMSE of 6.46 it clearly outperforms all considered models, except the EXAA model with an
MAE of 3.26 and RMSE of 5.23. In the night hour from 0:00 to 5:00 as well at 23:00 the X-Model seems {to be at the same error magnitude as the 
EXAA-model and sometimes} outperforms every other model under consideration.

The out-of-sample MAE proportion of the X-Model in comparison to the persistent model is 
about $40.6\%$. The second best model which uses the same information as the X-Model is the AIC-selected univariate AR with 
a relative MAE proportion of $55.4\%$. Here the MAE is in absolute value 1.58 larger than the MAE of the X-Model.

\section{Summary and conclusion}

We present a model for the day-ahead electricity spot price by directly modeling the supply and demand curves. We call our model the X-Model, as we estimate the market clearing price as the intersection of the sale and purchase curve of the German-Austrian day-ahead electricity market of the EPEX.
Simple dimension reduction techniques and high-dimensional statistical methods allow us to deal with the huge amount of bid data.
We group the possible bid prices to price classes and assume a linear model for the bid volume {for each price class}.
Afterwards we forecast the bid volumes in the price classes, reconstruct the sale and purchase curves and receive the corresponding 
market clearing price.

Our empirical results show that it is possible to model the electricity prices using such an approach in a very promising way. We can capture known stylized effects of the electricity price, like daily and weekly seasonalities, very well and are also able to model the newly elaborated stylized facts of price bids. The complex bidding structure for day-ahead prices allows us to model and predict extreme and rare price events by estimating realistic prediction densities for the market clearing price. The conducted out-of-sample study shows
that the introduced model clearly outperforms standard methods and even very well performing methods of the recent literature in terms of densities as well as error measures like MAE and RMSE. 
Especially the latter was stunning and remarkable to us, as the model approach is relatively simple in its core and mainly developed for the purpose of modeling extreme price events.

The provided X-Model approach opens the door to many other different applications, especially those related to policy making. One very important issue is for example the impact of market regularizations. Many
countries provide subsidies for renewable energy. This causes automatically the so called merit-order effect on
the corresponding electricity markets. There are many papers (e.g. \cite{sensfuss2008merit}, \cite{mcconnell2013retrospective}, \cite{cludius2014merit}, \cite{dillig2016impact}) that aim for estimating these effects.
With a sell and purchase curve based approach we can directly model the impact of the renewable energy. The only condition which must be met is the availability of data, like for the German and Austrian market. The advantage is that the sale and purchase curve based approach directly takes
into account the market behavior with all its complex dependencies and non-linear properties. Common model approach are hardly able to cover such behavior. 

Another important application could be the evaluation of the price effect by closing a power plant, e.g. due to a phase-out of nuclear or lignite based power plants.
By proposing just a few assumptions for the bidding behavior and the fuel costs it is possible to postulate a proper model for the electricity market.
This would allow the researcher to get realistic price forecasts that can be utilized from decision-makers.
Note that such forecasts could be achieved for more than one day ahead. Given a proper model design even long run studies of several months and years are possible. This could be combined with different scenarios for related indicators like GDP growth or fuel costs.

Moreover, the paper with its proposed model can support the dialog of two model disciplines in electricity price modeling. 
At the moment there are classical statistical, time series and machine learning techniques that forecast the market clearing price based on observations and related time series. 
The other model approaches are {mainly fundamental or multi-agent based} electricity price models, which analyze the electricity market from a theoretical point of view {and usually ignore real auction data. Even though both disciplines may differ in their targeted goal of e.g. forecasting the electricity price versus understanding the market relationships, they have a major similarity. Overall they are both modeling the electricity price and aiming to approximate it as close as possible - they just take different perspectives on it. Our approach, which is indeed based on econometric approaches, took one step towards the fundamental way of modeling and was therefore able to gain new insights which were crucial for our approach. Hence, we are convinced that} this paper may provide a good starting point for increased communication between representatives of both model disciplines. 

Future research should also improve the considered model for the time series of bids. 
First of all {our simplification of linear interpolating the bids without explicitly including more complex bids like block bids should be replaced by a more realistic algorithm which provides a closer approximation to the EUPHEMIA (see e.g. \cite{dourbois2015european}).} {In this sense, the market coupling in Europe as well as the influence of import and export to electricity prices can be incorporated as well (see e.g. \cite{wehinger2013modeling}).} Also, more investigation could also be done in terms of optimizing the way of classifying the bids. Applying other methods of dimension reduction techniques for the bid data might grant great improvements as well. 

Another important issue concerns other relevant data used for modeling {the bid volume of} the price classes. For example, we ignored so far the impact of public holidays.
On holidays like Christmas Eve, Christmas Day or New Year's Day 
the model performs relatively poorly. Here improvement is relatively easy possible.
Moreover, the inclusion of market price time series of different markets like the intra-day price as well as auction results of related markets,
such as those from neighboring countries, could be beneficiary for the model quality. Other useful regressors could be different fuel costs or $\text{CO}_2$ allowances.
Also the restructuring procedure that was used for mapping the local price behavior provides a lot of space for further improvement. The probabilities that a certain price is traded or not could be modeled time-varying.

 \clearpage
\bibliographystyle{apalike}
\bibliography{sd-model}

\begin{thebibliography}{}

\bibitem[Aneiros et~al., 2013]{aneiros2013functional}
Aneiros, G., Vilar, J.~M., Cao, R., and San~Roque, A.~M. (2013).
\newblock Functional prediction for the residual demand in electricity spot
  markets.
\newblock {\em IEEE Transactions on Power Systems}, 28(4):4201--4208.

\bibitem[Barlow, 2002]{barlow2002diffusion}
Barlow, M.~T. (2002).
\newblock A diffusion model for electricity prices.
\newblock {\em Mathematical Finance}, 12(4):287--298.

\bibitem[Boogert and Dupont, 2008]{boogert2008supply}
Boogert, A. and Dupont, D. (2008).
\newblock When supply meets demand: the case of hourly spot electricity prices.
\newblock {\em Power Systems, IEEE Transactions on}, 23(2):389--398.

\bibitem[Bowden and Payne, 2008]{bowden2008short}
Bowden, N. and Payne, J.~E. (2008).
\newblock {Short term forecasting of electricity prices for MISO hubs: Evidence
  from ARIMA-EGARCH models}.
\newblock {\em Energy Economics}, 30(6):3186--3197.

\bibitem[Bowsher, 2004]{bowsher2004modelling}
Bowsher, C.~G. (2004).
\newblock {Modelling the dynamics of cross-sectional price functions: an
  econometric analysis of the bid and ask curves of an automated exchange}.
\newblock Technical report, Nuffield Economics Working Paper.

\bibitem[Buzoianu et~al., 2005]{buzoianu2005dynamic}
Buzoianu, M., Brockwell, A., and Seppi, D.~J. (2005).
\newblock A dynamic supply-demand model for electricity prices.
\newblock Technical report, Carnegie Mellon University.

\bibitem[Carmona and Coulon, 2014]{carmona2014survey}
Carmona, R. and Coulon, M. (2014).
\newblock A survey of commodity markets and structural models for electricity
  prices.
\newblock In {\em Quantitative Energy Finance}, pages 41--83. Springer.

\bibitem[Carmona et~al., 2013]{carmona2013electricity}
Carmona, R., Coulon, M., and Schwarz, D. (2013).
\newblock {Electricity price modeling and asset valuation: a multi-fuel
  structural approach}.
\newblock {\em Mathematics and Financial Economics}, 7(2):167--202.

\bibitem[Christensen et~al., 2012]{christensen2012forecasting}
Christensen, T., Hurn, A., and Lindsay, K. (2012).
\newblock {Forecasting spikes in electricity prices}.
\newblock {\em International Journal of Forecasting}, 28(2):400--411.

\bibitem[Cludius et~al., 2014]{cludius2014merit}
Cludius, J., Hermann, H., Matthes, F.~C., and Graichen, V. (2014).
\newblock {The merit order effect of wind and photovoltaic electricity
  generation in Germany 2008--2016: Estimation and distributional
  implications}.
\newblock {\em Energy Economics}, 44:302--313.

\bibitem[Conejo et~al., 2005]{conejo2005day}
Conejo, A.~J., Plazas, M.~A., Espinola, R., and Molina, A.~B. (2005).
\newblock {Day-ahead electricity price forecasting using the wavelet transform
  and ARIMA models}.
\newblock {\em IEEE Transactions on Power Systems}, 20(2):1035--1042.

\bibitem[Coulon et~al., 2014]{coulon2014hourly}
Coulon, M., Jacobsson, C., and Str{\"o}jby, J. (2014).
\newblock Hourly resolution forward curves for power: statistical modeling
  meets market fundamentals.
\newblock In {\em Energy pricing models: recent advances, methods and tools}.
  Palgrave Macmillan.

\bibitem[Dillig et~al., 2016]{dillig2016impact}
Dillig, M., Jung, M., and Karl, J. (2016).
\newblock The impact of renewables on electricity prices in germany--an
  estimation based on historic spot prices in the years 2011--2013.
\newblock {\em Renewable and Sustainable Energy Reviews}, 57:7--15.

\bibitem[Dourbois and Biskas, 2015]{dourbois2015european}
Dourbois, G.~A. and Biskas, P.~N. (2015).
\newblock European market coupling algorithm incorporating clearing conditions
  of block and complex orders.
\newblock In {\em PowerTech, 2015 IEEE Eindhoven}, pages 1--6. IEEE.

\bibitem[Eichler et~al., 2012]{eichler2012new}
Eichler, M., Sollie, J., and Tuerk, D. (2012).
\newblock {A New Approach for Modelling Electricity Spot Prices Based on Supply
  and Demand Spreads}.
\newblock In {\em Conference on Energy Finance 2012, Trondheim, Norway}, pages
  1--4.

\bibitem[Eichler and Tuerk, 2013]{eichler2013fitting}
Eichler, M. and Tuerk, D. (2013).
\newblock {Fitting semiparametric Markov regime-switching models to electricity
  spot prices}.
\newblock {\em Energy Economics}, 36:614--624.

\bibitem[Escribano et~al., 2011]{escribano2011modelling}
Escribano, A., Ignacio~Pe{\~n}a, J., and Villaplana, P. (2011).
\newblock {Modelling Electricity Prices: International Evidence}.
\newblock {\em Oxford Bulletin of Economics and Statistics}, 73(5):622--650.

\bibitem[Eydeland and Wolyniec, 2003]{eydeland2003energy}
Eydeland, A. and Wolyniec, K. (2003).
\newblock {\em {Energy and Power Risk Management: New Developments in Modeling,
  Pricing, and Hedging}}.
\newblock John Wiley \& Sons, NJ, USA.

\bibitem[Friedman et~al., 2007]{friedman2007pathwise}
Friedman, J., Hastie, T., H{\"o}fling, H., and Tibshirani, R. (2007).
\newblock {Pathwise coordinate optimization}.
\newblock {\em The Annals of Applied Statistics}, 1(2):302--332.

\bibitem[Friedman et~al., 2010]{friedman2010regularization}
Friedman, J., Hastie, T., and Tibshirani, R. (2010).
\newblock Regularization paths for generalized linear models via coordinate
  descent.
\newblock {\em Journal of statistical software}, 33(1):1.

\bibitem[Gaillard et~al., 2016]{gaillard2016additive}
Gaillard, P., Goude, Y., and Nedellec, R. (2016).
\newblock Additive models and robust aggregation for gefcom2014 probabilistic
  electric load and electricity price forecasting.
\newblock {\em International Journal of Forecasting}.

\bibitem[Hastie et~al., 2015]{hastie2015statistical}
Hastie, T., Tibshirani, R., and Wainwright, M. (2015).
\newblock {\em Statistical learning with sparsity: the lasso and
  generalizations}.
\newblock CRC Press.

\bibitem[Hendricks and Ehrhardt, 2013]{hendricks2013clean}
Hendricks, C. and Ehrhardt, M. (2013).
\newblock Clean spread options in the german electricity market.
\newblock {\em Working Paper, Bergische Universit\"at Wuppertal}.

\bibitem[Hildmann et~al., 2015]{hildmann2015empirical}
Hildmann, M., Ulbig, A., and Andersson, G. (2015).
\newblock Empirical analysis of the merit-order effect and the missing money
  problem in power markets with high res shares.
\newblock {\em Power Systems, IEEE Transactions on}, 30(3):1560--1570.

\bibitem[Hirth, 2013]{hirth2013market}
Hirth, L. (2013).
\newblock The market value of variable renewables: The effect of solar wind
  power variability on their relative price.
\newblock {\em Energy economics}, 38:218--236.

\bibitem[Hortacsu and Puller, 2008]{hortacsu2008understanding}
Hortacsu, A. and Puller, S.~L. (2008).
\newblock Understanding strategic bidding in multi-unit auctions: a case study
  of the texas electricity spot market.
\newblock {\em The RAND Journal of Economics}, 39(1):86--114.

\bibitem[Howison and Coulon, 2009]{howison2009stochastic}
Howison, S. and Coulon, M.~C. (2009).
\newblock {Stochastic behaviour of the electricity bid stack: from fundamental
  drivers to power prices}.
\newblock {\em The Journal of Energy Markets}, 2(1):1--27.

\bibitem[Janczura and Weron, 2012]{janczura2012efficient}
Janczura, J. and Weron, R. (2012).
\newblock {Efficient estimation of Markov regime-switching models: An
  application to electricity spot prices}.
\newblock {\em AStA Advances in Statistical Analysis}, 96(3):385--407.

\bibitem[Karakatsani and Bunn, 2008]{karakatsani2008forecasting}
Karakatsani, N.~V. and Bunn, D.~W. (2008).
\newblock {Forecasting electricity prices: The impact of fundamentals and
  time-varying coefficients}.
\newblock {\em International Journal of Forecasting}, 24(4):764--785.

\bibitem[Keles et~al., 2012]{keles2012comparison}
Keles, D., Genoese, M., M{\"o}st, D., and Fichtner, W. (2012).
\newblock {Comparison of extended mean-reversion and time series models for
  electricity spot price simulation considering negative prices}.
\newblock {\em Energy Economics}, 34(4):1012--1032.

\bibitem[Ketterer, 2014]{ketterer2014impact}
Ketterer, J.~C. (2014).
\newblock {The impact of wind power generation on the electricity price in
  Germany}.
\newblock {\em Energy Economics}, 44:270--280.

\bibitem[Liu and Shi, 2013]{liu2013applying}
Liu, H. and Shi, J. (2013).
\newblock {Applying ARMA--GARCH approaches to forecasting short-term
  electricity prices}.
\newblock {\em Energy Economics}, 37:152--166.

\bibitem[Liu et~al., 2012]{liu2012multi}
Liu, Z., Yan, J., Shi, Y., Zhu, K., and Pu, G. (2012).
\newblock {Multi-agent based experimental analysis on bidding mechanism in
  electricity auction markets}.
\newblock {\em International Journal of Electrical Power \& Energy Systems},
  43(1):696--702.

\bibitem[Lockhart et~al., 2014]{lockhart2014significance}
Lockhart, R., Taylor, J., Tibshirani, R.~J., and Tibshirani, R. (2014).
\newblock A significance test for the lasso.
\newblock {\em Annals of statistics}, 42(2):413.

\bibitem[Ludwig et~al., 2015]{ludwig2015putting}
Ludwig, N., Feuerriegel, S., and Neumann, D. (2015).
\newblock Putting big data analytics to work: Feature selection for forecasting
  electricity prices using the lasso and random forests.
\newblock {\em Journal of Decision Systems}, 24(1):19--36.

\bibitem[Maciejowska et~al., 2016]{maciejowska2016probabilistic}
Maciejowska, K., Nowotarski, J., and Weron, R. (2016).
\newblock Probabilistic forecasting of electricity spot prices using factor
  quantile regression averaging.
\newblock {\em International Journal of Forecasting}, 32(3):957--965.

\bibitem[McConnell et~al., 2013]{mcconnell2013retrospective}
McConnell, D., Hearps, P., Eales, D., Sandiford, M., Dunn, R., Wright, M., and
  Bateman, L. (2013).
\newblock {Retrospective modeling of the merit-order effect on wholesale
  electricity prices from distributed photovoltaic generation in the Australian
  National Electricity Market}.
\newblock {\em Energy Policy}, 58:17--27.

\bibitem[Portela et~al., 2016]{portela2016residual}
Portela, J., Munoz, A., Sanchez-Ubeda, E., Garcia-Gonzalez, J., and Gonzalez,
  R. (2016).
\newblock Residual demand curves for modeling the effect of complex offering
  conditions on day-ahead electricity markets.
\newblock {\em Power Systems, IEEE Transactions on}.

\bibitem[Sensfu{\ss} et~al., 2008]{sensfuss2008merit}
Sensfu{\ss}, F., Ragwitz, M., and Genoese, M. (2008).
\newblock {The merit-order effect: A detailed analysis of the price effect of
  renewable electricity generation on spot market prices in Germany}.
\newblock {\em Energy Policy}, 36(8):3086--3094.

\bibitem[Swider and Weber, 2007]{swider2007extended}
Swider, D.~J. and Weber, C. (2007).
\newblock {Extended ARMA models for estimating price developments on day-ahead
  electricity markets}.
\newblock {\em Electric Power Systems Research}, 77(5):583--593.

\bibitem[Tibshirani, 1996]{tibshirani1996regression}
Tibshirani, R. (1996).
\newblock {Regression shrinkage and selection via the lasso}.
\newblock {\em Journal of the Royal Statistical Society: Series B (Statistical
  Methodology)}, 58(1):267--288.

\bibitem[V{\'a}zquez et~al., 2014]{vazquez2014residual}
V{\'a}zquez, S., Rodilla, P., and Batlle, C. (2014).
\newblock Residual demand models for strategic bidding in european power
  exchanges: revisiting the methodology in the presence of a large penetration
  of renewables.
\newblock {\em Electric power systems research}, 108:178--184.

\bibitem[Ventosa et~al., 2005]{ventosa2005electricity}
Ventosa, M., Ba{\i}llo, A., Ramos, A., and Rivier, M. (2005).
\newblock {Electricity market modeling trends}.
\newblock {\em Energy Policy}, 33(7):897--913.

\bibitem[Wagner et~al., 2014]{wagner2014residual}
Wagner, A. et~al. (2014).
\newblock Residual demand modeling and application to electricity pricing.
\newblock {\em Energy Journal}, 35(2):45--73.

\bibitem[Wehinger et~al., 2013]{wehinger2013modeling}
Wehinger, L.~A., Hug-Glanzmann, G., Galus, M.~D., and Andersson, G. (2013).
\newblock Modeling electricity wholesale markets with model predictive and
  profit maximizing agents.
\newblock {\em Power Systems, IEEE Transactions on}, 28(2):868--876.

\bibitem[Weron, 2006]{weron2006modeling}
Weron, R. (2006).
\newblock {\em {Modeling and Forecasting Electricity Loads and Prices: A
  Statistical Approach}}.
\newblock John Wiley \& Sons, Chichester, England.

\bibitem[Weron, 2008]{weron2008market}
Weron, R. (2008).
\newblock {Market price of risk implied by Asian-style electricity options and
  futures}.
\newblock {\em Energy Economics}, 30(3):1098--1115.

\bibitem[Weron, 2014]{weron2014electricity}
Weron, R. (2014).
\newblock {Electricity price forecasting: A review of the state-of-the-art with
  a look into the future}.
\newblock {\em International Journal of Forecasting}, 30(4):1030--1081.

\bibitem[Weron and Misiorek, 2008]{weron2008forecasting}
Weron, R. and Misiorek, A. (2008).
\newblock {Forecasting spot electricity prices: A comparison of parametric and
  semiparametric time series models}.
\newblock {\em International Journal of Forecasting}, 24(4):744--763.

\bibitem[Ziel, 2016a]{ziel2016forecasting}
Ziel, F. (2016a).
\newblock Forecasting electricity spot prices using lasso: On capturing the
  autoregressive intraday structure.
\newblock {\em IEEE Transactions on Power Systems}.

\bibitem[Ziel, 2016b]{ziel2016iteratively}
Ziel, F. (2016b).
\newblock Iteratively reweighted adaptive lasso for conditional heteroscedastic
  time series with applications to ar--arch type processes.
\newblock {\em Computational Statistics \& Data Analysis}, 100:773--793.

\bibitem[Ziel and Liu, 2016]{ziel2016lasso}
Ziel, F. and Liu, B. (2016).
\newblock {Lasso estimation for GEFCom2014 probabilistic electric load
  forecasting}.
\newblock {\em International Journal of Forecasting}, 32(3):1029--1037.

\bibitem[Ziel et~al., 2015a]{ziel2015efficient}
Ziel, F., Steinert, R., and Husmann, S. (2015a).
\newblock {Efficient modeling and forecasting of electricity spot prices}.
\newblock {\em Energy Economics}, 47:98--111.

\bibitem[Ziel et~al., 2015b]{ziel2015forecasting}
Ziel, F., Steinert, R., and Husmann, S. (2015b).
\newblock {Forecasting day ahead electricity spot prices: The impact of the
  EXAA to other European electricity markets}.
\newblock {\em Energy Economics}, 51:430--444.

\end{thebibliography}

\end{document}